\documentclass[aps,pra,longbibliography,11pt,reprint,floatfix,superscriptaddress,urlname]{revtex4-2}
\usepackage{multirow}
\usepackage[utf8]{inputenc}
\usepackage{amsmath}
\usepackage{color}
\usepackage{graphicx}
\usepackage{comment}
\usepackage{caption}
\usepackage{subcaption}
\usepackage[normalem]{ulem}
\usepackage{multirow}
\usepackage[autostyle]{csquotes}

\begin{document}

\title{An Analysis of First- and Quasi-Second-Order Optimization Algorithms in Variational Monte Carlo}

\author{Ruojing Peng}
\affiliation{Division of Chemistry and Chemical Engineering, California Institute of Technology, Pasadena,
California 91125, USA}

\author{Garnet Kin-Lic Chan}
\affiliation{Division of Chemistry and Chemical Engineering, California Institute of Technology, Pasadena,
California 91125, USA}
\affiliation{Institute for Quantum Information and Matter, California Institute of Technology, Pasadena,
California 91125, USA}

\begin{abstract}
Many quantum many-body wavefunctions, such as Jastrow-Slater, tensor network, and neural quantum states, are studied with the variational Monte Carlo technique, where stochastic optimization is usually performed to obtain a faithful approximation to the ground-state of a given Hamiltonian. While first-order gradient descent methods are commonly used for such optimizations, quasi-second-order optimization formulations offer the potential of faster convergence under certain theoretical conditions, but with a similar cost per sample to first-order methods. However, the relative performance of first-order and second-order optimizers is influenced in practice by many factors, including the sampling requirements for a faithful optimization step, the influence of wavefunction quality, as well as the wavefunction parametrization and expressivity. Here we analyze these performance characteristics of first-order and quasi-second-order optimization methods for a variety of Hamiltonians, with the additional context of understanding the scaling of these methods (for good performance) as a function of system size. Our findings help clarify the role of first-order and quasi-second-order methods in variational Monte Carlo calculations and the conditions under which they should respectively be used. In particular, we find that unlike in deterministic optimization, where closeness to the variational minimum determines the suitability of second-order methods, in stochastic optimization the main factor is the overall expressivity of the wavefunction: quasi-second-order methods lead to an overall reduction in cost relative to first-order methods when the wavefunction is sufficiently expressive to represent the ground-state, even when starting far away from the ground state. This makes quasi-second-order methods an important technique when used with wavefunctions with arbitrarily improvable accuracy.
\end{abstract}

\maketitle

\section{Introduction}

The variational Monte Carlo (VMC)~\cite{Foulkes2001,becca_sorella_2017,doi:10.1021/cr2001564,https://doi.org/10.1002/wcms.40,Rubenstein2017} method forms the basis for many numerical computations with a large variety of quantum many-body parametrized states. In its simplest form, VMC replaces a deterministic expectation value computation with a stochastic estimate based on sampling the wavefunction amplitudes from the probability distribution $|\Psi(\vec{x})|^2$ (where $\Psi(\vec{x})$ denotes the wavefunction and $\vec{x}$ labels the sampling, or computational, basis). Because $\Psi(\vec{x})$ can be efficiently evaluated for a wide variety of mathematical forms, and sampling from  $|\Psi(\vec{x})|^2$ is also efficient in most applications using the Metropolis algorithm~\cite{GBhanot_1988,814660,bj/1080222083,10.1063/1.1887186,10.1214/aos/1033066201}, VMC is used to compute observables for many different variational states, including Jastrow-Slater wavefunctions~\cite{10.1063/1.1604379,PhysRevB.78.115117,10.1063/1.2743972,10.1063/1.1752881,PhysRevE.74.066701,10.1063/1.480839,umrigar2005,sorella2005,PhysRevLett.98.110201,toulouse2007,toulouse2008,Sabzevari2020}, tensor network states and their relatives~\cite{sandvik2007,schush2008,wang2011,liu2017,liu2021,xie2014,orus2009,orus2019,ORUS2014117,SCHUCH20102153,levin2007,gu2008,gu2009,jiang2008,Sfondrini2010,PhysRevX.8.011006,PhysRevB.100.155148,Mezzacapo_2009,PhysRevB.96.054405,PhysRevB.94.155120,Mezzacapo_2010,PhysRevLett.133.260404}, and neural network states~\cite{Chen2024,Rende2024,webber2022,doi:10.1126/science.aag2302,PhysRevB.106.205136,PhysRevX.7.021021,Choo2020,Zhao_2023,https://doi.org/10.1002/qute.201800077,PhysRevB.107.195115,NEURIPS2023_68efc144}, to mention only a few.

The most common application of VMC is ground-state determination where the energy $\langle \Psi|\hat{H}|\Psi\rangle/\langle \Psi|\Psi\rangle$ is minimized with respect to the wavefunction parameters. Gradient based methods, also called first-order methods as they use only first derivatives of the energy with respect to the parameters, have been adapted to such high-dimensional optimization in the presence of stochastic noise. The simplest version of this is stochastic gradient descent (SGD), which follows the (stochastically evaluated) gradient with a stochastic step size~\cite{becca_sorella_2017}. One problem of SGD is its slow convergence, especially as the wavefunction approaches the ground state. Various improved update schemes have been developed to accelerate the convergence using only first-order information, such as momentum based approaches  \cite{duchi2011,mcmahan2010,kingma2015,zeiler2012adadelta,dozat2016,reddi2019convergence,Sabzevari2018} and stochastic reconfiguration (SR) \cite{PhysRevLett.80.4558,becca_sorella_2017,sorella2001,10.1063/1.1794632,neuscamman2012,Stokes2020quantumnatural}. Nonetheless, such methods also slow down near the ground state when the gradient is inherently small. Despite their slow convergence, first-order methods remain the state-of-art in large scale applications with more than a few thousand parameters. For instance, SR has been successfully applied to quantum many body problems with hundreds of thousands of wavefunction parameters, 
where the sample size can be orders of magnitude smaller than the number of parameters~\cite{neuscamman2012,Chen2024,Rende2024}.

In deterministic non-linear optimization, the Newton update and its variants~\cite{NumericalOptimization}, also called second-order methods since they are based on second derivatives of the energy (the Hessian), are widely used to achieve fast convergence near the variational minimum. Many developments have also been made for its adaptation to VMC that aim to reduce the stochastic error in the update and improve the stability and update efficiency~\cite{10.1063/1.480839,umrigar2005,toulouse2007,sorella2005}. However, these second-order implementations have mainly been used with types of variational wavefunctions with up to a {few thousand} parameters, and usually in a setting where the number of samples exceeds the number of parameters \cite{10.1063/1.480839,umrigar2005,toulouse2007,sorella2005}. A major bottleneck in these applications is the high cost of computing (and storing) the Hessian, which requires the wavefunction second derivative.

The use of approximate Hessian information to accelerate VMC convergence, which avoids the expensive computation of the wavefunction second derivatives, has motivated the development of quasi-second-order methods, such as the Linear Method (LM)~\cite{PhysRevLett.98.110201,toulouse2007,toulouse2008,Sabzevari2020,zhao2017,becca_sorella_2017}, closely related Rayleigh-Gauss-Newton (RGN) scheme~\cite{webber2022}, and other inexact Hessian approximations, such as the Kronecker-Factored-Approximate-Curvature method~\cite{kfac,martens2020optimizingneuralnetworkskroneckerfactored,deepmindvmc}. However, their relative power compared to first-order methods is still not well understood. On the one hand, by excluding the wavefunction second derivatives, and using iterative methods~\cite{neuscamman2012,Sabzevari2020} to solve for the corresponding step, the (per sample) cost of quasi-second-order methods is in principle only a small constant multiple of that of SR. On the other hand, one may expect first- and quasi-second-order methods to have different sample size requirements for an accurate update, which could negate any improved convergence from the use of second-order information. Indeed, the cost to accurately sample the approximate Hessian has been raised as a bottleneck in prior work~\cite{becca_sorella_2017,Sabzevari2020}. 

In this work we attempt to gain an understanding of the performance of first- and quasi-second-order methods for a set of variational quantum many-body states in several different models and realistic quantum many-body problems. We focus on lattice Hamiltonians, because these have a natural scaling to larger sizes, which allows us to discuss the scaling of the optimization procedures, and we focus on wavefunctions that have a product structure, i.e. tensor network states. The latter choice was made partly to define a finite scope of study, but also because a product structure of the wavefunction is common to wavefunctions that can be scaled to large quantum systems as it is necessary to preserve the extensivity of the energy. We take SR as our representative first-order method, and RGN as our representative quasi-second-order method of study (although we study also the traditional Newton update for a subclass of problems). We study the theoretical performance of first-order and quasi-second-order methods without noise, then provide a theoretical analysis of sample size requirements, followed by an empirical numerical evaluation of the influence of wavefunction quality, wavefunction expressivity, and scaling with system size on the sample size requirements and performance of SR and RGN. 
From the results of these examples, we provide some answers to the questions we raise, summarized in recommendations on how to use first- and quasi-second-order optimization methods in VMC. 

\section{Theory and Methods}\label{sec:theory}

\subsection{Recap of ideal first- and second-order stochastic optimization methods}\label{sec:exact_formulation}

We briefly describe the first- and quasi-second-order optimization methods considered in this work, following the general formulation in Ref.~\onlinecite{webber2022}. We also assume real wavefunctions and parameters for convenience. Consider the variational energy 
\begin{align}\label{eq:variational_energy}
E=\frac{\langle\psi|\hat{H}|\psi\rangle}{\langle\psi|\psi\rangle},
\end{align}
and a small perturbation to the parameters of the current normalized wavefunction $\hat{\psi}=\psi/\langle\psi|\psi\rangle^{1/2}$,
\begin{align}\label{eq:new_wfn}
\psi_{\text{new}}
&=\hat{\psi}
+\sum_ip_i\hat{\psi}_i
\end{align}
where we note that the first derivatives of $\hat{\psi}$
\begin{align}
\hat{\psi}_i=\partial_i\hat{\psi}=\frac{\partial_i\psi}{|\psi|}-\hat{\psi}\frac{\langle\psi|\partial_i\psi\rangle}{\langle\psi|\psi\rangle}
\end{align}
are all orthogonal to $\hat{\psi}$
\begin{align}
\langle\hat{\psi}|\hat{\psi}_i\rangle=0.
\end{align}
The perturbed energy from substituting the perturbed wavefunction Eq.~\ref{eq:new_wfn} into the variational energy Eq.~\ref{eq:variational_energy} is, up to second-order in the change in parameters, 
\begin{align}\label{eq:new_energy}
E_{\text{new}}
&=E(\hat{\psi})+2p^Tg+p^THp+D(p) +  O(p^3)
\end{align}
where
\begin{align}\label{eq:grad}
g_i=\langle\hat{\psi}_i|\hat{H}|\hat{\psi}\rangle
\end{align}
is half the energy gradient and 
\begin{align}\label{eq:hess}
H_{ij}=\langle\hat{\psi}_i|\hat{H}|\hat{\psi}_j\rangle-E\langle\hat{\psi}_i|\hat{\psi}_j\rangle
\end{align}
is an approximation to half the analytical energy Hessian $\mathcal{H}$, and $D$ is a term of $O(p^2)$ which depends on the second derivatives of the wavefunction $\partial^2_{ij}\hat{\psi}$. The true Hessian $\mathcal{H}$
involves $D$, which is expensive to compute \cite{10.1063/1.480839,10.1063/1.1924690,umrigar2005}. $H$, however, does not, and we refer to it as the approximate Hessian.  

First-order optimization methods use only the first-order terms in the energy expansion Eq.~\ref{eq:new_energy}. For example, gradient descent minimizes the perturbed energy Eq.~\ref{eq:new_energy} up to first-order plus a penalty on the norm of $p$
\begin{align}
\min_{p}\left[2p^Tg+\frac{p^Tp}{\epsilon}\right]
\end{align}
or equivalently, updates the wavefunction by 
\begin{align}
p=-\epsilon g. 
\end{align}
Alternatively, one could minimize the perturbed energy Eq.~\ref{eq:new_energy} up to first-order plus a penalty
\begin{align}
\min_{p}\left[2p^Tg+\frac{p^T(S+\delta I)p}{\epsilon}\right]
\end{align}
where the penalty $p^TSp$ keeps the overlap of $\hat{\psi}$ and $\psi_{\text{new}}$ 
\begin{align}\label{eq:ovlp}
\left|\frac{\langle\hat{\psi}|\psi_{\text{new}}\rangle}{|\psi_{\text{new}}|}\right|^2=1-p^TSp+O(p^4),\quad S_{ij}=\langle\hat{\psi}_i|\hat{\psi}_j\rangle
\end{align}
sufficiently large, and a small diagonal shift $\delta$ is added to avoid singularities in $S$. Equivalently, the wavefunction is updated by
\begin{align}\label{eq:invert_sr}
p=-\epsilon (S+\delta I)^{-1}g.
\end{align}
The $S$ matrix is sometimes called the overlap, but is also referred to as the quantum information metric~\cite{Stokes2020quantumnatural}. When it is stochastically sampled, this first-order method is known as stochastic reconfiguration (SR), and it is found to be a significant improvement over gradient descent~\cite{sorella2005,10.1063/1.1794632,neuscamman2012,Stokes2020quantumnatural,PhysRevResearch.2.023232}, thus it will be the first-order method of interest in this work.

Second-order optimization methods additionally use the quadratic terms in the energy expansion Eq.~\ref{eq:new_energy}. The original second-order method is the Newton method, which takes the step
\begin{align}
    p = -\mathcal{H} g
\end{align}
This update has provably superior convergence properties near the minimum of the objective function assuming the basin is nearly quadratic~\cite{NumericalOptimization}, but away from this the step size can also diverge (or fail to decrease the energy) if the Hessian has zero (or negative) eigenvalues, thus requiring some form of step-size control. In the VMC setting, early applications of the Newton optimization scheme can be found in Refs.~\cite{10.1063/1.480839,umrigar2005,toulouse2007,sorella2005}.  

Because of the cost and complexity of evaluating $D$ in Eq.~\ref{eq:new_energy}, the Rayleigh-Gauss-Newton (RGN) method utilizes $H$ rather than $\mathcal{H}$. In particular, it minimizes the perturbed energy Eq.~\ref{eq:new_energy} up to second-order with a penalty on the overlap
\begin{align}
\min_{p}\left[2p^Tg+p^THp+\frac{p^T(S+\delta I)p}{\epsilon}\right]
\end{align}
which leads to the update vector
\begin{align}\label{eq:invert_rgn}
p=-\left(H+\frac{S+\delta I}{\epsilon}\right)^{-1}g.
\end{align}
The $\epsilon \to \infty$ limit corresponds to the Newton method with $\mathcal{H}$ approximated by $2H$. RGN will be used as the representative quasi-second-order method in our numerical studies.

A closely related earlier approach is the Linear Method (LM)\cite{PhysRevLett.98.110201,toulouse2007,toulouse2008,Umrigar2015,Sabzevari2020,zhao2017,becca_sorella_2017}, which computes the update vector $p$ by solving a generalized eigenvalue problem 
\begin{align}\label{eq:lm}
\left[\begin{array}{cc}
    0 & g^T \\
    g & H + \delta I
\end{array}\right]\left[
\begin{array}{c}
    1 \\
    p
\end{array}\right]=\lambda\left[\begin{array}{cc}
    1 & 0 \\
    0 & S
\end{array}\right]\left[
\begin{array}{c}
    1 \\
    p
\end{array}\right]
\end{align}
in the linear subspace spanned by the semi-orthogonal basis $\{\hat{\psi},\hat{\psi}_i\}$, where $\lambda$ is the difference between the energy of the linear approximation (Eq.~\ref{eq:new_wfn}) and the energy of the current wavefunction. As shown in Eq. 33 of Ref.~\onlinecite{toulouse2008}, the linear method has the advantage that it can be thought of as an approximate Newton method, with a built-in stabilization that automatically vanishes at the energy minimum.
Hence the value of $\delta$ needed to provide additional stabilization can be very small.

Although quasi-second-order updates such as RGN and LM are expected to give fast convergence when the energy is well represented by the second-order expansion in Eq.~\ref{eq:new_energy} (assuming $D$ is small), they can also fail to decrease the energy, for example, when $H$ has negative eigenvalues. This can be handled by increasing the penalty parameter $\frac{1}{\epsilon}$, where it can be seen that if $\epsilon \to 0$, RGN reduces to SR, while LM reduces to gradient descent. Additional more sophisticated stepsize controls, for example, utilizing the freedom in the normalization of the wavefunction to rescale the update, or to compute optimal penalty parameters via a quadratic model, have been introduced in conjunction with the LM~\cite{PhysRevLett.98.110201,toulouse2007,toulouse2008}.
In this work, for simplicity we do not introduce these additional controls, but add the fallback that if the RGN update does not lower the energy, we use an SR update instead. 

\subsection{Stochastic formulation}\label{sec:vmc}
The VMC energy (Eq.~\ref{eq:variational_energy}), gradient (Eq.~\ref{eq:grad}), information metric (Eq.~\ref{eq:ovlp}) and approximate Hessian (Eq.~\ref{eq:hess}) are stochastically sampled according to the following expressions\cite{ceperley1977,Umrigar2015,Foulkes2001,Kolorenc2011,nightingale1998quantum,Hammond1994,becca_sorella_2017,gubernatis_kawashima_werner_2016,webber2022}
\begin{align}
&E\approx\langle E_L(\vec{x})\rangle_{|\psi^2|}\label{eq:energy_sampled}\\
&g_i\approx\text{Cov}_{|\psi^2|}\left(E_L(\vec{x}),\nu_i(\vec{x})\right)\label{eq:grad_sampled}\\
&S_{ij}\approx\text{Cov}_{|\psi^2|}\left(\nu_i(\vec{x}),\nu_j(\vec{x})\right)\label{eq:ovlp_sampled}\\
&H_{ij}\approx\text{Cov}_{|\psi^2|}\left(\nu_i(\vec{x}),h_j(\vec{x})\right)-g_i\langle\nu_j(\vec{x})\rangle_{|\psi^2|}-ES_{ij}\label{eq:hess_sampled}
\end{align}
where 
\begin{align}
&E_L(\vec{x})=\frac{\langle \vec{x}|\hat{H}|\psi\rangle}{\langle \vec{x}|\psi\rangle}\\
&\nu_i(\vec{x})=\frac{\langle \vec{x}|\partial_i\psi\rangle}{\langle \vec{x}|\psi\rangle}=\frac{\partial_i\langle \vec{x}|\psi\rangle}{\langle \vec{x}|\psi\rangle}\label{eq:nu}\\
&h_i(\vec{x})=\frac{\langle \vec{x}|\hat{H}|\partial_i\psi\rangle}{\langle \vec{x}|\psi\rangle}=\frac{\partial_i\langle \vec{x}|\hat{H}|\psi\rangle}{\langle \vec{x}|\psi\rangle}\label{eq:Hnu}.
\end{align}
and the subscript $|\psi^2|$ indicates that the $\vec{x}$ samples are drawn with respect to the probability distribution $|\psi(\vec{x})|^2$.

The vectors $\nu(\vec{x})$ and $h(\vec{x})$, whose components are given in Eq.~\ref{eq:nu} and Eq.~\ref{eq:Hnu} respectively, can be obtained with the same computational scaling as $\langle \vec{x}|\psi\rangle$ and $\langle \vec{x}|\hat{H}|\psi\rangle$. This result can be seen through the lens of automatic differentiation (AD). For each configuration $\vec{x}$, $\langle \vec{x}|\psi\rangle=f_1(\{\theta_i\})$ and $\langle \vec{x}|\hat{H}|\psi\rangle=f_2(\{\theta_i\})$ can be viewed as computational graphs where the wavefunction parameters $\{\theta_i\}$ appear on the leaf nodes. A single pass of back-propagation through each graph, which has the same scaling as the forward pass for computing the value of the graph, generates derivatives with respect to all nodes in the graph. 

The sampled $S$ and $H$ matrices in Eq.~\ref{eq:ovlp_sampled} and Eq.~\ref{eq:hess_sampled} are often singular or near singular. This can be due to linear dependence in the wavefunction parameters, for instance in over-parameterized neural network states~\cite{Rende2024,Chen2024,PhysRevResearch.2.023232,Stokes2020quantumnatural,10.21468/SciPostPhys.10.6.147}, or from the gauge freedom in tensor network states~\cite{ORUS2014117,jiang2008,wang2011,Verstraete2008,lubasch2014}. In addition, in some computations the sample size is less than the wavefunction parameter size, creating an exact rank-deficiency in the sampled $S$ and $H$ matrices. (We note that this latter problem can be removed in some recent reformulations of SR~\cite{Chen2024,Rende2024}). The small diagonal shift $\delta$ in Eq.~\ref{eq:invert_sr} and Eq.~\ref{eq:invert_rgn} can be used to remove these singularities when necessary. 

In the SR and RGN procedures we need to solve a linear equation problem to obtain the step, namely Eq.~\ref{eq:invert_sr} and Eq.~\ref{eq:invert_rgn} respectively. As discussed in Ref.~\cite{neuscamman2012}, these equations can be solved by an iterative solver, thus avoiding the cost of building the $S$ and $H$ matrices. In the following calculations, we use the MINRES~\cite{doi:10.1137/0712047,doi:10.1137/100787921} and LGMRES~\cite{doi:10.1137/0907058,doi:10.1137/S0895479803422014} algorithms to solve for the update vector in Eq.~\ref{eq:invert_sr} and Eq.~\ref{eq:invert_rgn}, respectively. In each case, we use a convergence threshold of $0.0001$ on the relative norm of the residue vector and perform a maximum of $500$ iterations. Furthermore, because $S$ and $H$ have a covariance structure, the cost of each iteration of the iterative solver is only linear in the number of variational parameters for both SR and RGN (see Appendix~\ref{sec:iterative_linea_eqn} for more details). In VMC calculations, SR (and related methods) have been applied to wavefunctions with up to $10^6$ parameters~\cite{neuscamman2012,Chen2024}.
Because of the same covariance structure, the RGN step is a small ($O(1)$) constant factor more expensive than the SR step for each sample. Thus, the sample cost of implementing the RGN update is not a problem in practice, at least up to the number of parameters that are accessible to SR (although the sample size requirements for an accurate RGN step may be different from SR).

\section{Results and Discussion}\label{sec:result}

\subsection{Questions to study}

We wish to investigate the relative performance of the first-order and quasi-second-order methods, SR and RGN, to answer some of the following questions:

\begin{enumerate}
\item The Newton update guarantees, under certain conditions, faster convergence over gradient updates. This underlies the expected convergence improvement of stabilized quasi-second-order methods, such as RGN, over first-order methods such as SR. To what extent is this fast convergence seen in real-world quantum-many body problems?

\item 
SR and RGN may have different sample size requirements to achieve an effective optimization step. What is the effect of noise on SR and RGN and what are their respective sample size requirements?

\item What is the effect of the wavefunction quality on the performance of SR and RGN? Here, two relevant facets of wavefunction quality include the closeness to the ground state (e.g. in terms of the energy) and the wavefunction expressivity (i.e. the quality of the best optimized state).

\item How do the sample size requirements and overall performance of SR and RGN scale with the system size (for problems that admit a natural scaling with system size)?

\item Is there a simple guide for when to use SR and RGN in VMC optimization?
\end{enumerate}

With respect to the above questions, we have a limited amount of theoretical information, especially with respect to the sample size requirements. 
The energy and gradient satisfy a zero-variance principle (i.e. their variance vanishes when $\psi$ is an eigenstate)~\cite{gubernatis_kawashima_werner_2016,becca_sorella_2017,webber2022} which reduces the sampling error of these inputs into SR and RGN. 
In addition, in the LM, the eigenvalue problem 
Eq.~\ref{eq:lm} using the non-Hermitian approximate Hessian Eq.~\ref{eq:hess_sampled} satisfies the strong zero-variance principle~\cite{toulouse2007,nightingale2001,neuscamman2016,becca_sorella_2017}, which means that the wavefunction is updated exactly (with zero variance) if the space spanned by $\{\hat{\psi},\hat{\psi}_i\}$ contains the ground state. It is thus reasonable to {hope} that the performance of the SR and RGN methods improves for a wavefunction of sufficient quality and expressivity. However, we expect differences in the sampling requirements for SR and RGN, because, for example, the elements in the overlap matrix which enters in SR are not extensive in the system size, while (some of) the approximate Hessian elements are. 

We thus seek to answer these questions by empirical numerical studies of three systems: (i) a non-interacting model problem with an exact simple product wavefunction ansatz where the sample variance for the quantities entering into the SR and RGN algorithms can be computed analytically, providing a foundation to understand the performance of SR and RGN in the stochastic setting, (ii) the 1D $J_1$-$J_2$ model at the exactly soluble point, where we optimize over a matrix product representation of the ground-state, which includes the exact ground-state in the optimization manifold, and (iii) the 2D $J_1$-$J_2$ model in the frustrated regime, where we use a projected entangled pair state ansatz, which for larger lattice sizes does not (effectively) include the exact ground-state in the optimization manifold.

\subsection{Model system}\label{sec:model}

We first consider a 1D model problem of $L$ sites with Hamiltonian
\begin{align}
    \hat{H}=\sum_{i=1}^{L}X_i,
\end{align}
where $X_i$ is the Pauli $X$ matrix acting on the $i$-th site. The exact ground-state of this model is a product wavefunction. We use the variational ansatz
\begin{align}
    \hat{\psi}&= \sum_{s_i \in \{0, 1\} } \prod_{i=1}^L \phi_i(s_i) \notag \\
    \phi_i(0) &= \sin \theta_i \notag\\
     \phi_i(1) &= \cos \theta_i 
    \end{align}
The energy is given by $E = \sum_i e_i$, $e_i=\sin(2\theta_i)$. The exact ground state wavefunction corresponds to $\theta_i=3\pi/4$ and has a per site energy $e_{\rm{g.s.}}=-1$. Since the ansatz does not contain a redundant parametrization, we use $\delta=0$ when inverting Eq.~\ref{eq:invert_sr} and Eq.~\ref{eq:invert_rgn}.

The gradient, overlap, and approximate Hessians can be computed exactly in this model giving
\begin{align}
&g_i=\cos(2\theta_i)\label{eq:model_grad}\\
&S_{ij}=\delta_{ij}\\
&H_{ij}=-2\delta_{ij}e_i\label{eq:model_hess}
\end{align}
Note that because the energy is a sum of terms, each involving a single $\theta_i$, the approximate Hessian is simply diagonal while the overlap matrix is just the identity. Thus the SR method without the small diagonal shift is just gradient descent. Furthermore, from the Taylor series for each $e_i$ at $\theta_i$
\begin{align}
e_i(\theta_i+p_i)&=\sum_{n=0}^\infty\frac{(-1)^n2^{2n}}{(2n)!}\sin(2\theta_i)p_i^{2n}\notag\\
&+\sum_{n=0}^\infty\frac{(-1)^n2^{2n+1}}{(2n+1)!}\cos(2\theta_i)p_i^{2n+1}\\
&=\sin(2\theta_i)+2\cos(2\theta_i)p_i\notag\\
&-2\sin(2\theta_i)p_i^2+O(p_i^3),
\end{align}
we can see that the true Hessian $\mathcal{H}_{ij}$ agrees with twice the approximate Hessian in Eq.~\ref{eq:model_hess}. Thus in this model, the $\epsilon \to \infty$ limit of RGN reduces to the true Newton method.

\subsubsection{Convergence properties of exact optimization algorithms}

To orient the discussion of the stochastic optimization, we first understand the performance of first- and (quasi-)second-order optimizers in the absence of noise. Since each site is decoupled, it suffices to consider the single-site problem. 
As already discussed above, in the absence of noise, SR is just gradient descent with step size $\epsilon/(1+\delta)$ and RGN with $\epsilon = \infty$ coincides with Newton's method. In deterministic optimization, Newton's method is known to have second-order convergence in the variational parameters for strongly convex and Lipschitz smooth functions \cite{NumericalOptimization,bookBoris}. In other words, $|\theta_{k+1}-\theta_*|/|\theta_k-\theta_*|^2\leq C$, where $\theta_*$ corresponds to the energy minimum and $\theta_k$ are the variational parameters at step $k$. The convergence rate of the function value $f(\theta_k)$ is, however, not guaranteed. Below, we consider different classes of convergence of the energy, where linear convergence means
\begin{align}
\frac{|E_{k+1}-E_{g.s.}|}{|E_{k}-E_{g.s.}|}\leq C
\end{align}
for some constant $0<C<1$, superlinear convergence
\begin{align}
\lim_{k\to\infty}\frac{|E_{k+1}-E_{g.s.}|}{|E_{k}-E_{g.s.}|}=0
\end{align}
and sublinear convergence
\begin{align}
\lim_{k\to\infty}\frac{|E_{k+1}-E_{g.s.}|}{|E_{k}-E_{g.s.}|}=1.
\end{align}
Defining the relative energy error at step $k$ as $\Delta_{\rm{rel}}(k)=|(E_k-E_{\rm{g.s.}})/E_{\rm{g.s.}}|$, and assuming an energy convergence form of
\begin{align}\label{eq:exp_fit}
{\Delta_{\rm{rel}}}(k)=C\exp(-ak^n)
\end{align}
then $n=1$, $n>1$ and $n<1$ correspond to linear, superlinear and sublinear convergence respectively.

\begin{figure}[htb]
    \centering
    \begin{subfigure}{0.85\linewidth}
        \centering
        \includegraphics[width=1\linewidth]{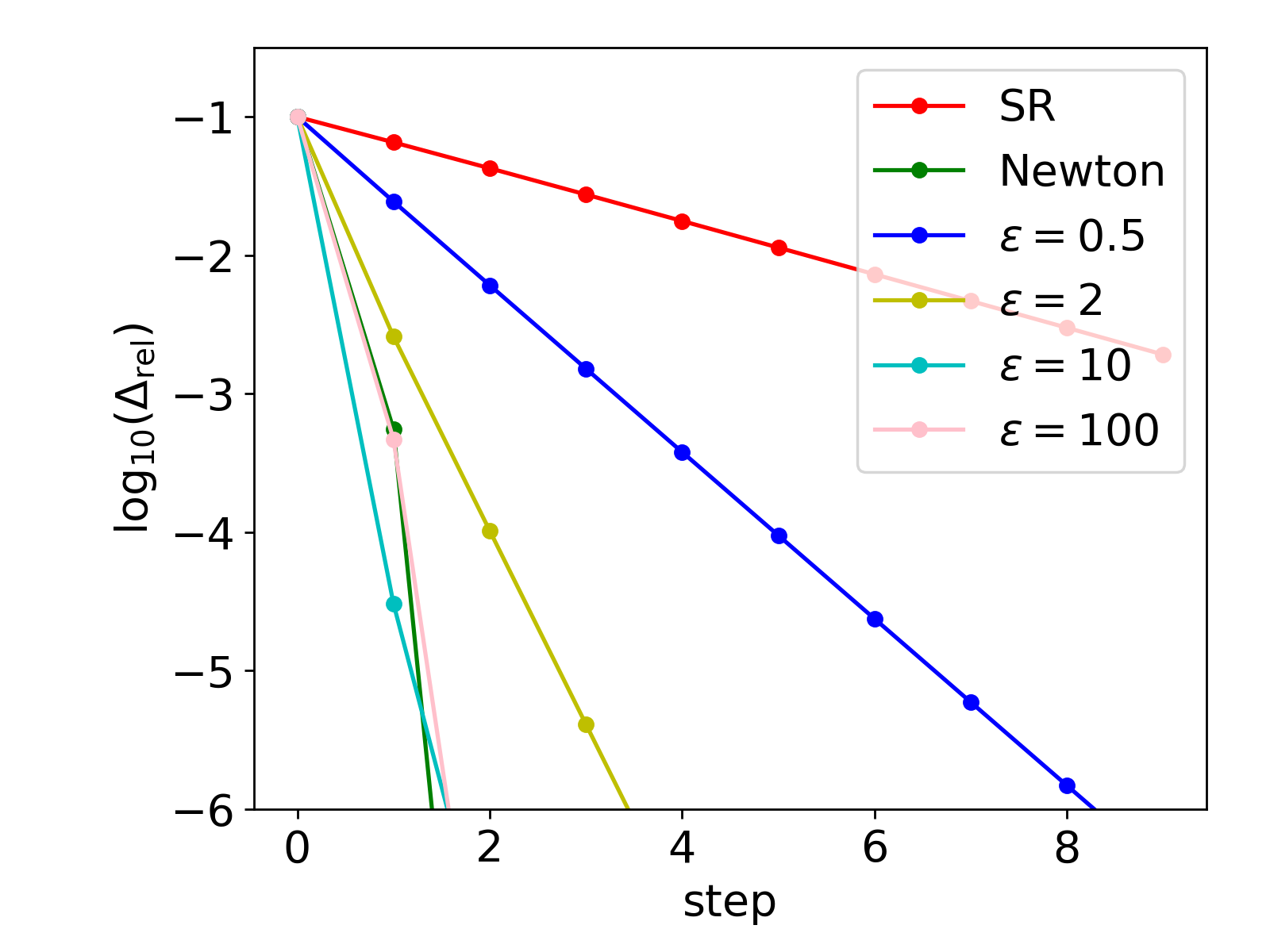}
        \caption{}
        \label{fig:1site_convergence_0.1}
    \end{subfigure}
    \centering
    \begin{subfigure}{0.85\linewidth}
        \centering
        \includegraphics[width=1\linewidth]{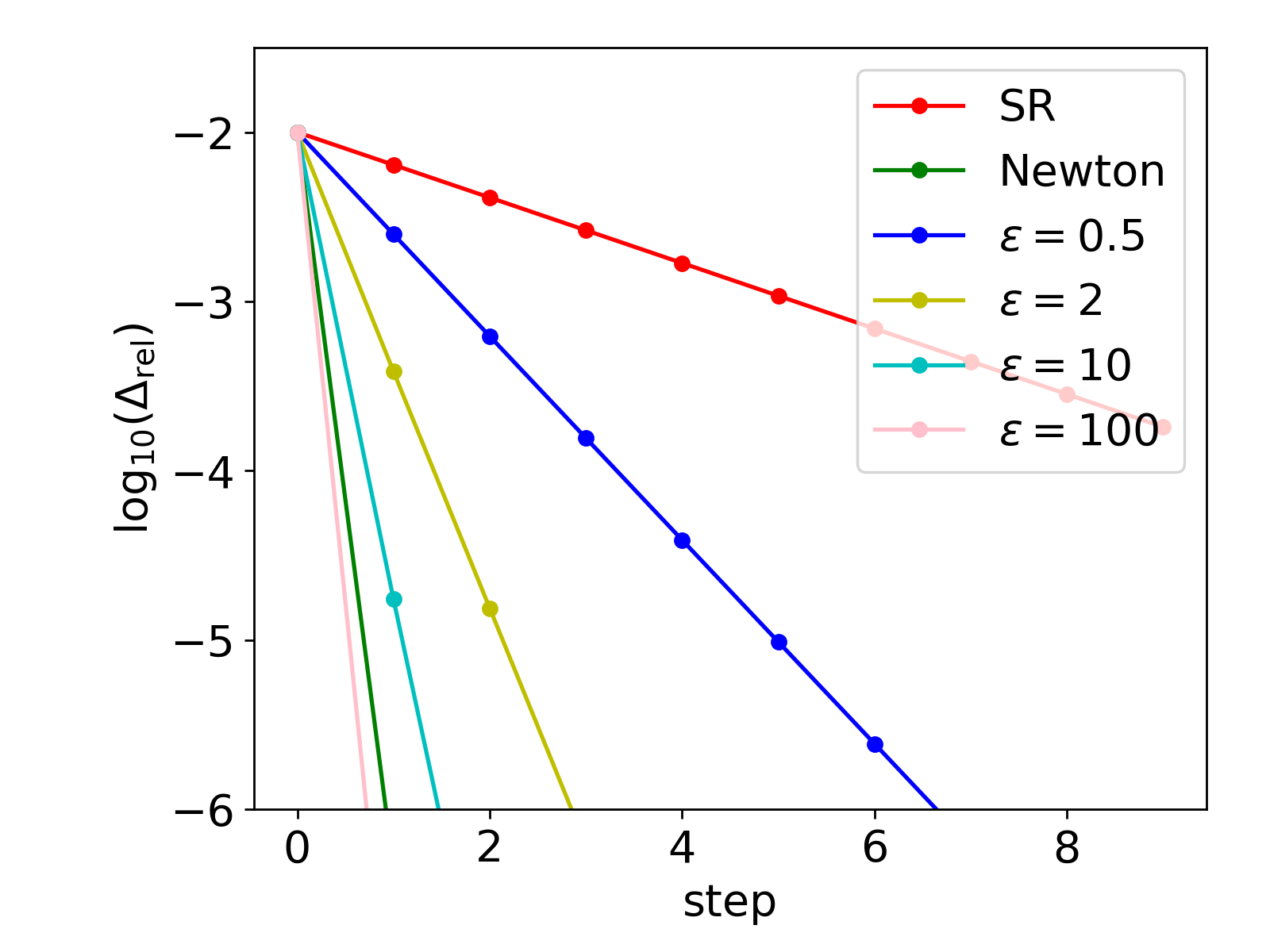}
        \caption{}
        \label{fig:1site_convergence_0.01}
    \end{subfigure}
    \caption{Optimization trajectories of the 1-site model problem without stochastic noise with initial energy error (a) $\Delta_{\rm{rel}}(0)=0.1$ and (b) $\Delta_{\rm{rel}}(0)=0.01$. The different values of $\epsilon$ refer to the RGN method. RGN with $\epsilon=\infty$ corresponds to the Newton step. }
    \label{fig:1site_convergence}
\end{figure}

Fig.~\ref{fig:1site_convergence} plots the error convergence of SR, Newton's method, and RGN with various regularization parameters $\epsilon$, at initial energy error $\Delta_{\rm{rel}}(0)=0.1$ and $\Delta_{\rm{rel}}(0)=0.01$. We see that Newton converges to numerical precision in one step when $\Delta_{\rm{rel}}$ is sufficiently small, and converges superlinearly for larger $\Delta_{\rm{rel}}(0)$. On the other hand, SR and RGN with different $\epsilon$ always display close to linear convergence behavior.

\subsubsection{Model stochastic properties}\label{sec:model_stochastic_properties}

The analytic tractability of this model allows us to compute certain statistical properties exactly. The SR and RGN algorithms involve the energy $E$, gradient $g$, and the $S$ and $H$ matrices. For $E$, the sample variance $\sigma^2_M\sim\sigma^2[E]/M$ where $M$ is sample size, and $\sigma^2[E]=\langle E_L(x)^2\rangle-E^2$ is the exact variance. For vector and matrix elements $g_i$, $S_{ij}$ and $H_{ij}$, the sample variance $\sigma^2$ can be estimated from error propagation, as described in Appendix~\ref{sec:model_exact_variance}.

The variances of all the quantities scale with both $M$ and the system size $L$. To simplify the analysis, we will focus on the leading contributions {(in sample size dependence, i.e. of $O(1/M)$)}, and analyze the behavior with respect to $L$. The expressions are given by
\begin{align}
\sigma^2[E]&=\sum_i\left(1-e_i^2\right)\notag\\
\sigma^2[g_i]&=\sum_k\left(1-e_k^2\right)+O(1)\notag\\
\sigma^2[S_{ij}]&=\Bigg\{\begin{array}{cc}
  4/e_i^2-4,  & i=j \\
   1,  & i\ne j
\end{array}\notag\\
\sigma^2[H_{ij}]&=\bigg\{\begin{array}{cc}
   \left(4/e_i^2-4\right)\sum_{k}\left(1-e_k^2\right) + O(1),  & i=j \\
   \sum_{k}\left(1-e_k^2\right) + O(1),  & i\ne j
\end{array}\notag.
\end{align}
where the $O(1)$ terms in $\sigma^2[H_{ij}]$ refer to scaling with $L$. Away from the ground state, assuming a constant energy error per site, $\sigma^2[E]$, $\sigma^2[g_i]$ and $\sigma^2[H_{ij}]$ all scale as $L$, while $\sigma^2[S_{ij}]$ is constant with respect to $L$. In this case, we expect the sample size requirement for RGN to scale more sharply with $L$ than for SR. On the other hand, as the wavefunction approaches the ground state, $\sigma^2[E]$, $\sigma^2[g_i]$ and $\sigma^2[H_{ij}]$ all decrease monotonically. In particular, while $\sigma^2[E]$ and $\sigma^2[g_i]$ vanish exactly at the ground state by the zero-variance principle \cite{toulouse2007,nightingale2001,neuscamman2016,becca_sorella_2017}, $\sigma^2[H_{ij}]$ does not vanish at the ground state but goes to a constant. Since $\sigma^2[S_{ij}]$ is size independent, we expect that close to the ground-state, the (scaling with system size of) the sample size requirement to be similar for SR and RGN. 

\begin{figure}[htb]
    \centering
    \begin{subfigure}{0.85\linewidth}
        \centering
        \includegraphics[width=1\linewidth]{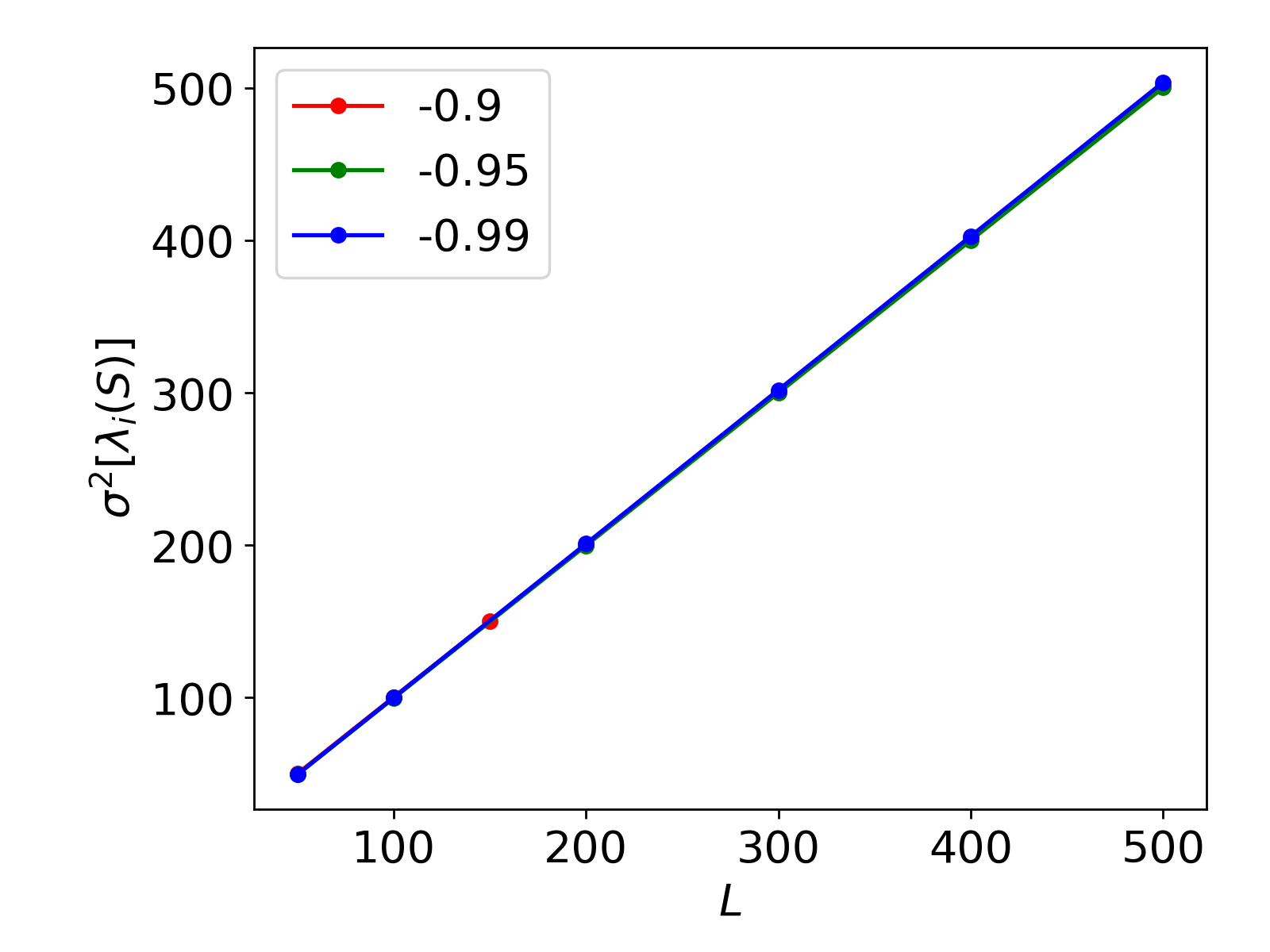}
        \caption{}
    \end{subfigure}
    \centering
    \begin{subfigure}{0.85\linewidth}
        \centering
        \includegraphics[width=1\linewidth]{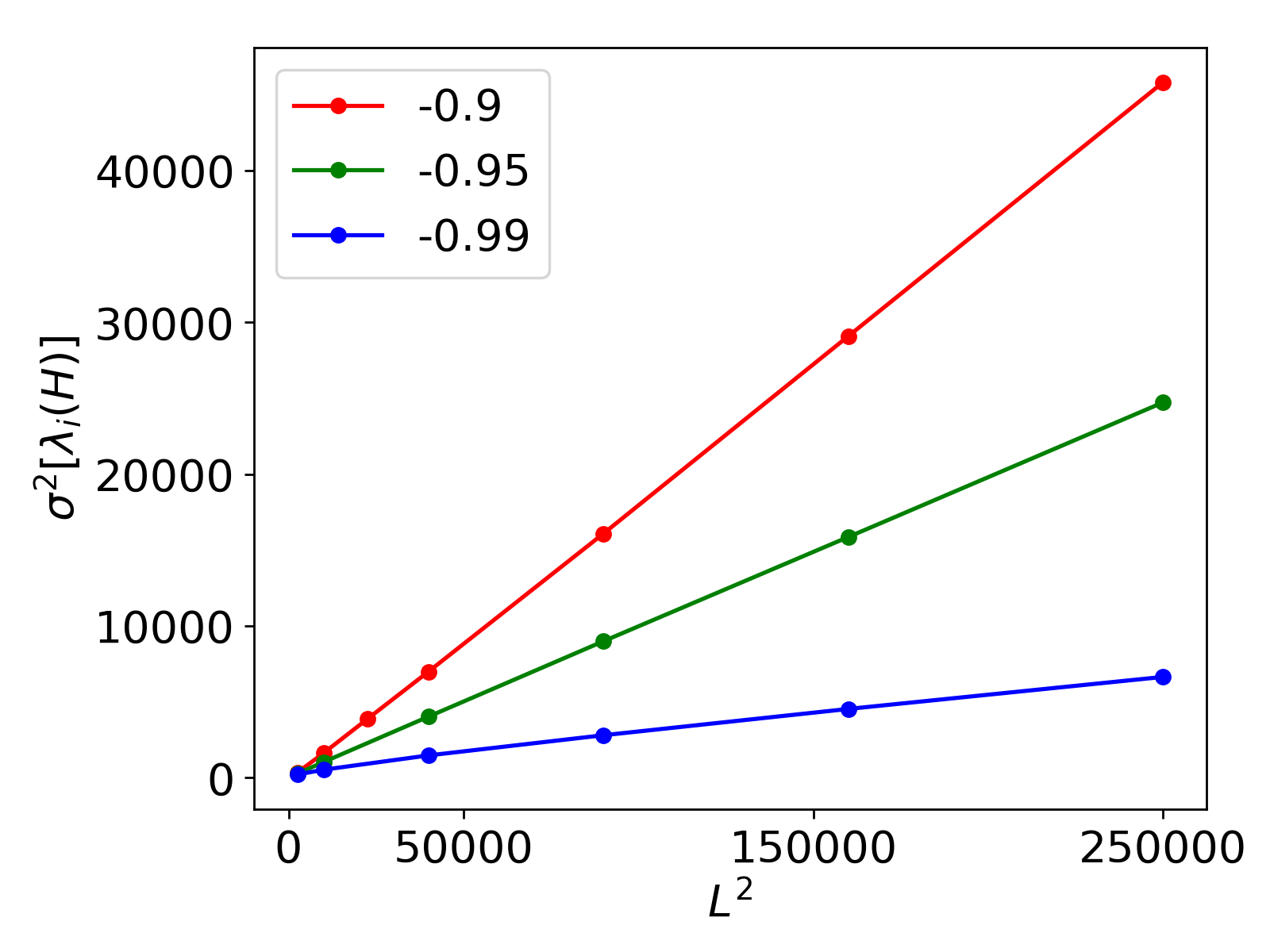}
        \caption{}
    \end{subfigure}
    \caption{Average variance over eigenvalues $\lambda_i$ for the model problem. (a) $\sigma^2[\lambda_i(S)]$ v.s. $L$. (b) $\sigma^2[\lambda_i(H)]$ v.s. $L^2$. Data corresponding to $e=-0.9\pm0.01$, $-0.95\pm0.005$ and $-0.99\pm0.002$ are labeled in red, green and blue. }
    \label{fig:eigenvalues}
\end{figure}

In addition, as the SR and RGN algorithms involve solving the matrix equations Eq.~\ref{eq:invert_sr} and Eq.~\ref{eq:invert_rgn}, the spectrum of $S$ and $H$ is relevant. Although these cannot be easily computed analytically, we can compute the sample variance and extract $\sigma^2[\lambda_i(S)]$, $\sigma^2[\lambda_i(H)]$
from the large $M$ limit. In Fig.~\ref{fig:eigenvalues}, we plot the average of these quantities as a function of $L$ at fixed per site energy $e=-0.9\pm0.01$, $-0.95\pm0.005$ and $-0.99\pm0.002$ (where the statistical fluctuation reflects our choice of random non-uniform $\theta_i$ in the state). We again observe that
\begin{align}
    \sigma^2[\lambda_i(S)] \sim \notag L\\ 
    \sigma^2[\lambda_i(H)] \sim L^2
\end{align}
away from the ground state. Furthermore, as the quality of the wavefunction improves, the eigenvalue variance for SR is unchanged, while that of RGN significantly decreases. Thus we similarly expect it to become easier to solve for an accurate RGN update when we are close to the ground-state. 

\subsubsection{Influence of statistical error on a single optimization step}\label{sec:model_1step}

To understand the effect of statistical noise on the optimization algorithms, we can consider the influence of statistical error on a given optimization step, for a few representative points along an optimization trajectory. First, consider the case of infinite samples. Then in the $k$-th step, a given optimization algorithm (e.g. SR, or RGN) leads to a change in the energy $\Delta E = E_{k+1} - E_k$. We can see the effect of noise then by computing $\Delta_M E$ as a function of sample size $M$. 

\begin{figure}[htb]
    \centering
    \begin{subfigure}{0.49\linewidth}
        \centering
        \includegraphics[width=1\linewidth]{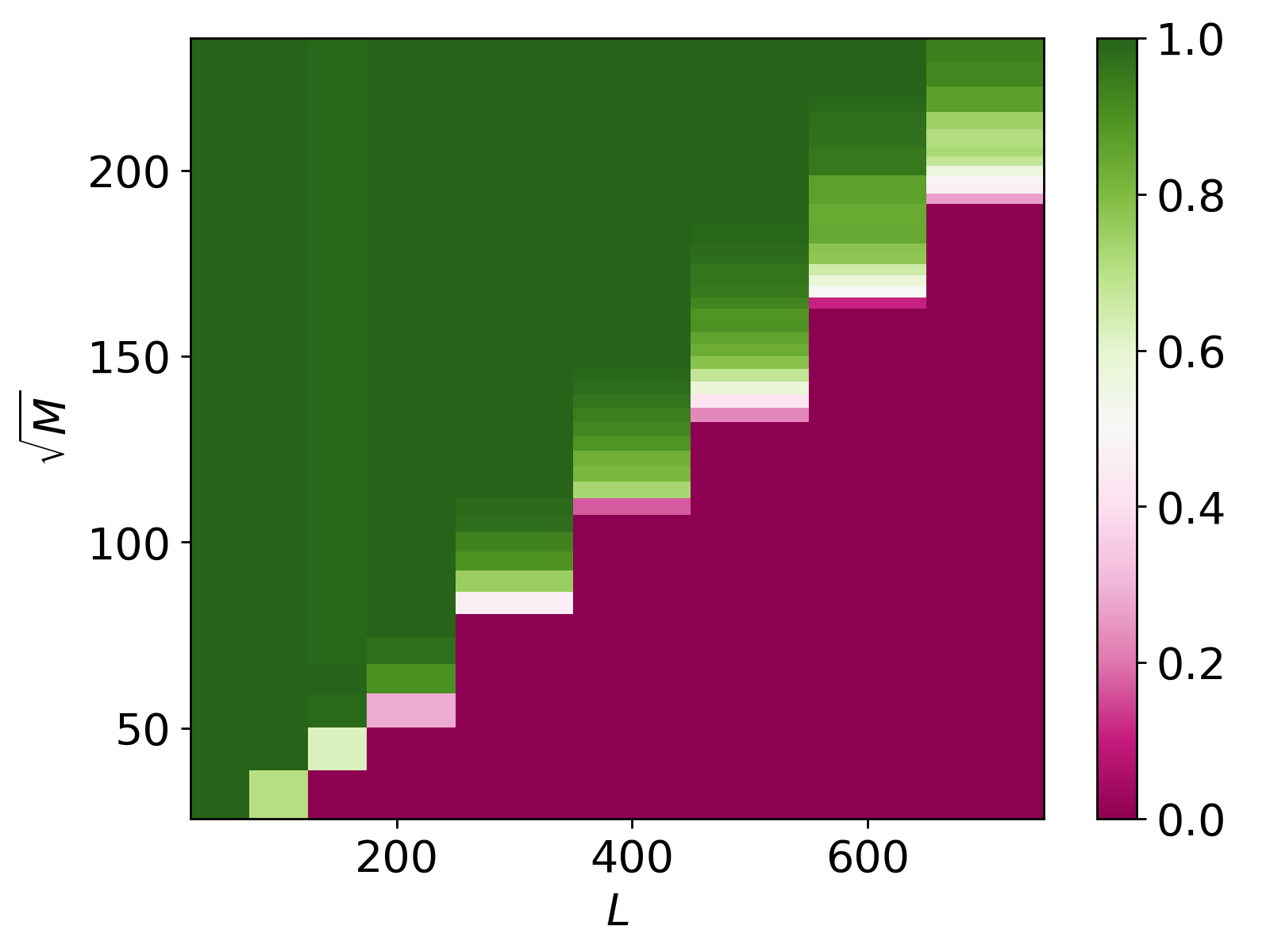}
        \caption{}
        \label{fig:perr10_scale_0.01_de_rgn}
    \end{subfigure}
    \centering
    \begin{subfigure}{0.49\linewidth}
        \centering
        \includegraphics[width=1\linewidth]{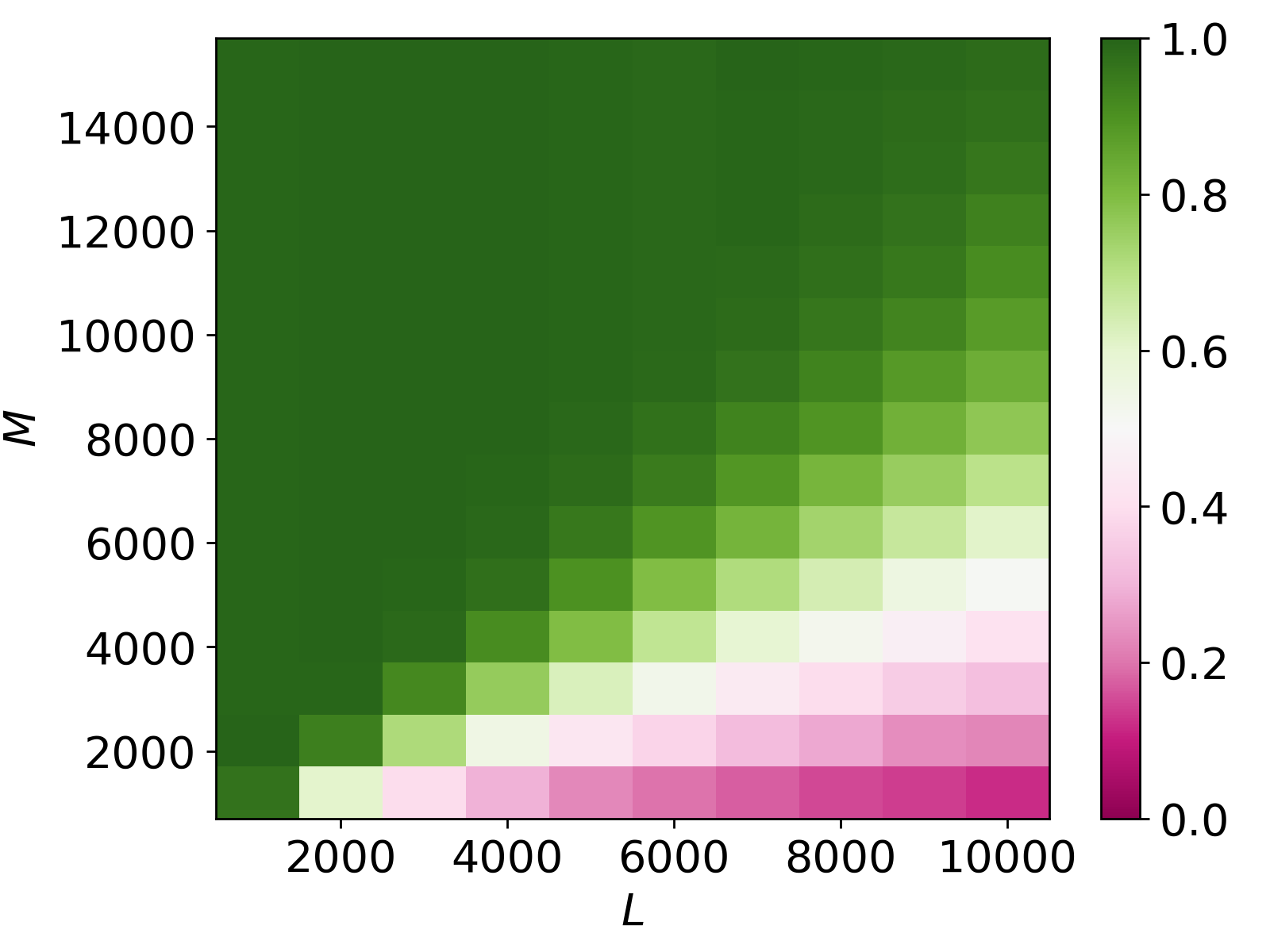}
        \caption{}
        \label{fig:perr10_scale_0.01_de_sr}
    \end{subfigure}
    \centering
    \begin{subfigure}{0.49\linewidth}
        \centering
        \includegraphics[width=1\linewidth]{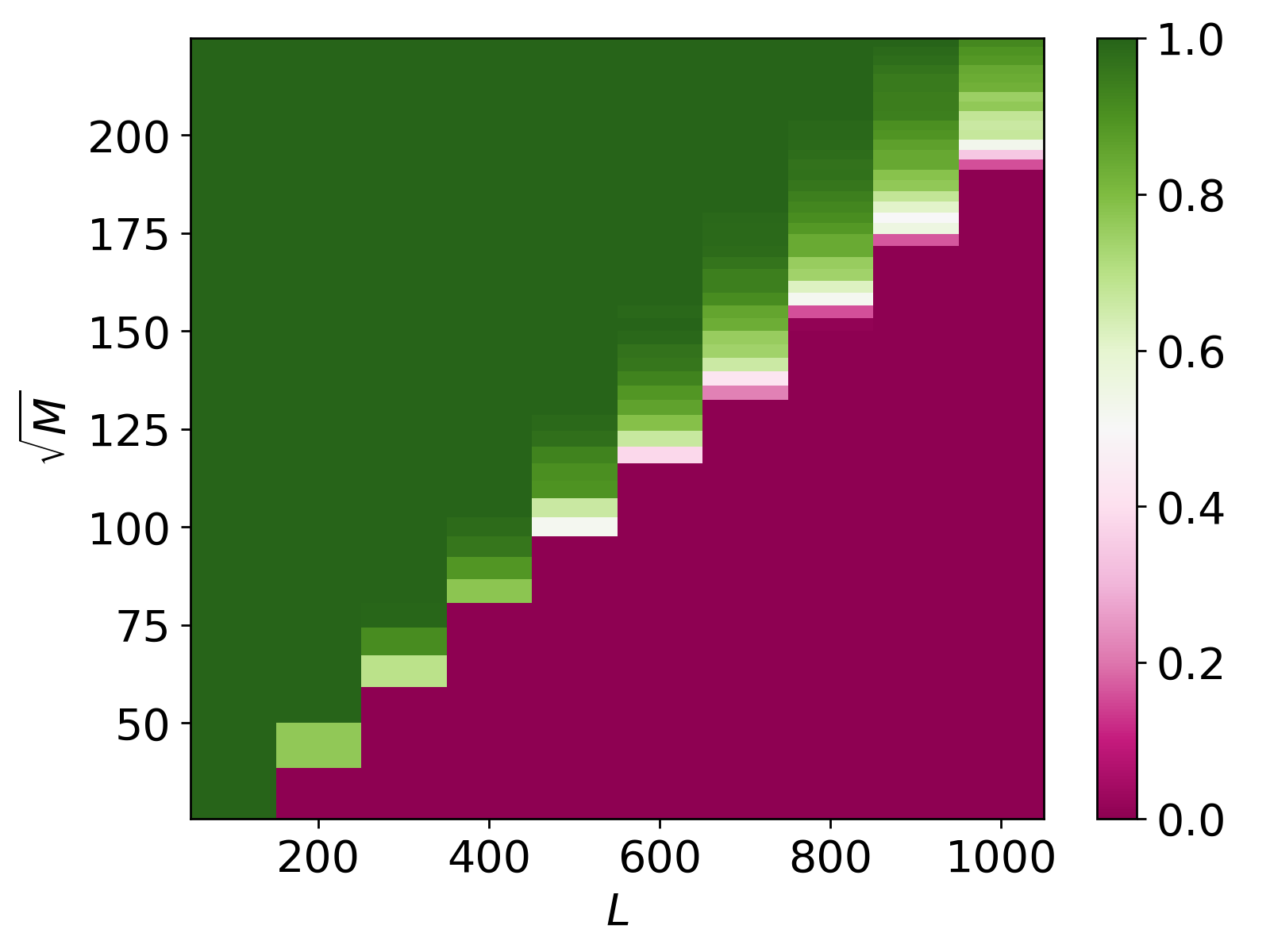}
        \caption{}
        \label{fig:perr05_scale_0.005_de_rgn}
    \end{subfigure}
    \centering
    \begin{subfigure}{0.49\linewidth}
        \centering
        \includegraphics[width=1\linewidth]{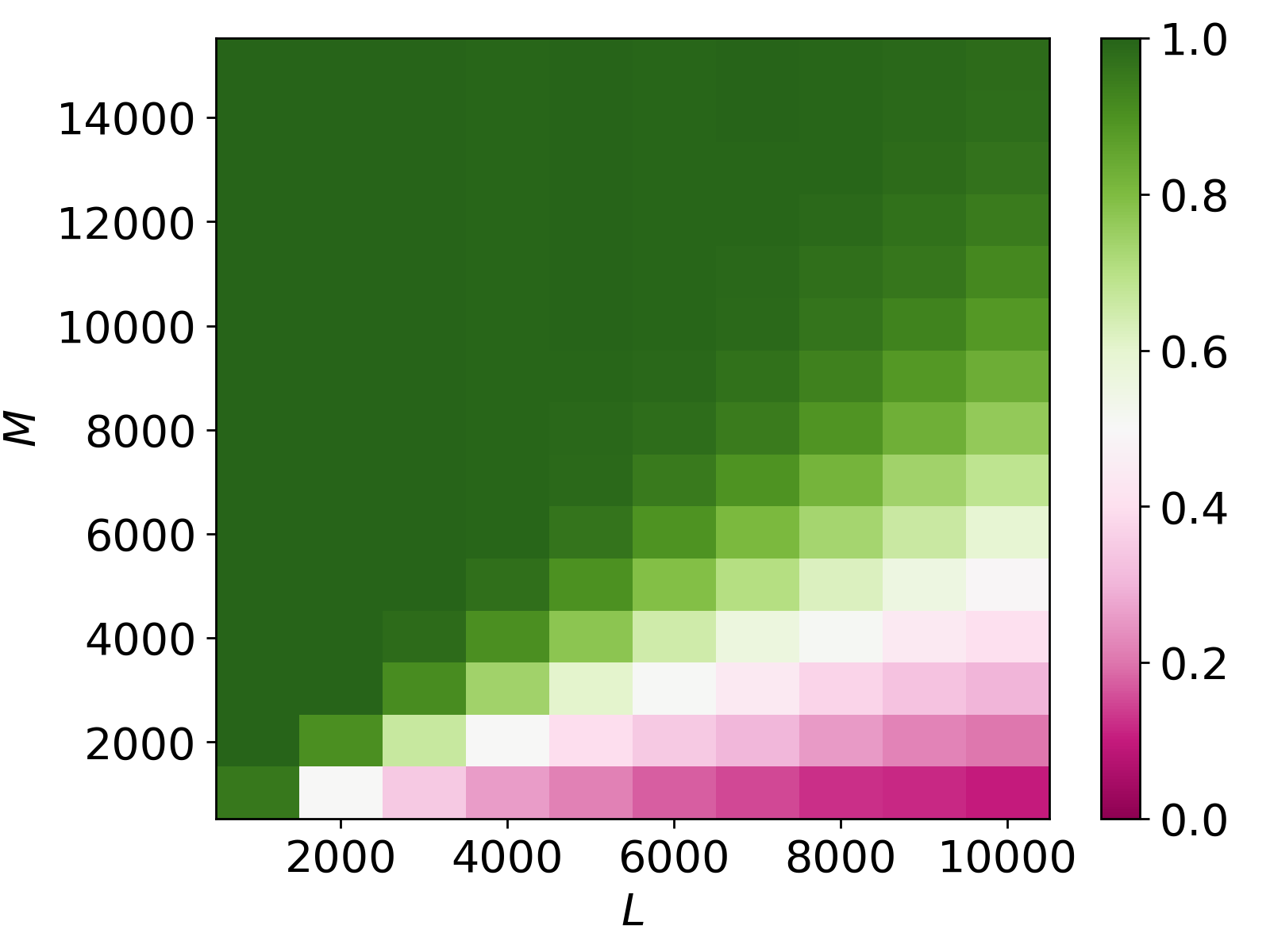}
        \caption{}
        \label{fig:perr05_scale_0.005_de_sr}
    \end{subfigure}
    \caption{1-step energy difference ratio $\Delta_ME/\Delta E$ (shown in the colorbar) for the model problem as a function of (a) $\sqrt{M}$ and $L$ for RGN at and (b) $M$ and $L$ for SR at $e=-0.9\pm0.01$, and (c) $\sqrt{M}$ and $L$ for RGN and (d) $M$ and $L$ for SR at $e=-0.95\pm0.005$.}
    \label{fig:1step_model}
\end{figure}

Fig.~\ref{fig:1step_model} plots the ratio $\Delta_ME/\Delta E$ as a function of $M$ and $L$ for RGN and SR  using wavefunctions with per site energy $e=-0.9\pm0.01$ and $e=-0.95\pm0.005$. We make the following observations: (i) By comparing Fig.~\ref{fig:perr10_scale_0.01_de_rgn} with Fig.~\ref{fig:perr10_scale_0.01_de_sr}, or Fig.~\ref{fig:perr05_scale_0.005_de_rgn} with Fig.~\ref{fig:perr05_scale_0.005_de_sr}, we see that to achieve a constant fraction $\Delta_ME/\Delta E$ at a fixed wavefunction quality (as given by a fixed per site energy $e$), RGN requires a sample size scaling of $M\sim L^2$, whereas SR requires $M\sim L$. (ii) For the RGN sample size scaling $M=c_{\rm{RGN}}L^2$, the constant $c_{\rm{RGN}}$ decreases as the wavefunction quality improves, as seen by comparing Fig.~\ref{fig:perr10_scale_0.01_de_rgn} with Fig.~\ref{fig:perr05_scale_0.005_de_rgn}; whereas for the SR sample size scaling $M=c_{\rm{SR}}L$, and the constant $c_{\rm{SR}}$ is almost unaffected by wavefunction quality, as seen by comparing Fig.~\ref{fig:perr10_scale_0.01_de_sr} with Fig.~\ref{fig:perr05_scale_0.005_de_sr}. We note that observation (ii) is also consistent with the discussion of the statistical variance of $S$ and $H$ in Section~\ref{sec:model_stochastic_properties}. 

\subsubsection{Optimization trajectories}\label{sec:model_traj}

\begin{figure}[htb]
    \centering
    \begin{subfigure}{0.85\linewidth}
        \centering
        \includegraphics[width=1\linewidth]{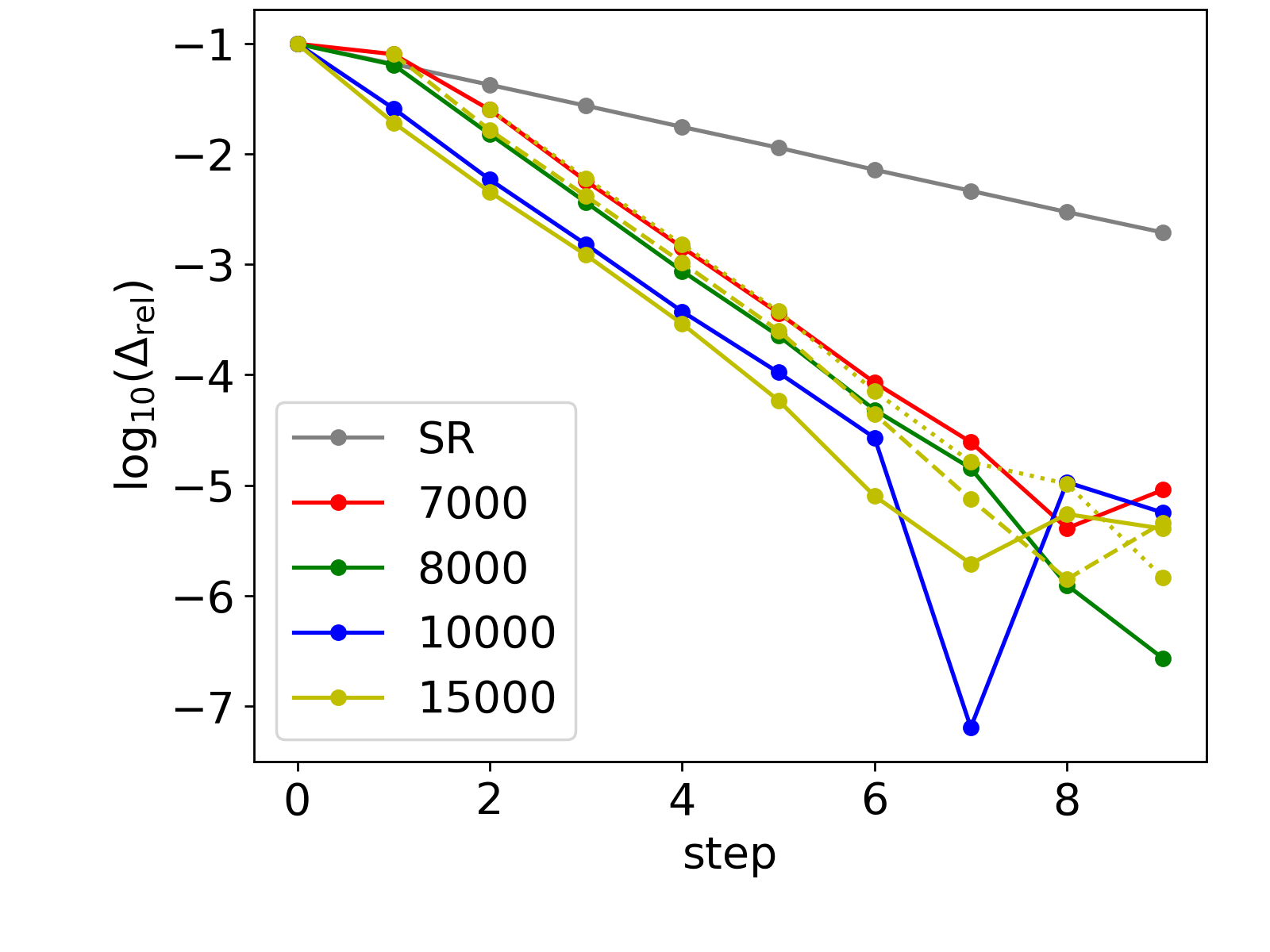}
        \caption{}
        \label{fig:model300_opt}
    \end{subfigure}
    \centering
    \begin{subfigure}{0.85\linewidth}
        \centering
        \includegraphics[width=1\linewidth]{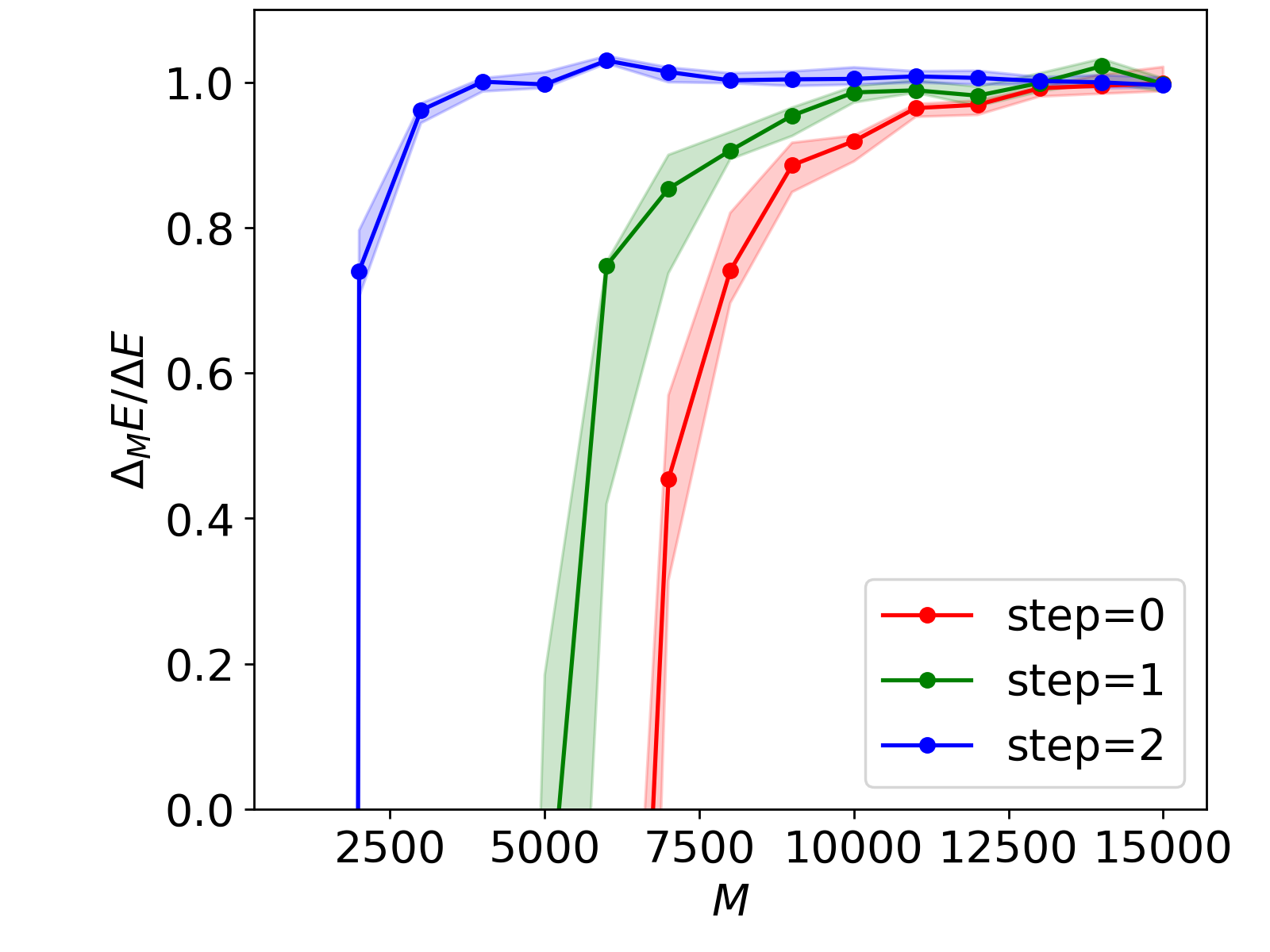}
        \caption{}
        \label{fig:model300_1step}
    \end{subfigure}
    \caption{(a) Optimization trajectory for the model problem at $L=300$. SR (grey) is computed with $M=7000$. Red, green, blue and yellow are RGN trajectories with different $M$. (b) RGN 1-step $\Delta_ME/\Delta E$ as a function of $M$ for different wavefunctions taken along the RGN trajectory with $M=7000$ in (a). Red, green and blue correspond to wavefunctions taken at step 0, 1 and 2 respectively.}
\end{figure}

We now consider the multi-step optimization trajectories produced by the first-order SR and (quasi-)second-order RGN updates in the model problem. Fig.~\ref{fig:model300_opt} plots the optimization trajectory for SR and RGN at $L=300$. The SR trajectory in grey is computed with $M=7000$ which we consider to have saturated its sample size requirements. The RGN trajectories computed with different $M$ are shown in different colors. We again make two observations: (i) At the first step of the optimization, the RGN trajectories show an increasing 1-step $\Delta_ME$ with increasing $M$, consistent with the observation in Section.~\ref{sec:model_1step}. (ii) Overall the RGN trajectories with different $M$ have parallel slopes. This is because even for the RGN trajectories with small $M$, the wavefunction quality quickly improves so that the small sample size also saturates the sample size requirement to achieve an exact 1-step update. For instance, along the RGN trajectory with $M=7000$, we take the wavefunction corresponding to steps 0, 1 and 2 as the initial states and recompute the RGN trajectory with $M=15000$, as shown by the solid, dashed and dotted yellow curves in Fig.~\ref{fig:model300_opt}. We see that the RGN trajectory with $M=7000$ (red solid) is very close to the $M=15000$ (yellow dashed) optimization trajectory that starts from the $M=7000$ wavefunction at step 1, while the RGN trajectories with $M=7000$ and $M=15000$ (yellow dotted) essentially overlap if we start with the $M=7000$ wavefunction at step 2. In Fig.~\ref{fig:model300_1step}, we plot the RGN 1-step $\Delta_ME/\Delta E$ as a function of $M$ for step 0 (red), 1 (green) and 2 (blue). For each fixed sample size $M$, the ratio $\Delta_ME/\Delta E$ for steps 0 to 2 quickly approach 1 from below.  

\subsection{1D $J_1$-$J_2$ model}\label{sec:1DJ1J2}

We next extend our observations on the model non-interacting problem from the previous section to an interacting problem. For this we choose the 1D $J_1$-$J_2$ spin $1/2$ Heisenberg model with open boundary conditions \cite{10.1063/1.1664978,10.1063/1.1664979}
\begin{align}
\hat{H}=J_1\sum_{i=0}^{L-2}\Vec{S}_i\cdot\Vec{S}_{i+1}+J_2\sum_{i=0}^{L-3}\Vec{S}_i\cdot\Vec{S}_{i+2}.
\end{align}
We use $J_1=1$ and $J_2=0.5$ (we report energies in units of $J_1$), which corresponds to a simple exactly soluble point for which the ground state per site energy is $e_{g.s.}=-0.375$~\cite{Lavarelo2014,PhysRevB.84.035130}. For the wavefunction, we use a matrix product state (MPS)~\cite{orus2019,Verstraete2008,ORUS2014117}, which can be written as 
\begin{align}
\label{eq:mps}
    \hat{\psi}(s_1, s_2, \ldots, s_L) = A(s_1) A(s_2) \ldots A(s_L)
\end{align}
where $A(s_i)$ are $D\times D$ matrices except for the first and last which are are $1 \times D$ and $D\times 1$ vectors. The ground state at the exactly soluble point can be written as an MPS of $D=2$. However, to reflect a more typical optimization procedure (where we do not know what the final structure of the ground-state) we will use an MPS of $D=5$ in the following calculations. Furthermore, as written in Eq.~\ref{eq:mps}, the MPS wavefunction contains redundant variational parameters \cite{orus2019,Verstraete2008,ORUS2014117}. Thus we expect $S$ and $H$ to have singularities, which are handled by the diagonal shift $\delta$ in Eq.~\ref{eq:invert_sr} and Eq.~\ref{eq:invert_rgn}.

\subsubsection{Convergence properties of exact optimization algorithms}

\begin{figure}[htb]
    \centering
    \begin{subfigure}{0.85\linewidth}
        \centering
        \includegraphics[width=1\linewidth]{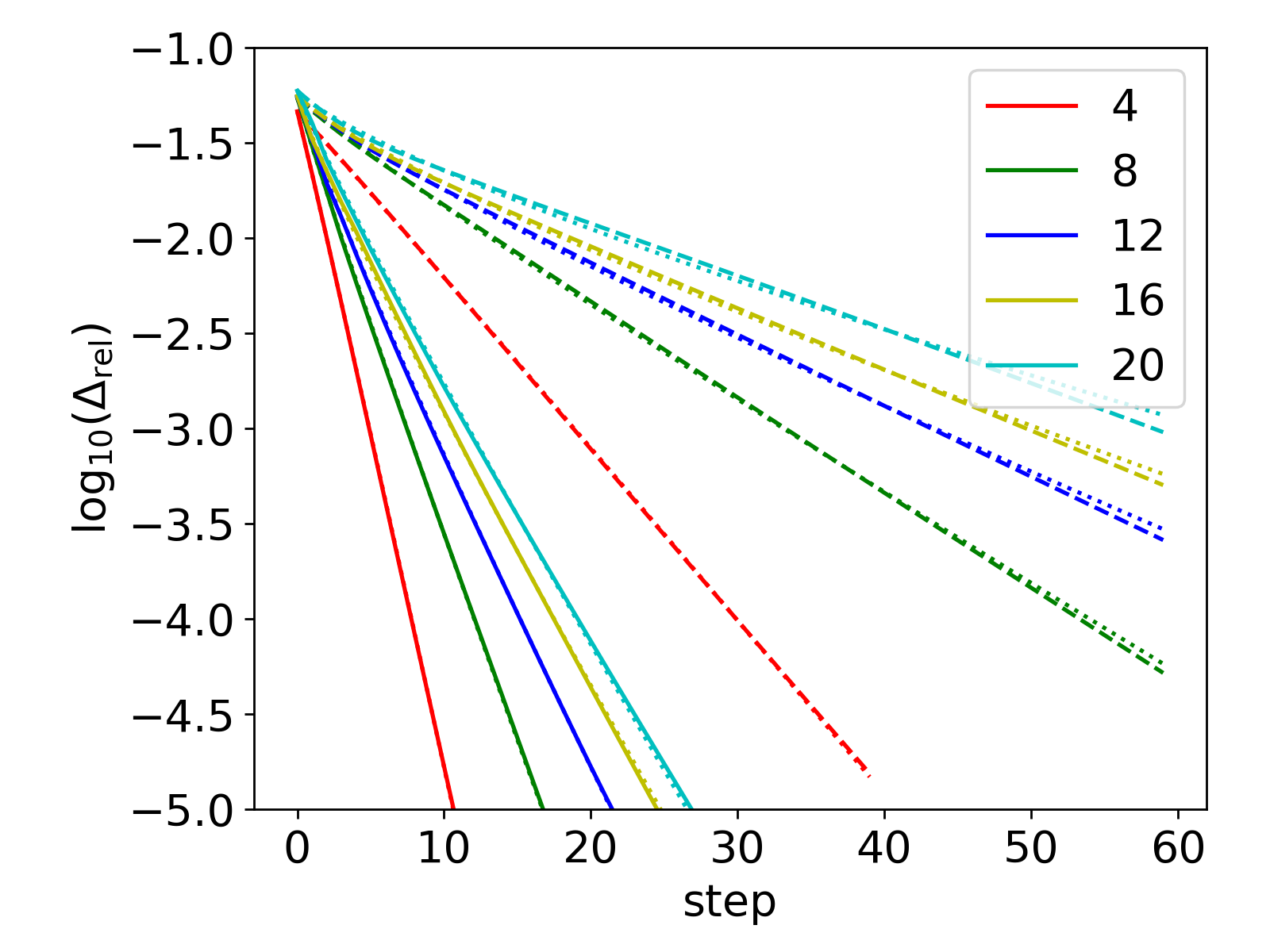}
        \caption{}
        \label{fig:J1J2_1D_exact_energy}
    \end{subfigure}
    \centering
    \begin{subfigure}{0.85\linewidth}
        \centering
        \includegraphics[width=1\linewidth]{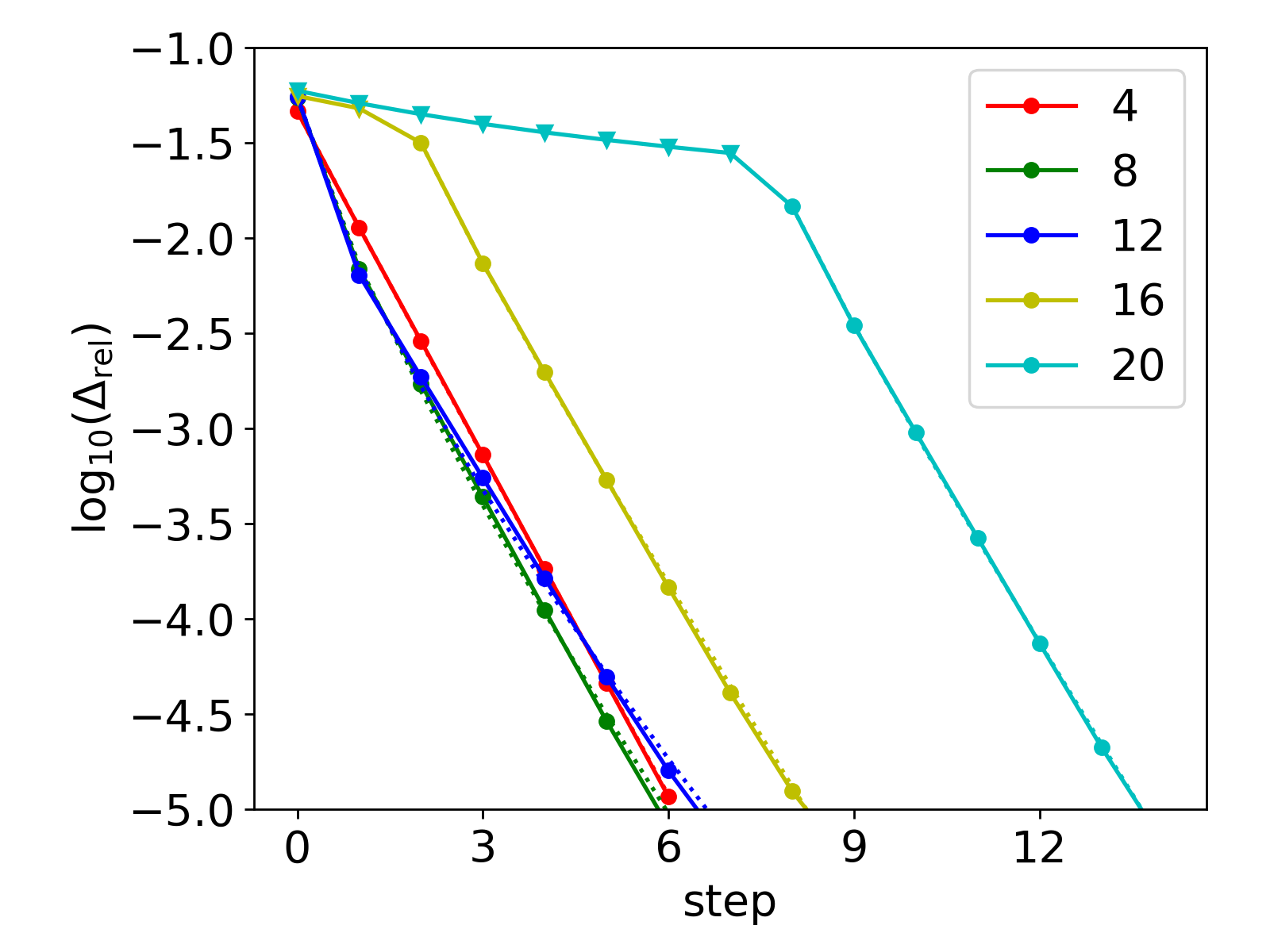}
        \caption{}
        \label{fig:J1J2_1D_exact_energy_newton}
    \end{subfigure}
    \caption{Optimization trajectories of the 1D $J_1$-$J_2$ model from exact sampling for (a) SR (dashed) and RGN (solid), and (b) Newton method with approximate Hessian. Exponential fit for each trajectory with parameters listed in Table~\ref{table:J1J2_1D_exact} are plotted in dotted curves. Colors label different system sizes $L$. }
    \label{fig:1d_opt_exact}
\end{figure}

We first demonstrate the convergence properties of SR, RGN and the approximate Newton method (equivalently, RGN with $\epsilon=\infty$) in the absence of noise for $L\leq 20$ where we can exactly sample the Hilbert space by summing over all configurations. Fig.~\ref{fig:J1J2_1D_exact_energy} plots the optimization trajectories of SR (dashed) and RGN (solid) and Fig.~\ref{fig:J1J2_1D_exact_energy_newton} plots the approximate Newton optimization trajectories (solid), for various $L$  with $\delta=0.001$. We note that the approximate Newton trajectories for $L=16$ (yellow) and $L=20$ (cyan) initially follow that of SR (marked by triangles). This is because the wavefunction is too far from the quadratic region and the approximate Hessian has negative eigenvalues, and our algorithm (see Section~\ref{sec:exact_formulation}) falls back to the SR step. On the other hand, the introduction of $S$ in the RGN matrix $H+S/\epsilon$ (dropping $\delta$ for simplicity) removes the negative eigenvalues and leads to a smoother convergence as seen by comparing Fig.~\ref{fig:J1J2_1D_exact_energy} and Fig.~\ref{fig:J1J2_1D_exact_energy_newton}. 

We note that, for this noiseless case, faster convergence is obtained by the approximate Newton method with SR fallback, than by the recommended RGN update with finite $\epsilon$ when targeting high accuracy. This illustrates the potential for improvement when more sophisticated stepsize control procedures (which can adjust $\epsilon$~\cite{toulouse2007,toulouse2008,PhysRevLett.94.150201}) are implemented. However, it should be noted that when stochastic noise is introduced, we expect to always require finite $\epsilon$ to regularize  the finitely sampled $H$ which can have negative eigenvalues even when all eigenvalues of the exactly sampled $H$ are non-negative. We also observe that even when the optimal $\epsilon$ is chosen (here $\epsilon = \infty$ corresponding to the approximate Newton update) and the quasi-second order step is taken, the improvement over RGN is a constant factor, rather than a scaling difference, since the convergence rate of both RGN and approximate Newton is at best linear. The fitted exponential parameters $a$ and $n$ from Eq.~\ref{eq:exp_fit} for the optimization trajectories of SR, RGN, and the approximate Newton's method are listed in Table~\ref{table:J1J2_1D_exact}. 

\begin{table}[h!]
\centering
\begin{tabular}{ |c|c|c|c|c|c|c| } 
\hline
&  \multicolumn{2}{|c|}{SR} & \multicolumn{2}{|c|}{RGN($\epsilon=0.5$)}  & \multicolumn{2}{|c|}{Approx. Newton}\\
 \hline
L & $n$ & $a$ & $n$ & $a$ & $n$ & $a$\\
\hline
4 & 1.020 & 0.0832  & 1.023 & 0.327 & 0.991 & 0.609\\
\hline
8 & 0.932 & 0.0666 & 0.954 & 0.255 & 0.812 & 0.878\\ 
\hline
12 & 0.864 & 0.0670 & 0.905 & 0.234 & 0.752 & 0.904\\ 
\hline
16 & 0.829 & 0.0675 & 0.895 & 0.212 & 0.934 & 0.634\\ 
\hline
20 & 0.789 & 0.0682 & 0.919 & 0.185 & 0.938 & 0.624\\
\hline
\end{tabular}
\caption{Exponential fit parameters of optimization trajectories (Fig.~\ref{fig:J1J2_1D_exact_energy} and Fig.~\ref{fig:J1J2_1D_exact_energy_newton}) for 1D $J_1$-$J_2$ model without noise.} 
\label{table:J1J2_1D_exact}
\end{table}

\subsubsection{Influence of statistical error on a single optimization step}\label{sec:j1j2_1step}

\begin{figure}[htb]
    \centering
    \begin{subfigure}{0.49\linewidth}
        \centering
        \includegraphics[width=1\linewidth]{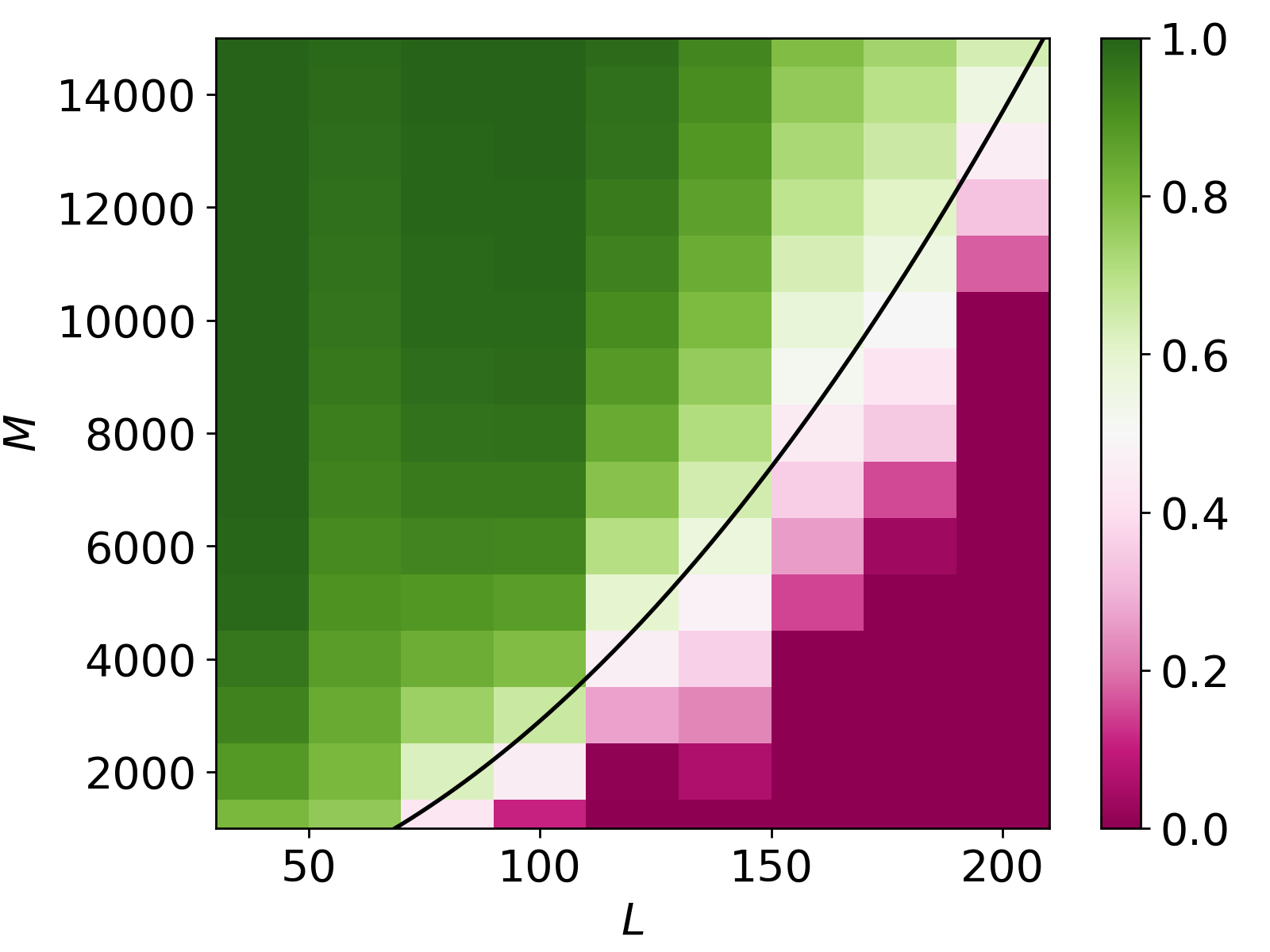}
        \caption{}
        \label{fig:p5_c2_rgn_0}
    \end{subfigure}
    \centering
    \begin{subfigure}{0.49\linewidth}
        \centering
        \includegraphics[width=1\linewidth]{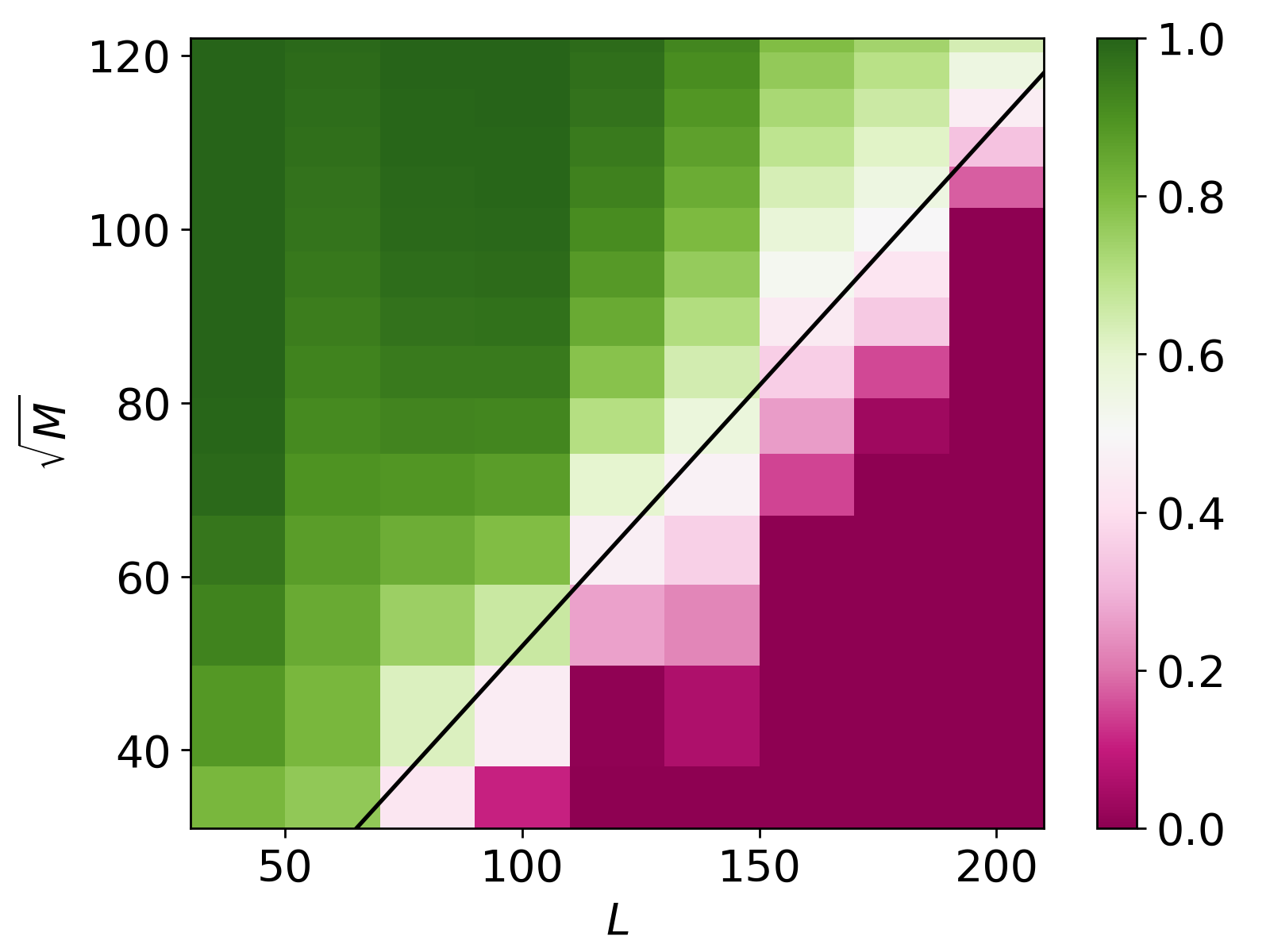}
        \caption{}
        \label{fig:p5_c2_rgn_2}
    \end{subfigure}
    \centering
    \begin{subfigure}{0.49\linewidth}
        \centering
        \includegraphics[width=1\linewidth]{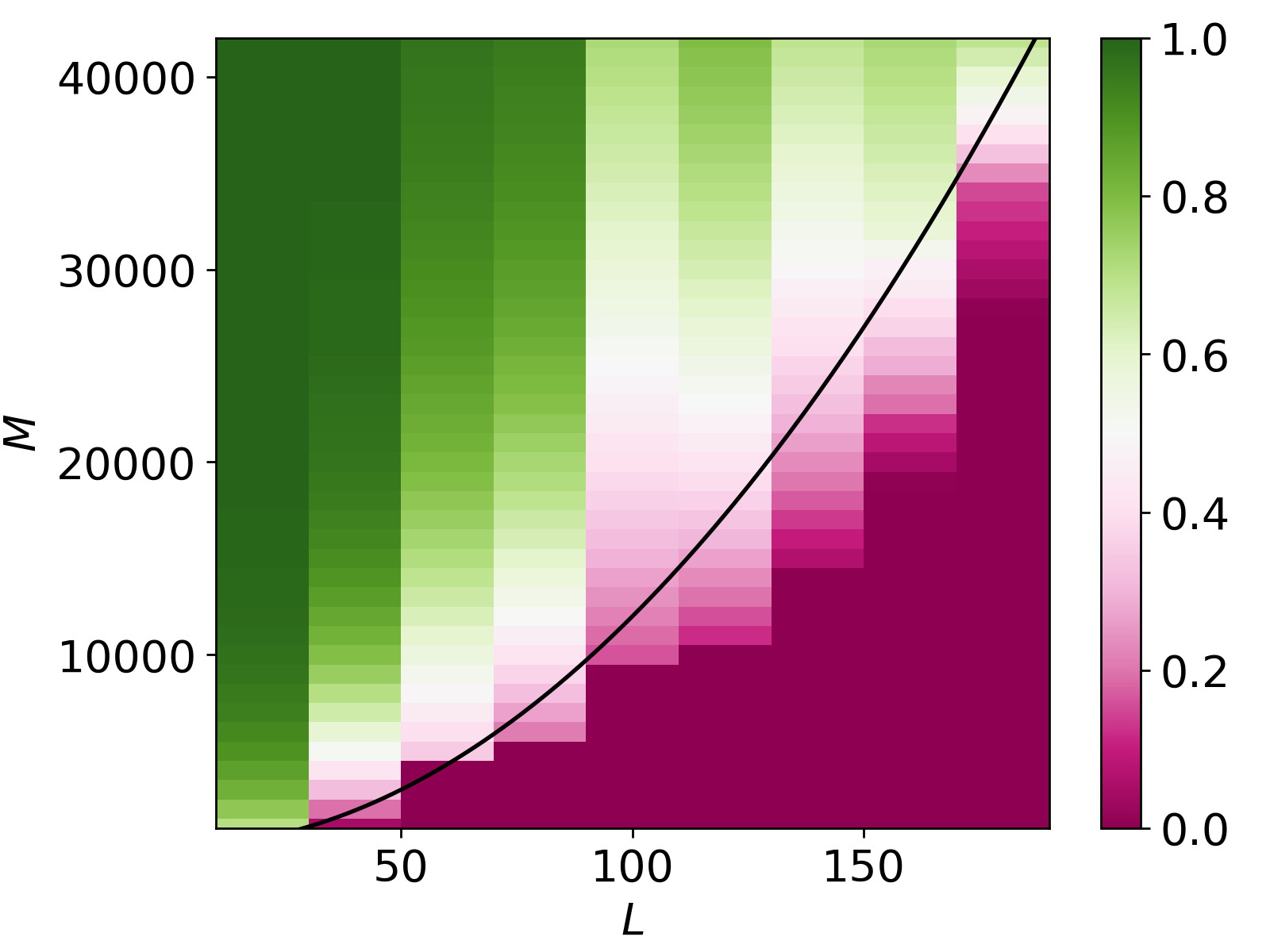}
        \caption{}
        \label{fig:p5_c3_rgn_0}
    \end{subfigure}
    \centering
    \begin{subfigure}{0.49\linewidth}
        \centering
        \includegraphics[width=1\linewidth]{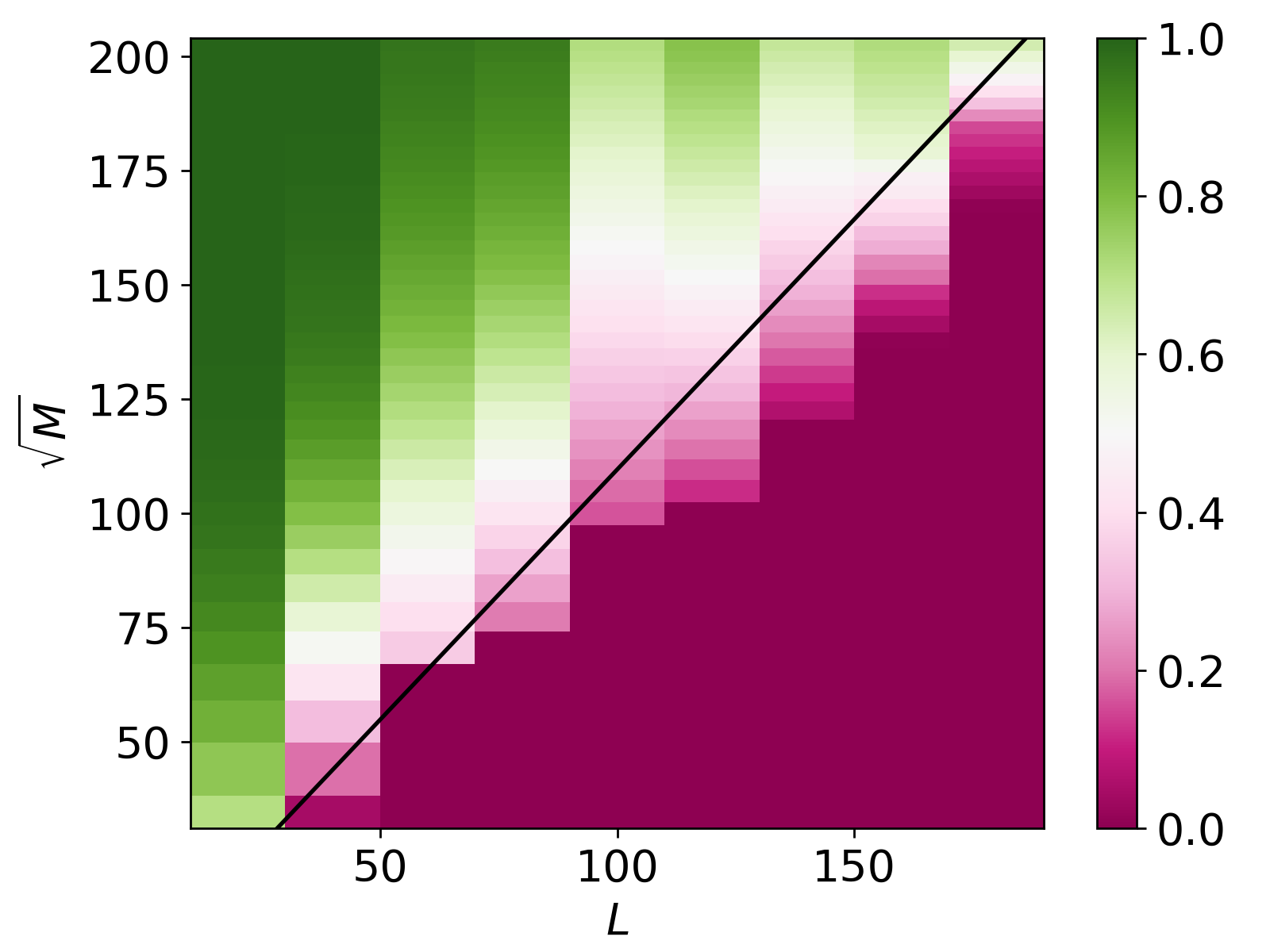}
        \caption{}
        \label{fig:p5_c3_rgn_2}
    \end{subfigure}
    \centering
    \begin{subfigure}{0.49\linewidth}
        \centering
        \includegraphics[width=1\linewidth]{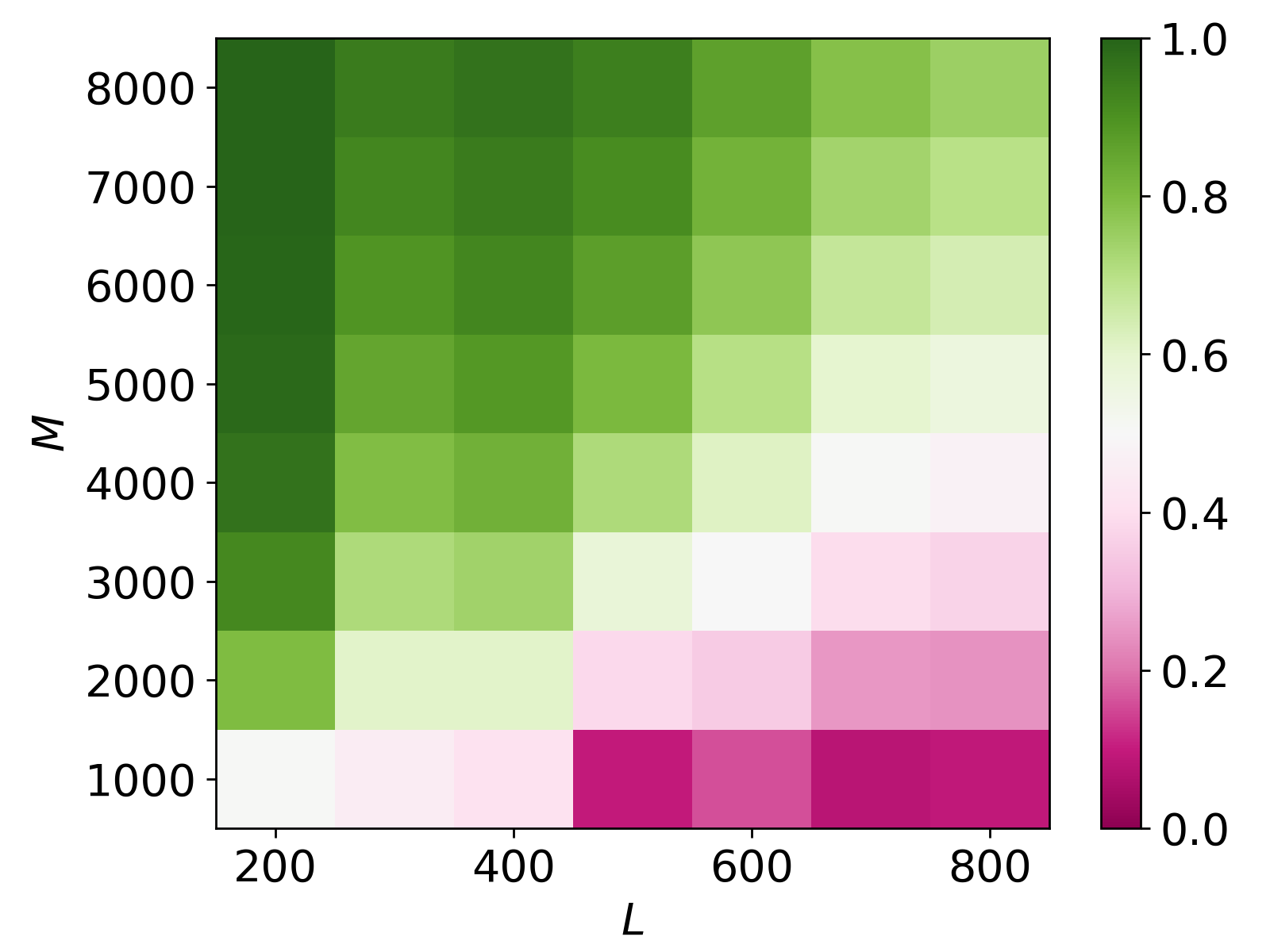}
        \caption{}
        \label{fig:p5_c2_sr_0}
    \end{subfigure}
    \centering
    \begin{subfigure}{0.49\linewidth}
        \centering
        \includegraphics[width=1\linewidth]{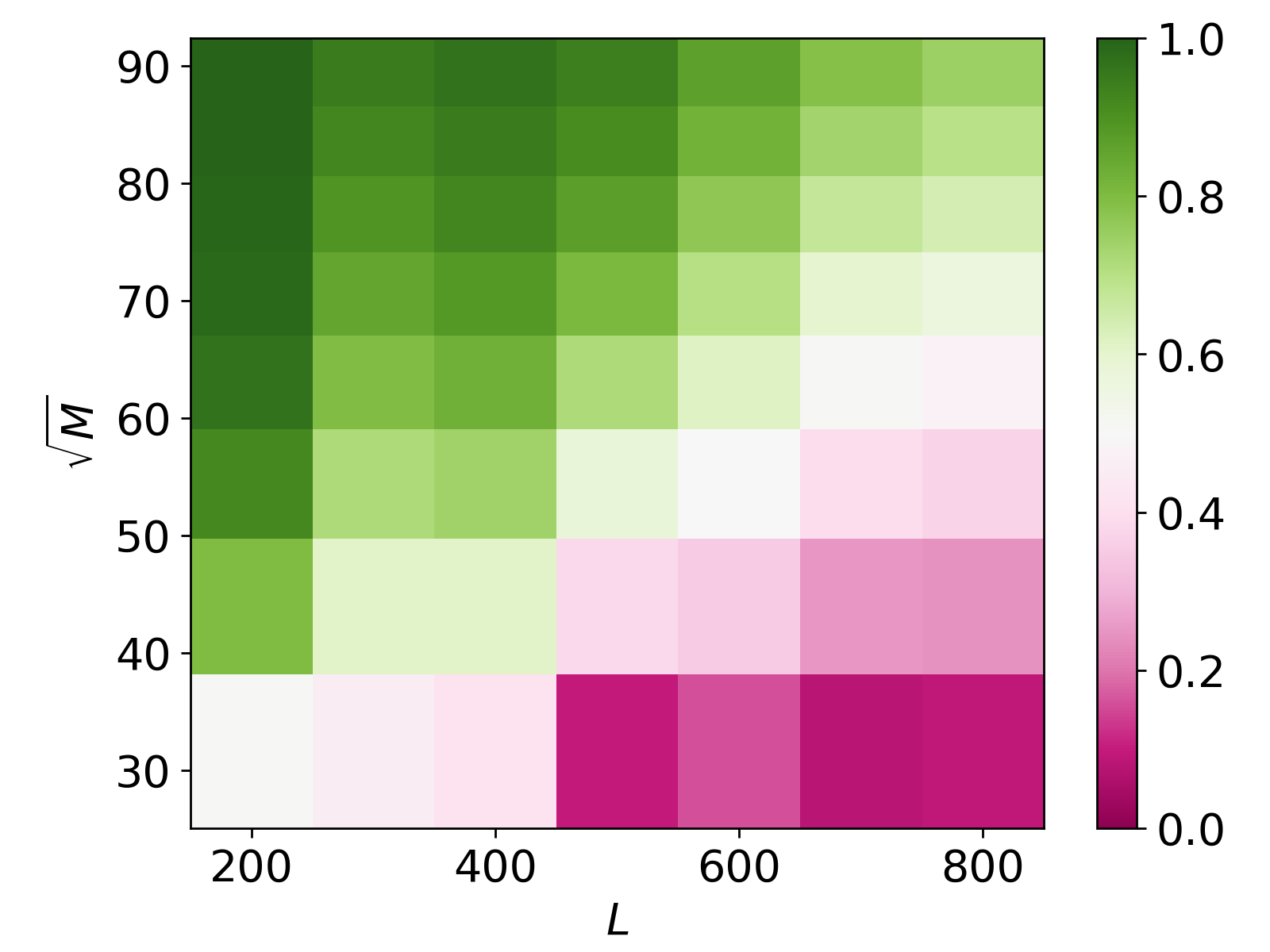}
        \caption{}
        \label{fig:p5_c2_sr_2}
    \end{subfigure}
    \caption{1-step energy difference ratio $\Delta_M E/\Delta E$ (shown in the colorbar) for the 1D $J_1$-$J_2$ model at $\Delta_{\rm{rel}}\approx5.15\pm0.05\%$. (a) and (b) are the ratio as a function of $(L,M)$ and  $(L,\sqrt{M})$, respectively, for RGN with $\delta=0.01$. (c) and (d) are the ratio as a function of $(L,M)$ and  $(L,\sqrt{M})$, respectively, for RGN with $\delta=0.001$. (c) and (d) are the ratio as a function of $(L,M)$ and $(L,\sqrt{M})$, respectively, for SR with $\delta=0.01$. }
    \label{fig:1step_J1J2}
\end{figure}

We again consider the effect of statistical noise on a single optimization step at fixed wavefunction quality. We first note that the near singularities in the exact $S$ and $H$ matrices lead to a large condition number in defining the update, and the statistical noise can be large compared to the small eigenvalues, which leads to a more noisy step than in the model problem. Therefore, we plot in Fig.~\ref{fig:1step_J1J2} the {\it fitted} 1-step ratio $\Delta_ME/\Delta E$ for RGN and SR at wavefunctions with $\Delta_{\rm{rel}}=5.15\pm0.05\%$, with the computational procedure described in Appendix~\ref{sec:1step_J1J2}. 

Fig.~\ref{fig:p5_c2_rgn_0} shows $\Delta_ME/\Delta E$ as a function of $(L,M)$ from the RGN update with $\delta=0.01$, where the black curve emphasizes the approximate transition from stable to unstable RGN updates, corresponding to $M\sim c_{\rm{RGN}}L^2$ with $c_{\rm{RGN}}\approx0.36$. Fig.~\ref{fig:p5_c2_rgn_2} shows the same data with rescaled axes $(L,\sqrt{M})$, where the approximate transition corresponding to $\sqrt{M}\sim\sqrt{c_{\rm{RGN}}}L$ is again plotted in black. Similarly, Fig.~\ref{fig:p5_c3_rgn_0} and Fig.~\ref{fig:p5_c3_rgn_2} show $\Delta_ME/\Delta E$ as a function of $(L,M)$ and $(L,\sqrt{M})$, respectively, for the RGN update with $\delta=0.001$, where $c_{\rm{RGN}}\approx1.2$ for the black curves. These plots confirm the scaling $M\sim L^2$ for stable RGN updates as observed in Section~\ref{sec:model}. 

Fig.~\ref{fig:p5_c2_sr_0} and Fig.~\ref{fig:p5_c2_sr_2} show $\Delta_ME/\Delta E$ as a function of $(L,M)$ and $(L,\sqrt{M})$, respectively, for the SR update with $\delta=0.01$. For SR, the required sampling size scaling with system size is less clear, although it is clear that the sample requirements grow with $L$, and that for a given $L$, the stable SR update requires a much smaller $M$ than the stable RGN update. 

\subsubsection{Optimization trajectory}\label{sec:opt_J1J2}

\begin{figure*}[htb]
    \centering
    \begin{subfigure}{0.43\linewidth}
        \centering
        \includegraphics[width=1\linewidth]{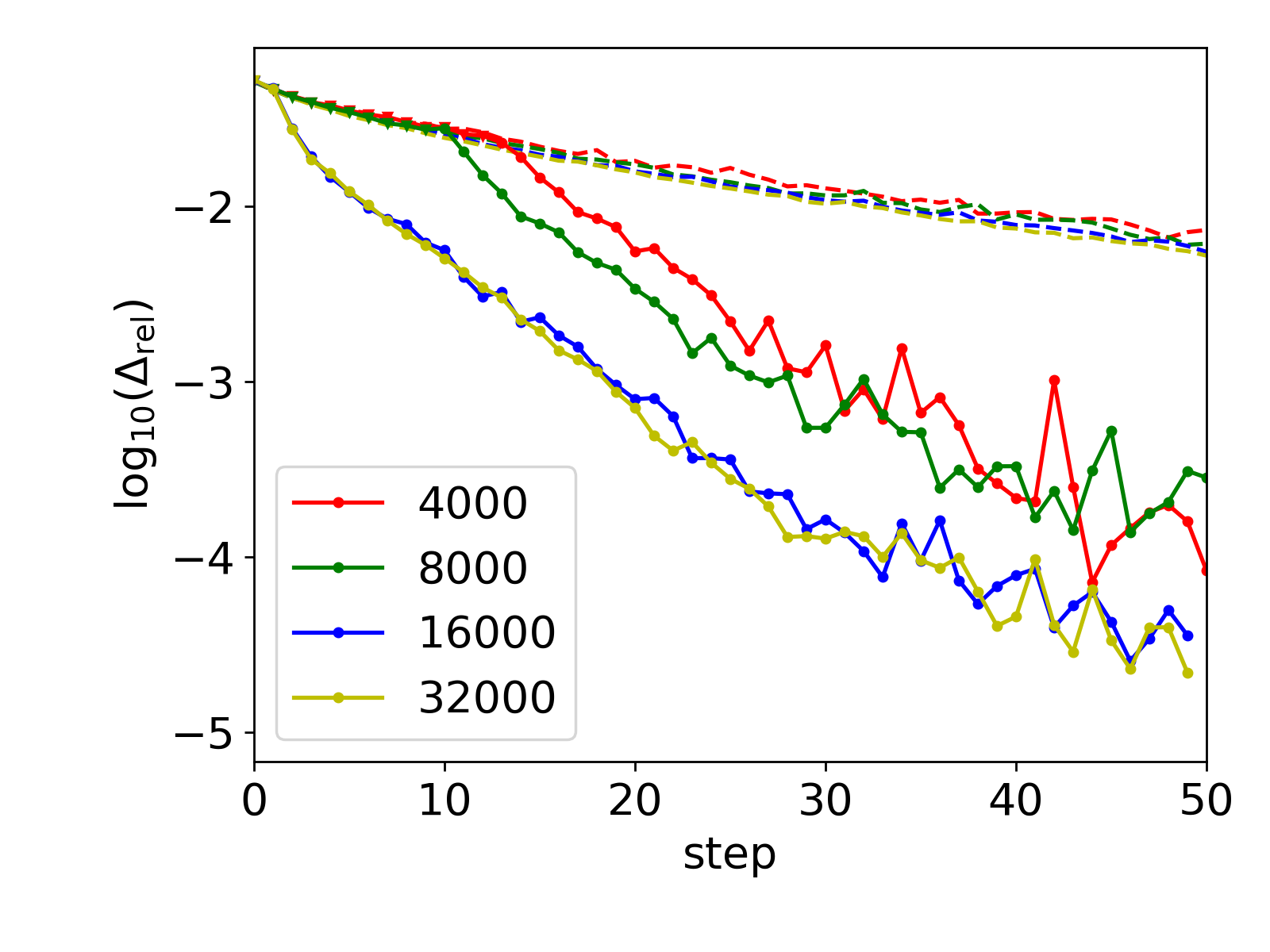}
        \caption{}
        \label{fig:J1J2_100_3}
    \end{subfigure}
    \centering
    \begin{subfigure}{0.43\linewidth}
        \centering
        \includegraphics[width=1\linewidth]{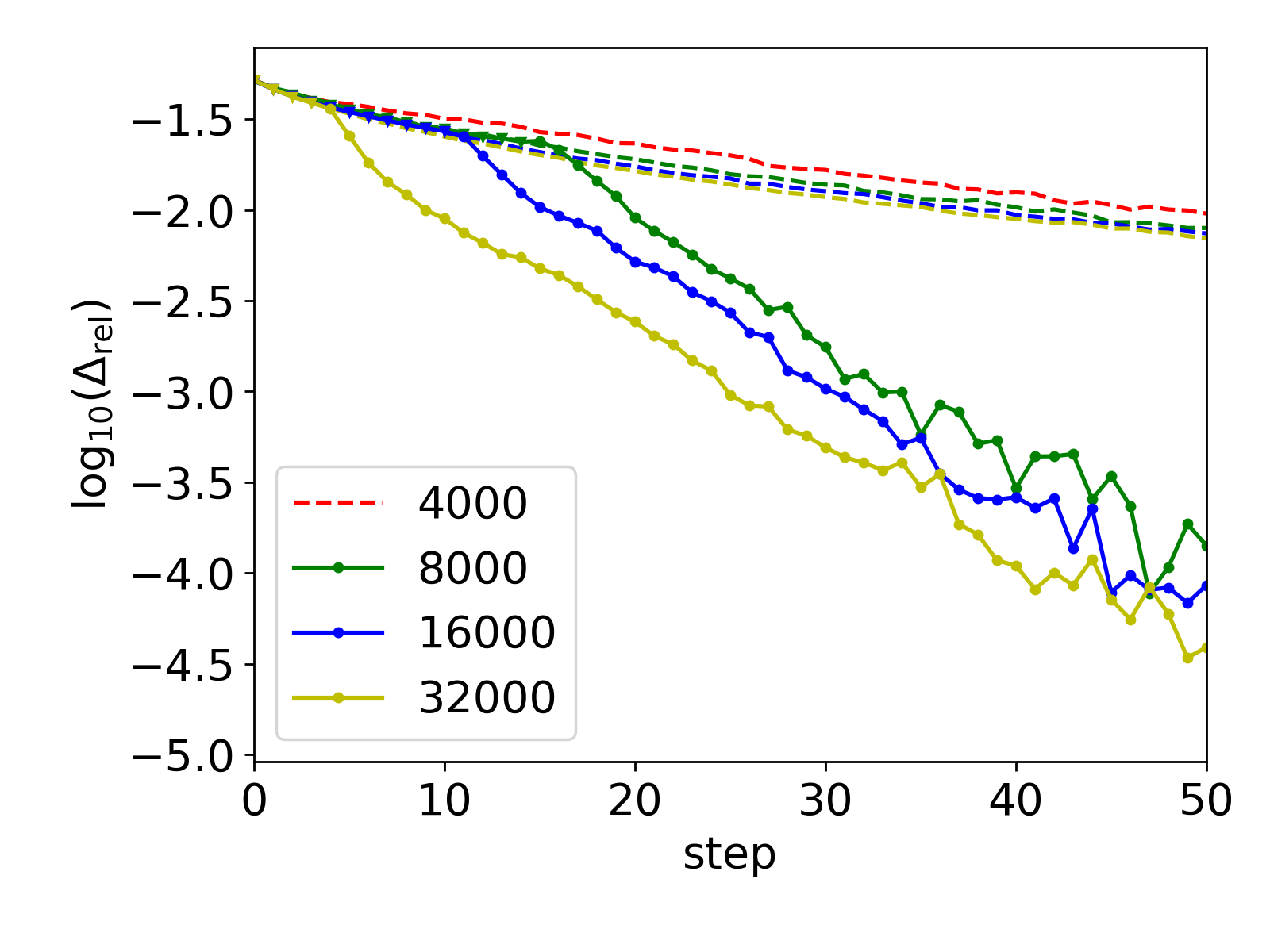}
        \caption{}
        \label{fig:J1J2_200_3}
    \end{subfigure}
        \centering
    \begin{subfigure}{0.43\linewidth}
        \centering
        \includegraphics[width=1\linewidth]{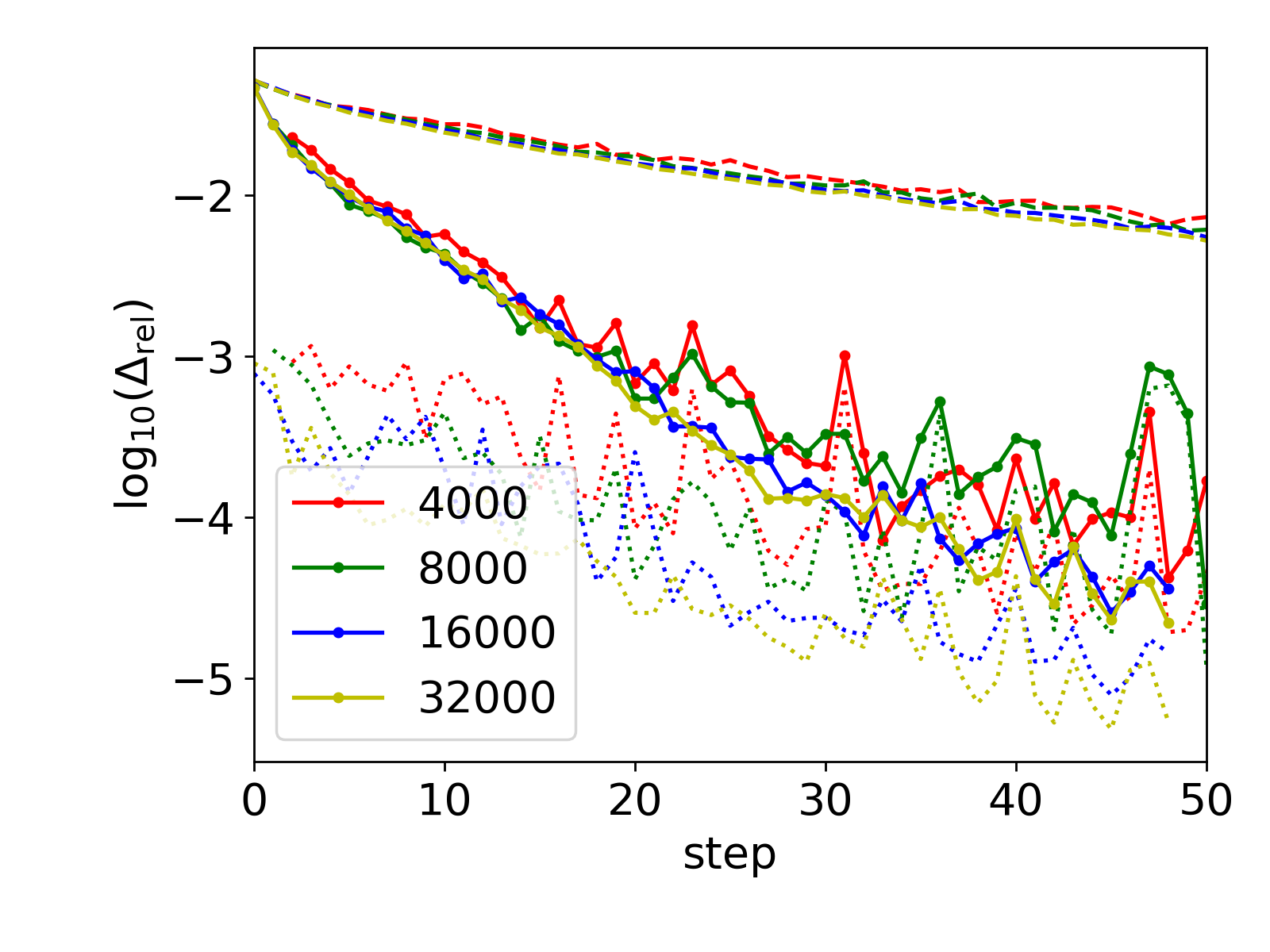}
        \caption{}
        \label{fig:J1J2_100_3_shift}
    \end{subfigure}
    \centering
    \begin{subfigure}{0.43\linewidth}
        \centering
        \includegraphics[width=1\linewidth]{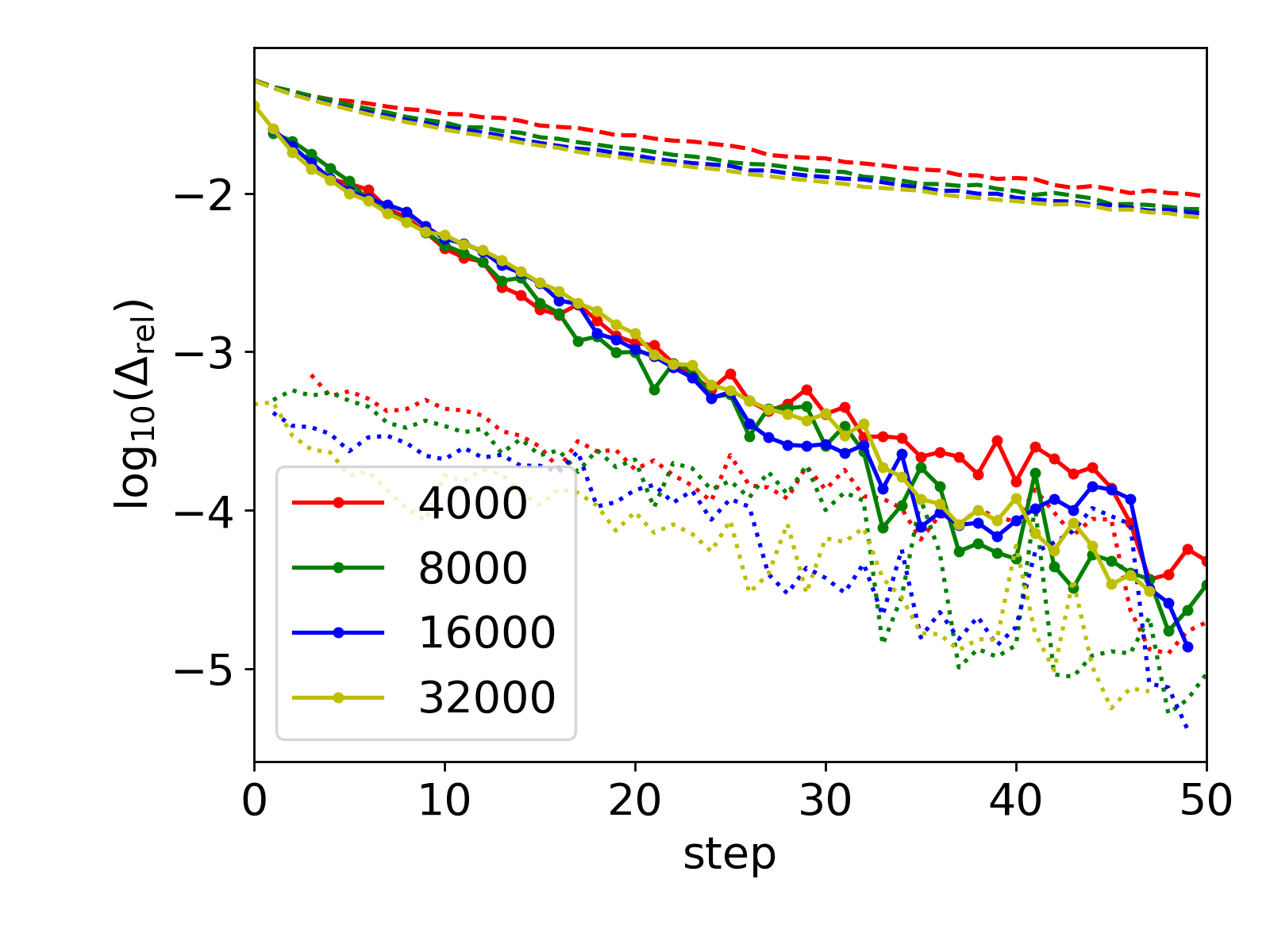}
        \caption{}
        \label{fig:J1J2_200_3_shift}
    \end{subfigure}
    \caption{Optimization trajectories for the 1D $J_1$-$J_2$ model at (a) $L=100$ and (b) $L=200$. In all plots, we use $\delta=0.001$, and each color corresponds to a sample size $M$. SR (RGN) results are plotted in dashed curves (solid curves connected with markers). (c) and (d) compares the slopes of the trajectories in (a) and (b) respectively, by plotting the same data as (a) and (b), but with the RGN trajectories shifted so that the first successful RGN step of each trajectory align. The relative energy stochastic error $\sigma_M/|E_{\rm{g.s.}}|$ for each RGN trajectory is also plotted in the corresponding color with dotted curves.}
    \label{fig:J1J2_opt_3}
\end{figure*}

We now consider the optimization trajectories of SR and RGN in the 1D $J_1$-$J_2$ model. Fig.~\ref{fig:J1J2_100_3} and Fig.~\ref{fig:J1J2_200_3} plot the optimization trajectories for SR (dashed) and RGN (solid, with markers) at $L=100$ and $L=200$ with $\delta=0.001$, where different sample sizes $M$ are shown in different colors.  We first observe that, when the wavefunction has large $\Delta_{\rm{rel}}$, insufficient sampling could cause the RGN matrix $H+S/\epsilon$ (dropping $\delta$ for simplicity) to have negative eigenvalues, and thus the updated wavefunction could have a higher energy. These steps correspond to triangle markers in the RGN trajectories, where our stabilizing mechanism uses instead the SR update, hence the trajectory follows that of SR. By comparing Fig.~\ref{fig:J1J2_100_3} and Fig.~\ref{fig:J1J2_200_3}, we see that the required sample size for a successful RGN update grows significantly from $L=100$ to $L=200$ at a fixed (large) $\Delta_{\rm{rel}}$. This is consistent with the observation in Section~\ref{sec:j1j2_1step} that at fixed (large) $\Delta_{\rm{rel}}$, the sample size requirement for good RGN performance scales sharply with system size. 

On the other hand, for a fixed $L$, the overall slopes of the RGN trajectories for different $M$ are very similar. This can be seen more clearly by shifting the RGN trajectories, so that the first successful RGN step in each of the trajectories align, as shown in Fig.~\ref{fig:J1J2_100_3_shift} and Fig.~\ref{fig:J1J2_200_3_shift}. In particular, the RGN trajectories (solid with markers) with different $M$ almost overlap for the range $-3.5\leq\log_{10}(\Delta_{\rm{rel}})\leq-2$ in both figures. After this intermediate error regime, the RGN trajectories with different $M$ start to diverge in the high accuracy regime, i.e. for $\log_{10}(\Delta_{\rm{rel}})<-3.5$. In Fig.~\ref{fig:J1J2_100_3_shift} and Fig.~\ref{fig:J1J2_200_3_shift}, we also plot $\sigma_M/E_{\rm{g.s.}}$ for each RGN trajectory in the corresponding color with dotted curves, where $\sigma_M$ is the standard deviation of the sampled energy. We see that the RGN trajectories with small $M$ each approach a $\Delta_{\rm{rel}}$ that saturates $\sigma_M/|E_{\rm{g.s.}}|$. We finally remark that the SR trajectories (dashed) with different $M$ start to diverge at relatively large $\Delta_{\rm{rel}}$ (note the logarithmic $y$ axis), suggesting that its performance also depends on the sample size. In Appendix~\ref{sec:delta001_J1J2}, we show the results for repeating the calculations with $\delta=0.01$, which confirm the observations above.

\subsubsection{Influence of variational expressivity on sample size scaling}\label{sec:D1}

\begin{figure*}[htb]
    \centering
    \begin{subfigure}{0.329\linewidth}
        \centering
        \includegraphics[width=1\linewidth]{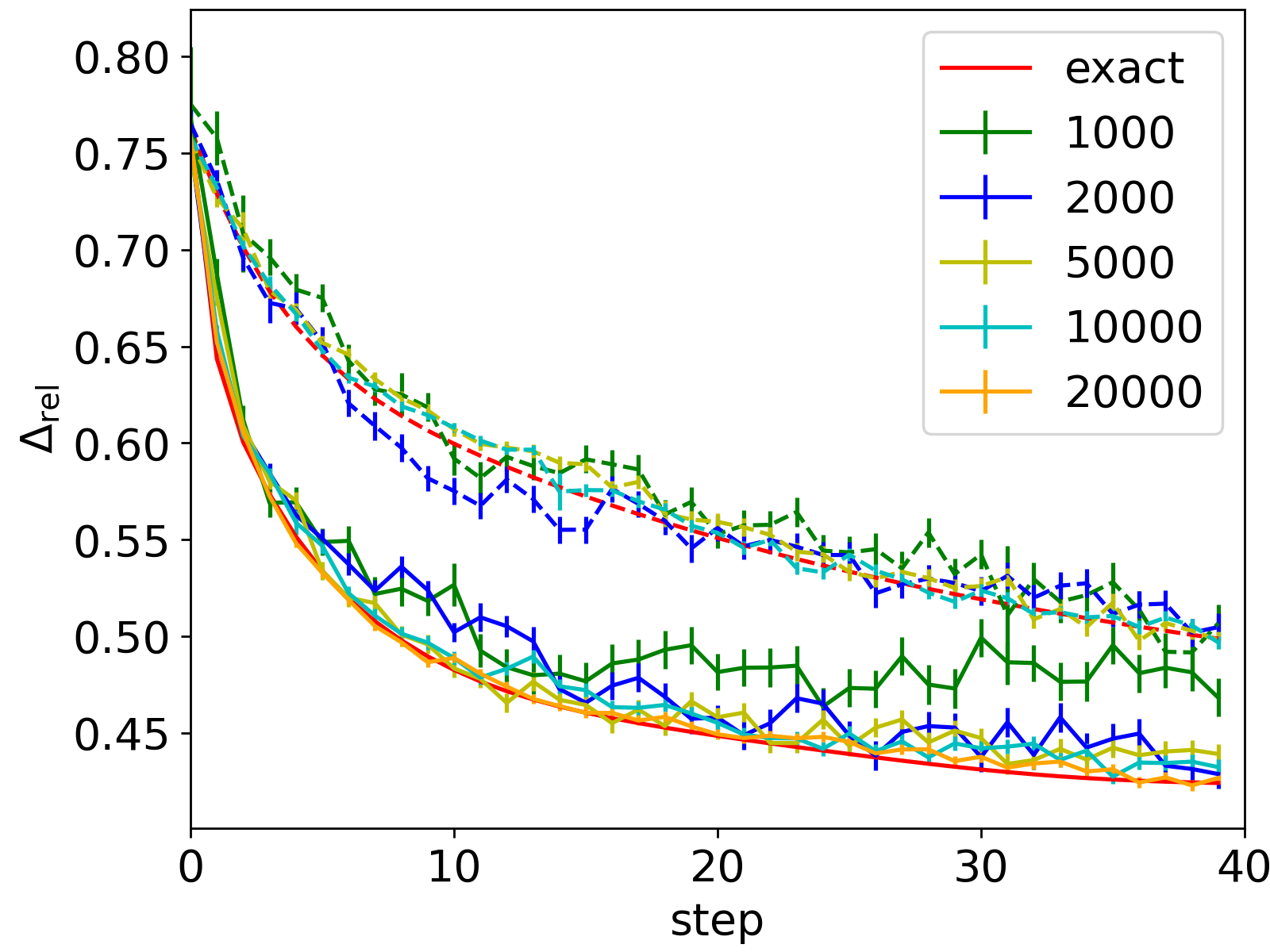}
        \caption{}
        \label{fig:L6_D1}
    \end{subfigure}
    \centering
    \begin{subfigure}{0.329\linewidth}
        \centering
        \includegraphics[width=1\linewidth]{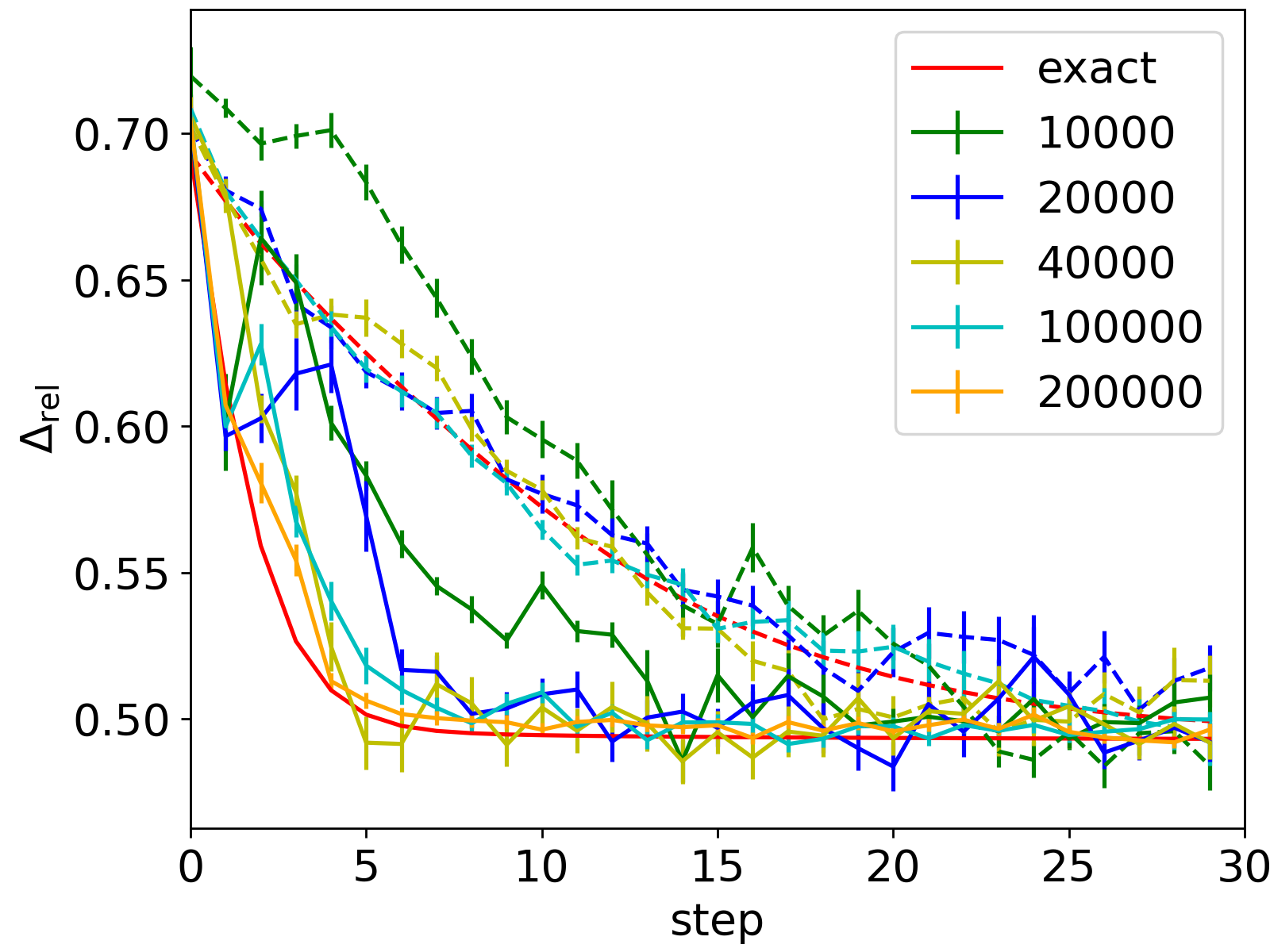}
        \caption{}
        \label{fig:L10_D1}
    \end{subfigure}
    \centering
    \begin{subfigure}{0.329\linewidth}
        \centering
        \includegraphics[width=1\linewidth]{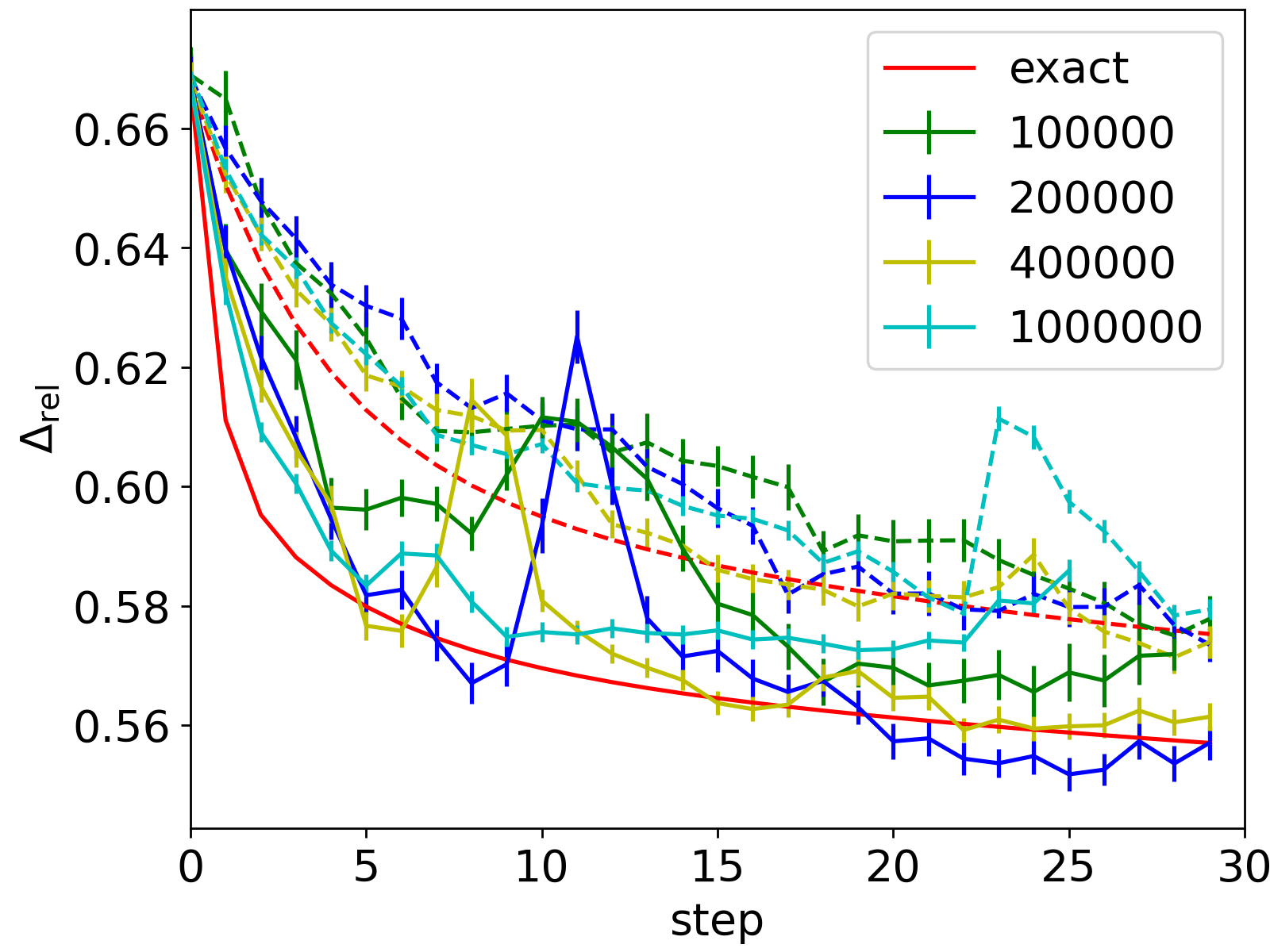}
        \caption{}
        \label{fig:L20_D1}
    \end{subfigure}
    \caption{Optimization trajectories for the 1D $J_1$-$J_2$ model  using an MPS of bond dimension $D=1$ at (a) $L=6$, (b) $L=10$ and (c) $L=20$. In each plot, each color corresponds to a sample size $M$ and red corresponds to the result from exact sampling. SR (RGN) results are plotted in dashed (solid) curves. $\sigma_M/|E_{\rm{g.s.}}|$ are plotted as error bar. }
    \label{fig:D1}
\end{figure*} 

We finally consider the effect of wavefunction expressivity on the sampling requirements of SR and RGN. While the  1-dimensional $J_1$-$J_2$ ground-state can be written as an MPS of small bond dimension, this is exceptional to the common VMC scenario, where the wavefunction ansatz does not contain the exact ground-state in the variational manifold. To emulate this, we consider an MPS with bond dimension $D=1$ as the variational state, which does not contain the ground-state. Fig.~\ref{fig:D1} plots the optimization trajectories of SR (dashed) and RGN (solid) for $L=6$, $L=10$ and $L=20$ at $\delta=0.001$. Different sample sizes are labeled by different colors, with $\sigma_M/|E_{\rm{g.s.}}|$ plotted as the error bar, and red denotes the result using exact sampling. 

We first note that RGN with a small sample size can occasionally obtain wavefunctions with low energy along the optimization trajectory. For instance, in Fig.~\ref{fig:L20_D1}, the lowest energy wavefunction corresponds to RGN optimization with 200000 samples (solid blue) at step 25. However, the optimization trajectory does not in fact follow that of the noiseless case, and has a large jump. This suggests that this case is best thought of as optimization from a different starting point (after the jump) and that it is not a reliable strategy to use small sample sizes, as corroborated by the data for other system sizes. {Nonetheless, it shows that it can be preferable to choose the optimized wavefunction corresponding to the lowest energy during the optimization run, taking into account the statistical error of the energy, rather than from the last optimization iteration.}

We also see that, for both RGN and SR, a stable update requires a much sharper scaling of sample size with $L$ for the $D=1$ MPS, than when the MPS contains the ground state (Section~\ref{sec:opt_J1J2}). In particular, with the $D=1$ MPS, the sample size requirement for RGN increases from $M=2000$ for $L=6$, to $M=40000$ for $L=10$, to $M>1000000$ for $L=20$; and for SR it increases from $M<1000$ for $L=6$, to $M=10000$ for $L=10$, to $M=100000$ for $L=20$. This is to be contrasted with the $D=5$ MPS, where the RGN sample size requirement scales mildly with $L$ if $\Delta_{\rm{rel}}$ is in the intermediate to high accuracy regime, and the SR trajectories show a modest dependence on sample size and never become unstable. Furthermore, the sharp sample size scaling with $L$ is also reflected in the increasing energy infidelity of the converged variational minimum with $L$ (i.e., $\Delta_{\rm{rel}}<0.45$, $\Delta_{\rm{rel}}\approx 0.5$, and $\Delta_{\rm{rel}}\approx0.55$ for $L=6$, $L=10$, and $L=20$, respectively).



\subsection{2D $J_1$-$J_2$ model}\label{sec:2D_J1J2}

We finally consider the non-trivial 2D $J_1$-$J_2$ spin $1/2$ model with open boundary conditions on a $L \times L$ lattice
\begin{align}
\hat{H}=J_1\sum_{\langle i,j\rangle}\Vec{S}_i\cdot\Vec{S}_j+J_2\sum_{\langle\langle i,j\rangle\rangle}\Vec{S}_i\cdot\Vec{S}_j
\end{align}
with $J_1=1$ and $J_2=0.5$ (we report energies in units of $J_1$). For the wavefunction, we use the projected entangled pair state (PEPS)~\cite{verstraete2006,orus2019,ORUS2014117,Verstraete2008}
\begin{align}
    \hat{\psi}(s_{1,1}, s_{1,2}, \ldots, s_{L,L}) = \mathrm{tr} [A(s_{1,1}) A(s_{1,2}) \ldots A(s_{L, L})]
\end{align}
where $A(s)$ are $ D\times D \times D\times D$ tensors at each site of the lattice and the trace is over all common tensor indices according to the bonds of the lattice.
Similar to the MPS wavefunction, the PEPS wavefunction also contains redundant parameters. Therefore, we use a diagonal shift of $\delta=0.001$ to remove singularities when inverting Eq.~\ref{eq:invert_sr} and Eq.~\ref{eq:invert_rgn} in all the calculations below. 

Although the quality of the PEPS ansatz increases as $D$ increases, the ground-state of this model cannot be exactly represented by a PEPS with finite $D$. Thus the VMC optimization is always over an approximate ground-state manifold, unlike (most of) the simulations in the earlier sections. Therefore, in the following we define $\Delta_{\rm{rel}}=|(E-E_{\rm{min}})/E_{\rm{min}}|$ where $E_{\rm{min}}$ is the variational minimum of the PEPS ansatz. 

\subsubsection{Influence of variational expressivity on optimization trajectory}\label{sec:2D_D}

As the PEPS we consider is an approximate ansatz (at finite $D$) we first consider the effect of wavefunction variational expressivity on the sampling requirement of SR and RGN. 

\begin{figure*}[htb]
    \centering
    \begin{subfigure}{0.43\linewidth}
        \centering
        \includegraphics[width=1\linewidth]{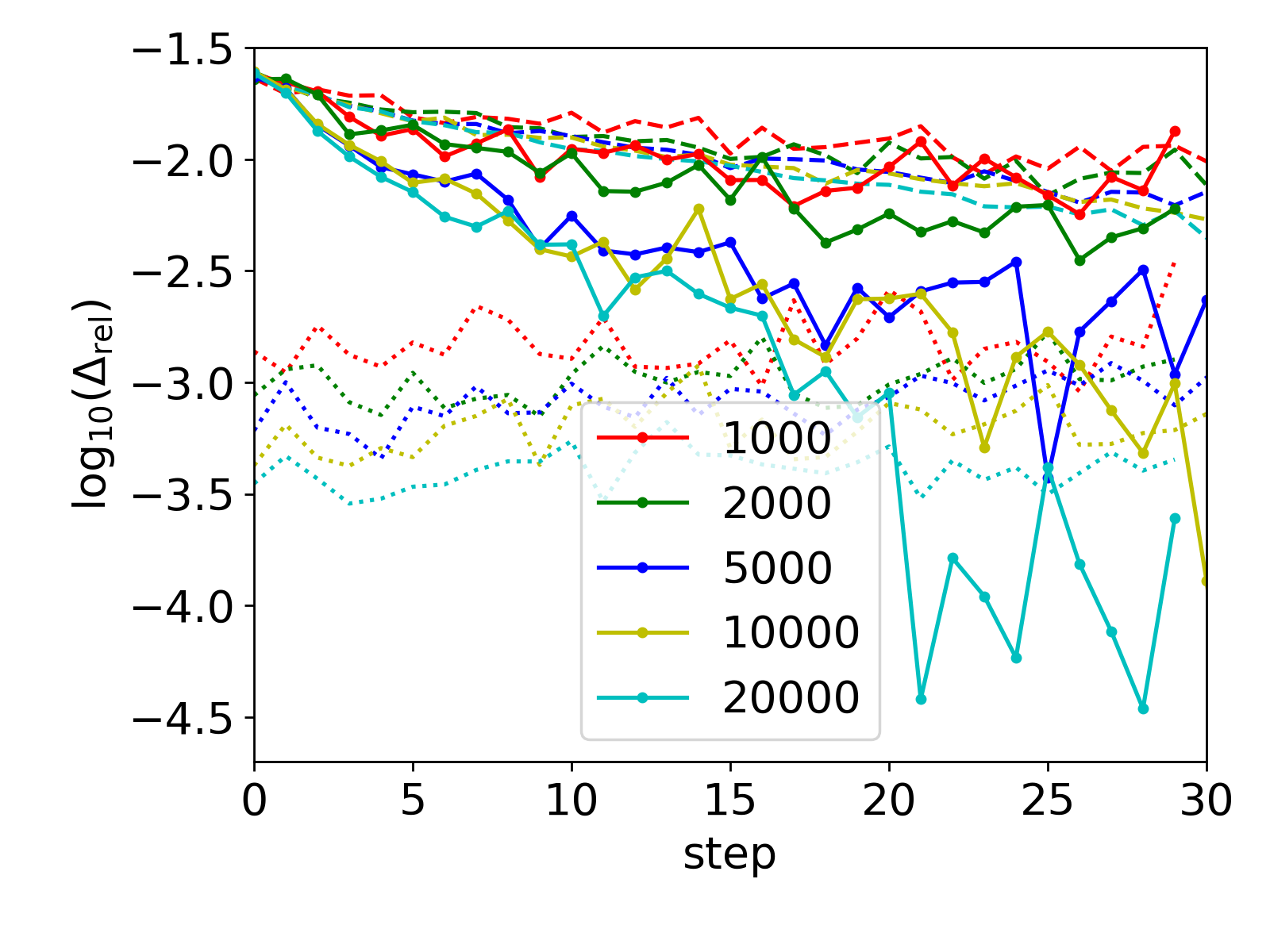}
        \caption{}
        \label{fig:6x6_D3}
    \end{subfigure}
    \centering
    \begin{subfigure}{0.43\linewidth}
        \centering
        \includegraphics[width=1\linewidth]{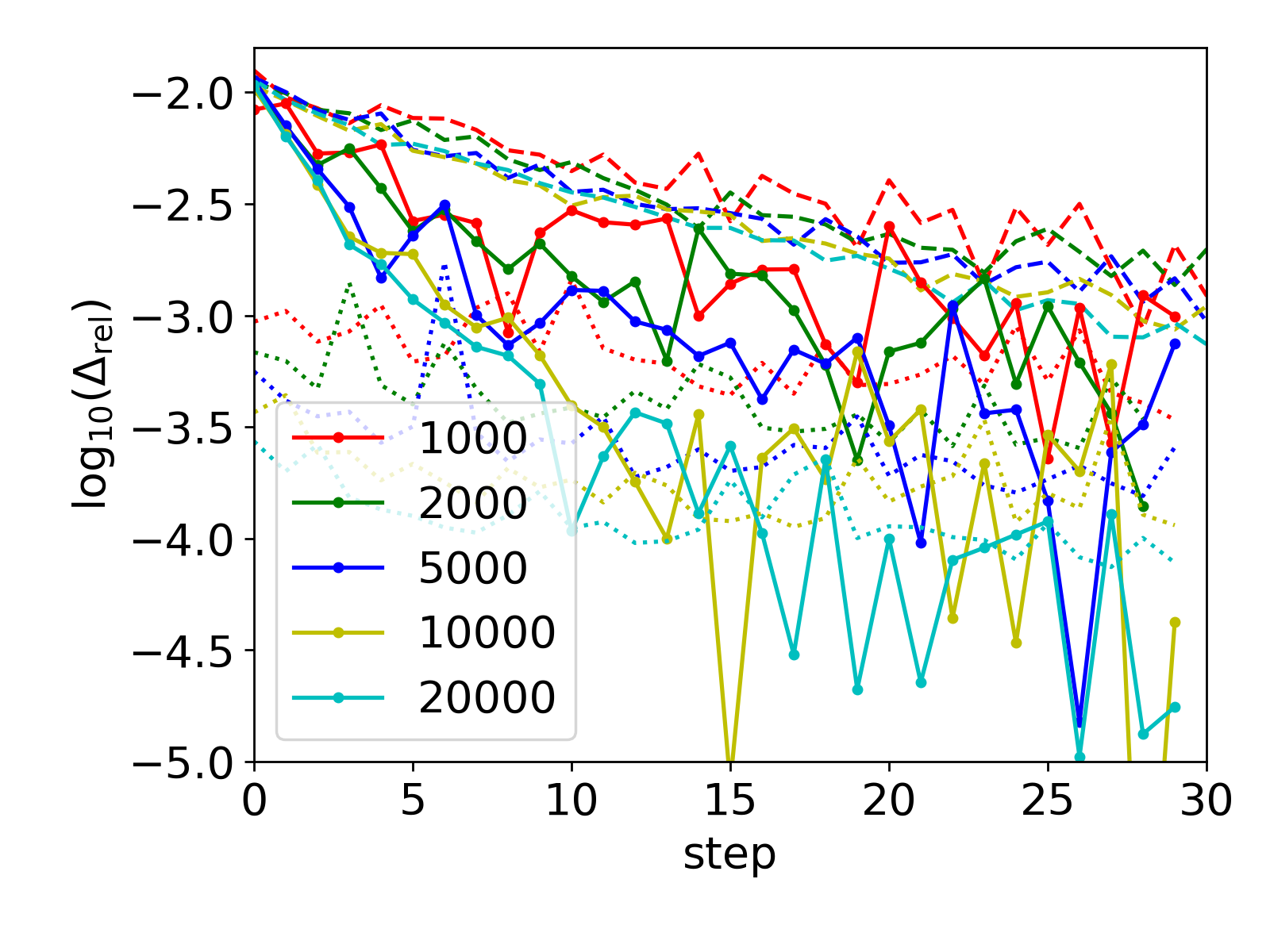}
        \caption{}
        \label{fig:6x6_D4}
    \end{subfigure}   
    \centering
    \begin{subfigure}{0.43\linewidth}
        \centering
        \includegraphics[width=1\linewidth]{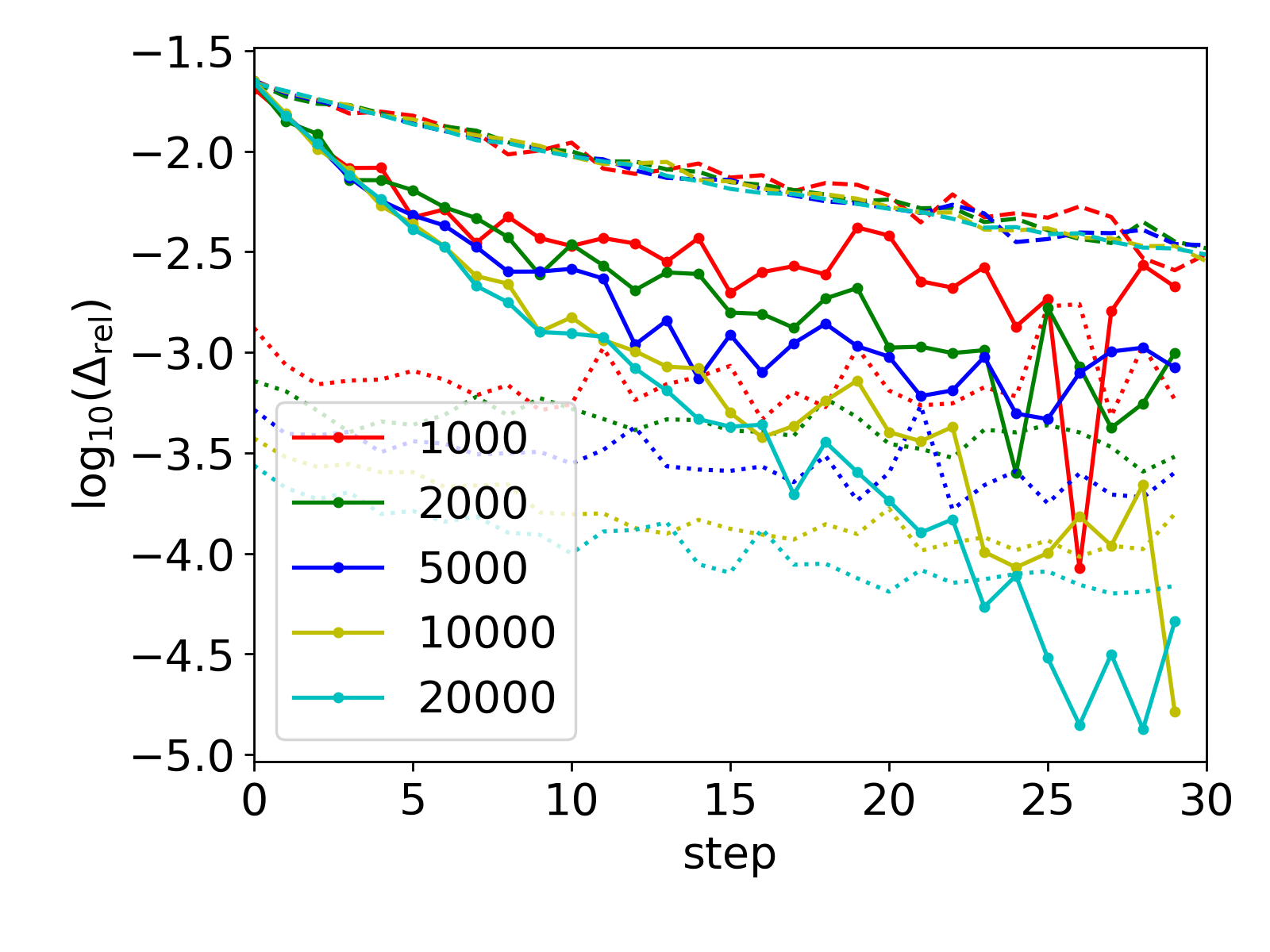}
        \caption{}
        \label{fig:6x6_D6}
    \end{subfigure}   
    \centering
    \begin{subfigure}{0.43\linewidth}
        \centering
        \includegraphics[width=1\linewidth]{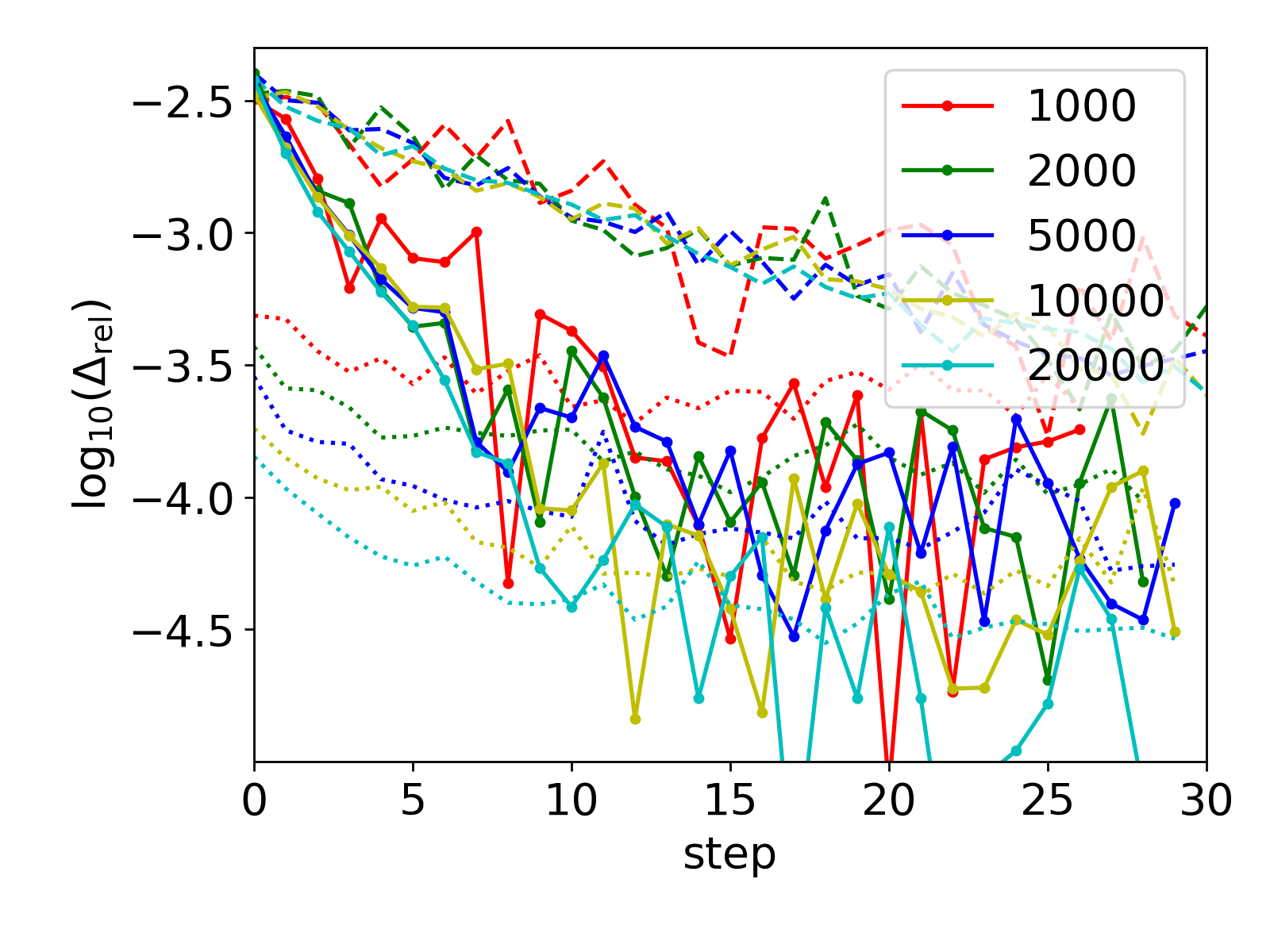}
        \caption{}
        \label{fig:6x6_D8}
    \end{subfigure}   
    \caption{$6\times6$ $J_1$-$J_2$ model optimization trajectories. (a) $D=3$, $e_{\rm{min}}=-0.47680$. (b) $D=4$, $e_{\rm{min}}=-0.47880$.  (c) $D=6$, $e_{\rm{min}}=-0.47892$. (d) $D=8$, $e_{\rm{min}}=-0.47906$. Different $M$ are labeled by different colors. SR (RGN) results are shown in dashed curves (solid curves with markers). The dotted curves are the relative energy stochastic error $\sigma_M/|E_{\rm{min}}|$ of the RGN trajectories. }
    \label{fig:6x6}
\end{figure*}

Fig.~\ref{fig:6x6} plots the optimization trajectories for a $6\times 6$ lattice with PEPS of $D=3$, $D=4$, $D=6$ and $D=8$, where $e_{\rm{min}}=E_{\rm{min}}/(L_x\times L_y)$ in each case is reported in the caption. The ground state per site energy $e_{\rm{g.s.}}=-0.47909$ from DMRG~\cite{10.1063/5.0180424} calculations. We first consider Fig.~\ref{fig:6x6_D8}, which plots the trajectories for $D=8$ PEPS, which for this system size has sufficient variational power to represent the ground state to within the stochastic errors that we achieve. Similarly to the observations made of the 1D $J_1$-$J_2$ model using the MPS wavefunction in Section~\ref{sec:opt_J1J2}, we again observe an intermediate accuracy range ($-3.5\leq\log_{10}(\Delta_{\rm{rel}})\leq-2.5$, first 10 optimization steps) where the RGN trajectories (solid, with markers) of different $M$ almost overlap, so that a sample size as small as $M=1000$ and $M=2000$ can achieve an almost optimal energy decrease. In the high accuracy regime $\log_{10}(\Delta_{\rm{rel}})\leq-3.5$, the $\Delta_{\rm{rel}}$ from the RGN trajectories for each $M$ saturates the corresponding relative energy stochastic error $\sigma_M/|E_{\rm{min}}|$, shown in the dotted curves of the corresponding color. 

If we consider Fig.~\ref{fig:6x6_D6}, Fig.~\ref{fig:6x6_D4} and Fig.~\ref{fig:6x6_D3} for the $D=6$, $D=4$ and $D=3$ PEPS, we see that the intermediate accuracy regime where RGN trajectories with different sample sizes overlap stops at a higher energy error; in Fig.~\ref{fig:6x6_D8} ($D=8$), this is $\log_{10}(\Delta_{\rm{rel}})\approx-3.5$; in Fig.~\ref{fig:6x6_D6} ($D=6$), this is $\log_{10}(\Delta_{\rm{rel}})\approx-2.2$; in Fig.~\ref{fig:6x6_D4} ($D=4$) and Fig.~\ref{fig:6x6_D3} ($D=3$) the overlapping region almost vanishes. Thus, for these more approximate wavefunctions, there is a strong dependence of RGN on sample size to optimize to within the statistical error in the energy. Similarly, for approximate PEPS (i.e. with small $D$), the SR trajectories (dashed) with different $M$ also show a much larger spread than for the more expressive wavefunctions with larger $D$, although the dependence on ($D$,$M$) seems less severe than for RGN. We therefore conclude that,  when optimizing PEPS with small $D$ using either SR or RGN, one needs larger sample sizes to obtain an optimal energy decrease. On the other hand, the performance degradation of RGN due to insufficient sample size is much more drastic than that of SR at small $D$. 

In Appendix~\ref{sec:6x8}, we provide the optimization trajectories for the $6\times 8$ lattice using PEPS of $D=4$, $D=6$, $D=8$ and $D=10$, which confirm the above observations. 

\subsubsection{Influence of system size on optimization trajectories}

\begin{figure*}[htb]
    \centering
    \begin{subfigure}{0.329\linewidth}
        \centering
        \includegraphics[width=1\linewidth]{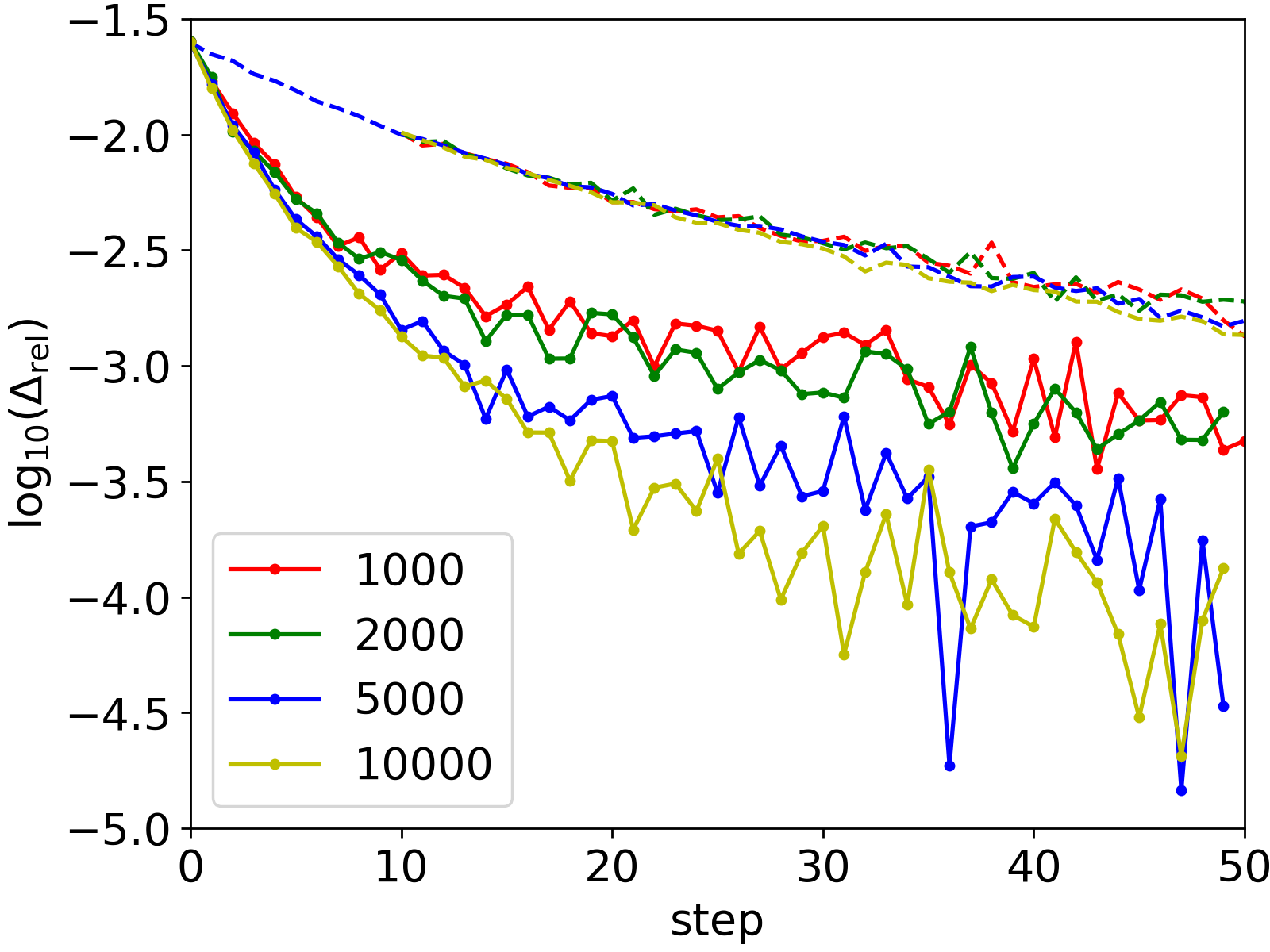}
        \caption{}
        \label{fig:8x8_D8}
    \end{subfigure}
    \centering
    \begin{subfigure}{0.329\linewidth}
        \centering
        \includegraphics[width=1\linewidth]{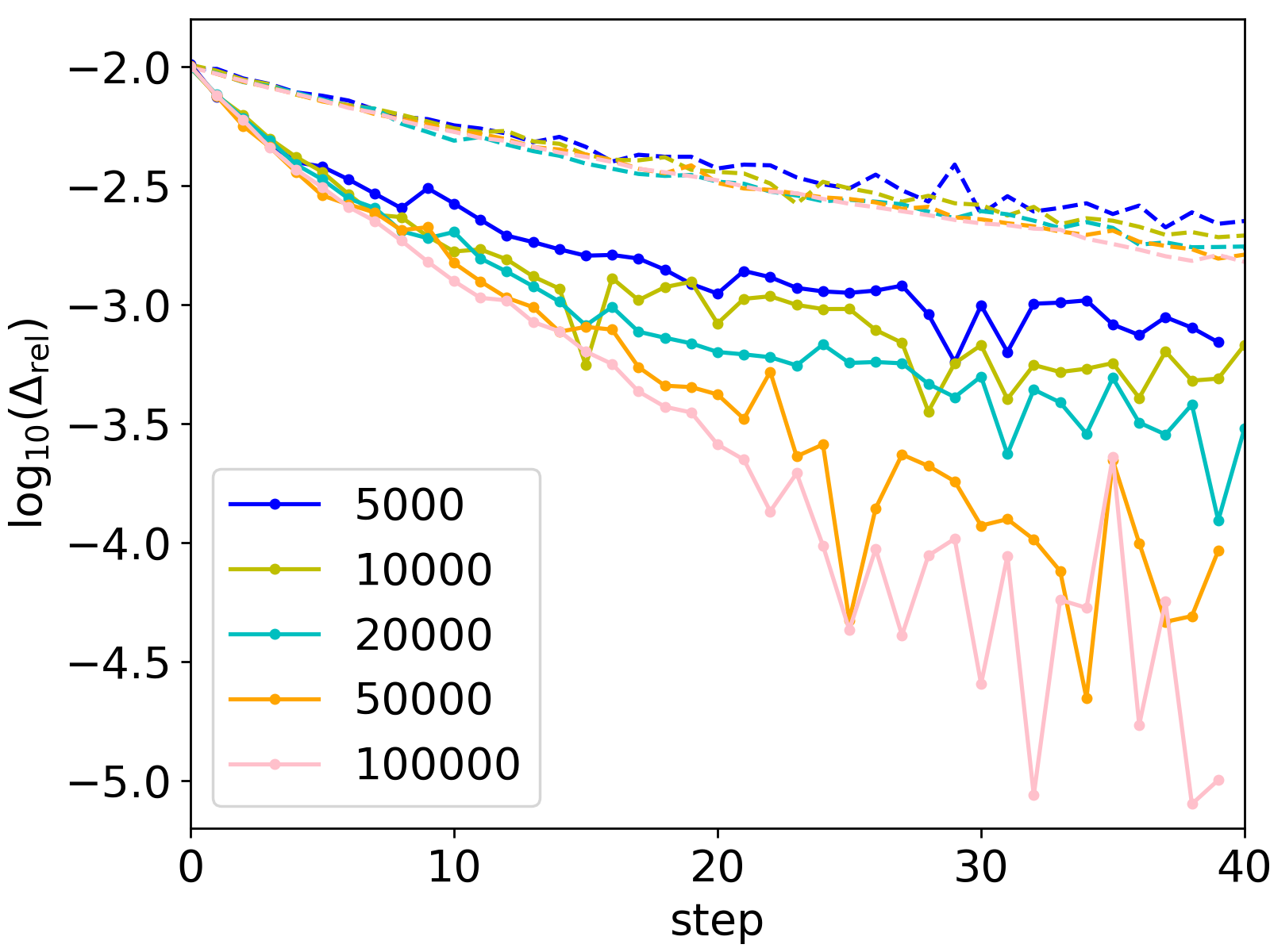}
        \caption{}
        \label{fig:10x8_D8}
    \end{subfigure}   
    \centering
    \begin{subfigure}{0.329\linewidth}
        \centering
        \includegraphics[width=1\linewidth]{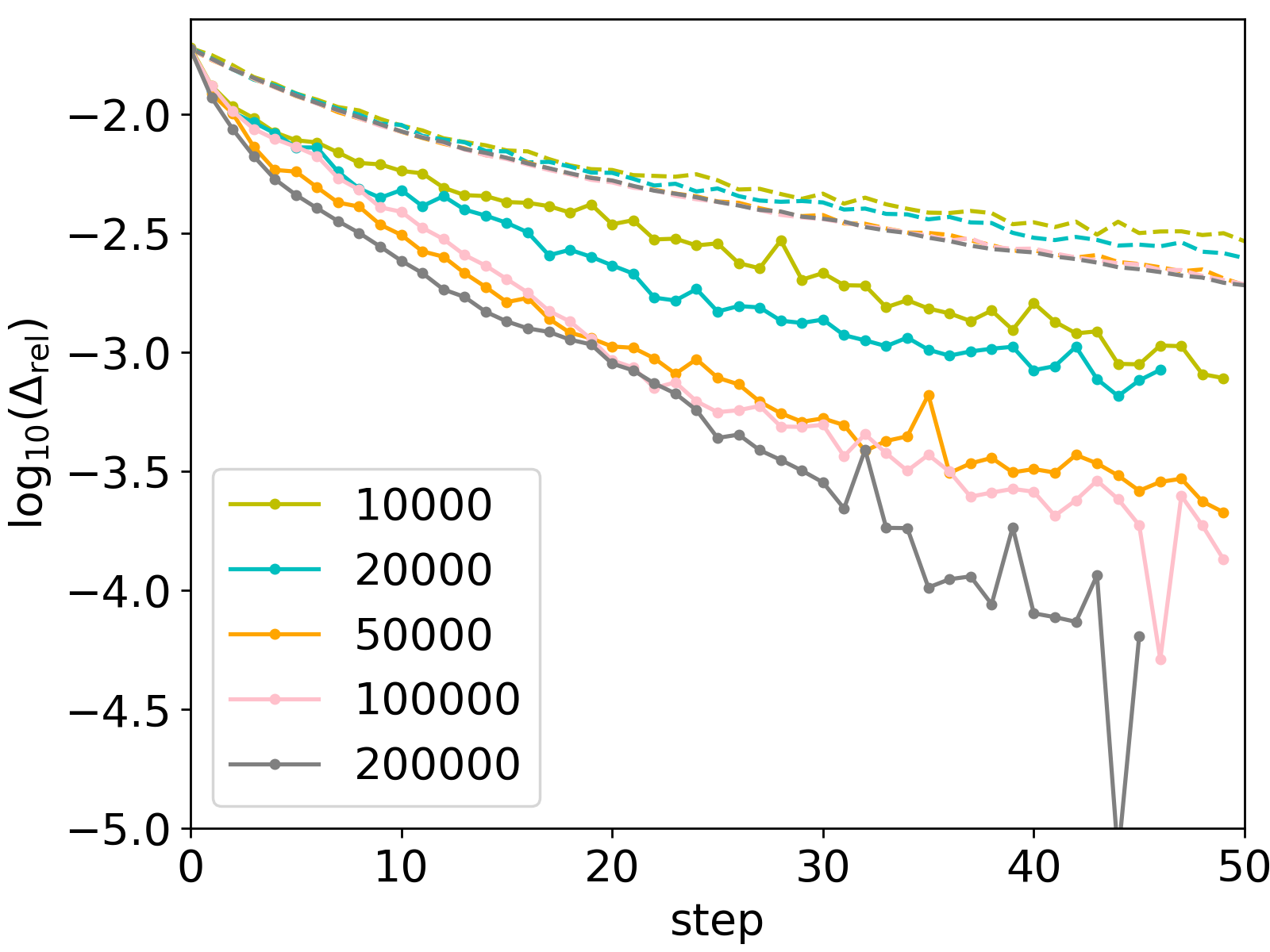}
        \caption{}
        \label{fig:10x10_D8}
    \end{subfigure} 
    \caption{$J_1$-$J_2$ model optimization trajectories of $D=8$ PEPS. (a) $8\times8$, $e_{\rm{min}}=-0.48373$. (b) $8\times10$, $e_{\rm{min}}=-0.48515$. (c) $10\times10$, $e_{\rm{min}}=-0.48654$. Different $M$ are labeled by color. SR (RGN) results are shown in dashed curves (solid curves with markers). }
    \label{fig:D8}
\end{figure*}

\begin{figure*}[htb]
    \centering
    \begin{subfigure}{0.329\linewidth}
        \centering
        \includegraphics[width=1\linewidth]{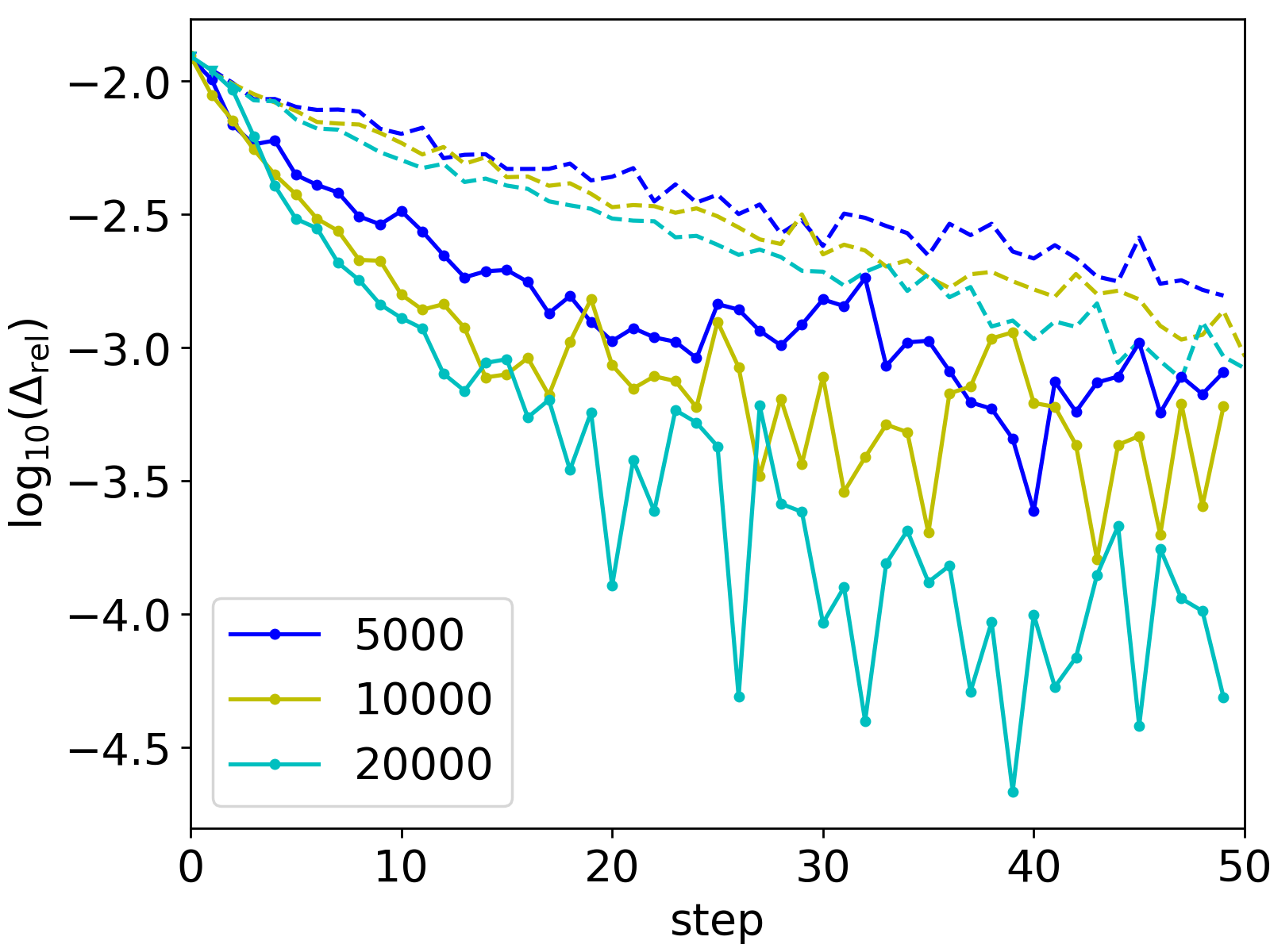}
        \caption{}
        \label{fig:8x8_D4}
    \end{subfigure}
    \centering
    \begin{subfigure}{0.329\linewidth}
        \centering
        \includegraphics[width=1\linewidth]{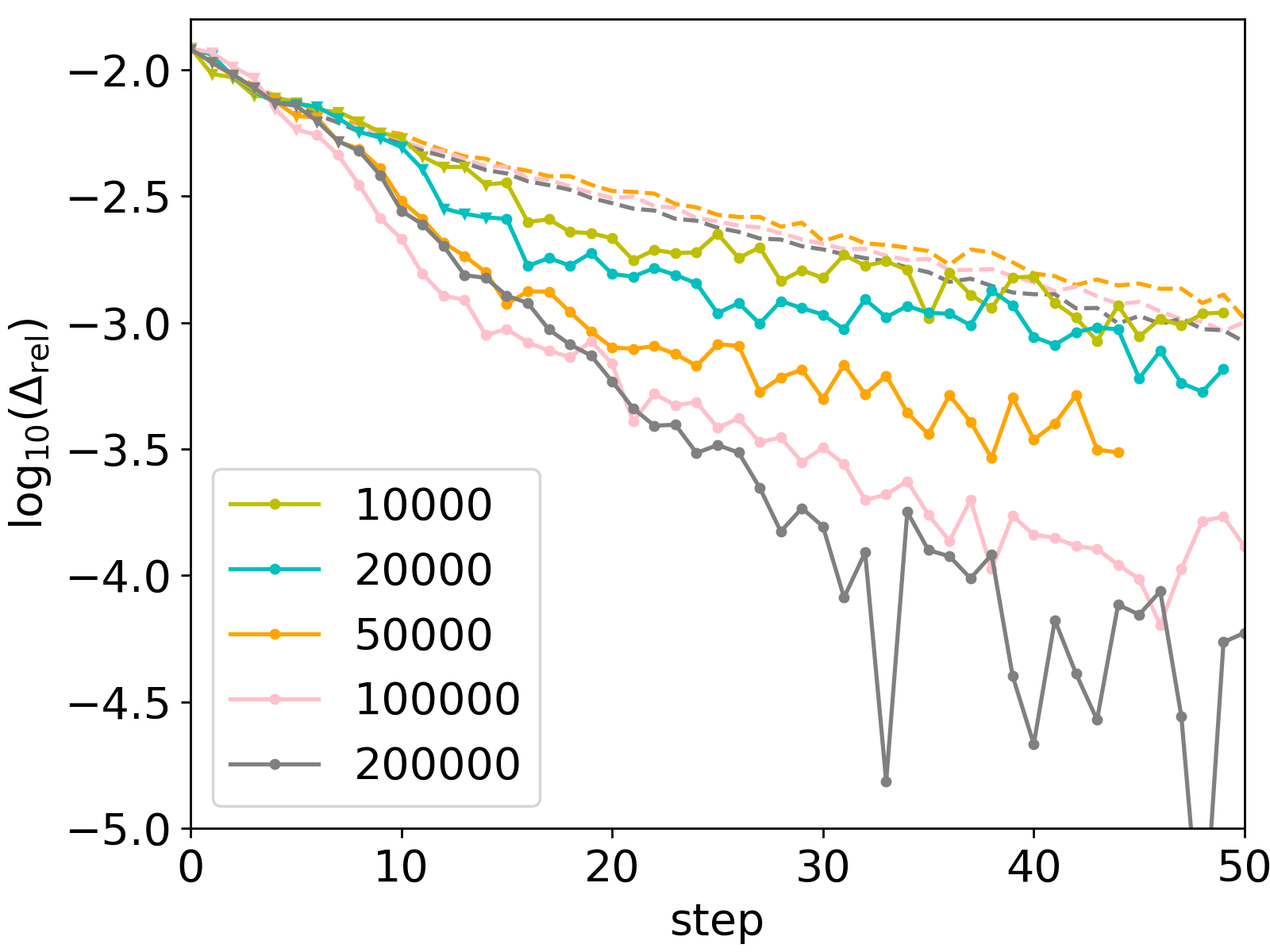}
        \caption{}
        \label{fig:8x10_D4}
    \end{subfigure}  
    \centering
    \begin{subfigure}{0.329\linewidth}
        \centering
        \includegraphics[width=1\linewidth]{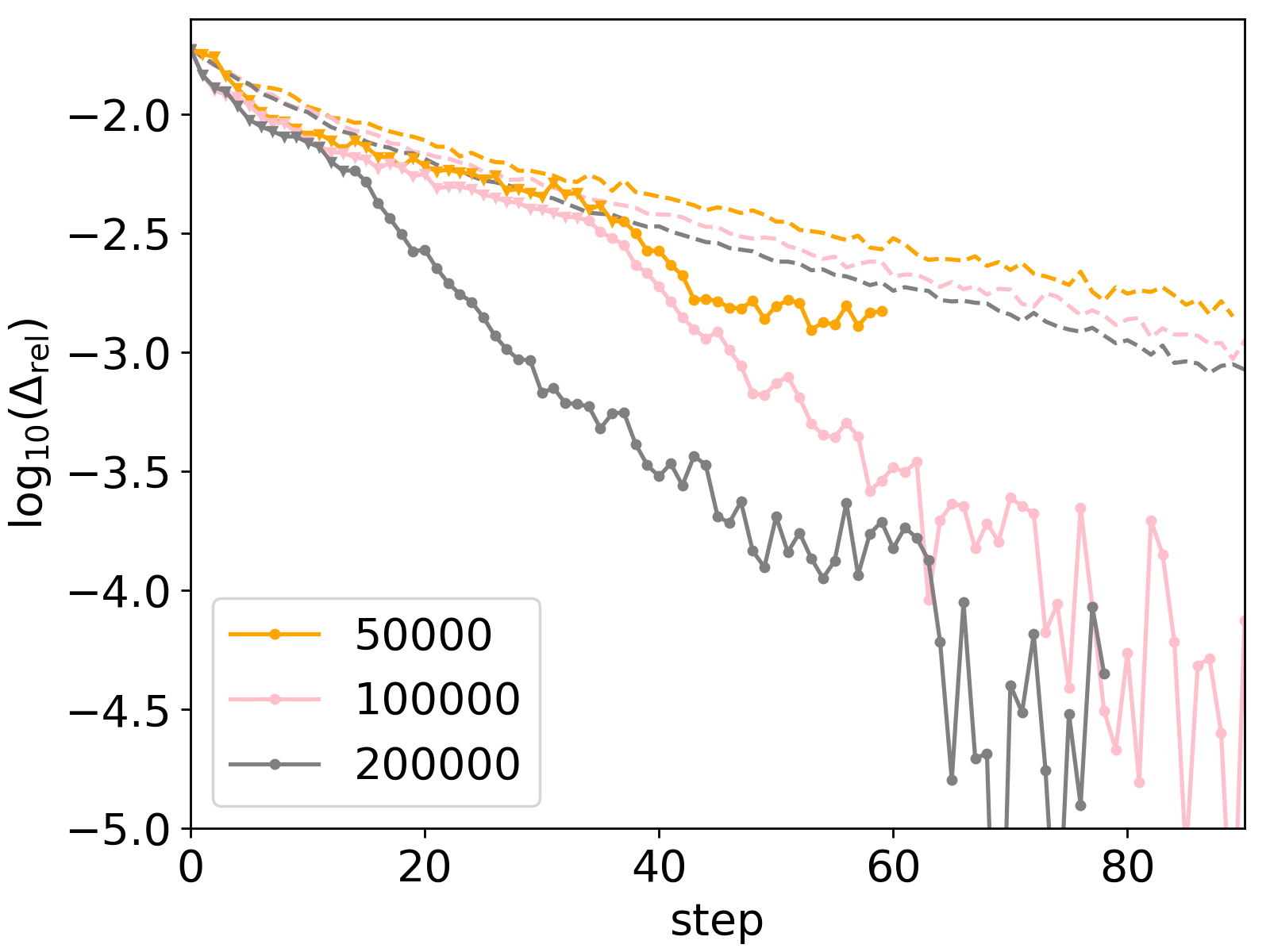}
        \caption{}
        \label{fig:10x10_D4}
    \end{subfigure}  
    \caption{$J_1$-$J_2$ model optimization trajectories of $D=4$ PEPS. (a) $8\times8$, $e_{\rm{min}}=-0.48330$. (b) $8\times10$, $e_{\rm{min}}=-0.48485$. (c) $10\times10$, $e_{\rm{min}}=-0.48610$. Different $M$ are labeled by color. SR (RGN) results are shown in dashed curves (solid curves with markers). }
    \label{fig:D4}
\end{figure*}

We finally consider optimizing a PEPS of a fixed $D$ for increasing lattice size. For the $8\times8$ and $10\times10$ lattices with $D=8$ PEPS, $E_{\rm{min}}$ are taken from Ref.~\cite{PhysRevB.98.241109}. Otherwise, $E_{\rm{min}}$ is taken as the converged result with the largest sample size. Note that unlike in the $6\times 6$ lattice, the $D=8$ PEPS does not contain the exact ground-state (to within the achieved statistical error) in these larger lattices. For example, as estimated from Ref.~\cite{PhysRevB.98.241109}, the difference between the $D=6$ and $D=8$ PEPS energies on the $8 \times 8$ lattice is $5 \times 10^{-5}$, and for the $10 \times 10$ lattice it is $2 \times 10^{-4}$. 

Fig.~\ref{fig:D8} plots the SR and RGN optimization trajectories for the $8\times8$, $8\times10$ and $10\times10$ lattices with PEPS of $D=8$. With sufficient samples, RGN optimization clearly converges to the variational minimum much faster than SR. On the other hand, to achieve significant convergence improvement over SR, the required sample size for RGN increases sharply with system size, e.g. 1000 for the $8\times8$ lattice, at least 5000 for the $8\times10$ lattice, and 50000 for $10\times10$ lattice. We note that this observation of sharp sample size scaling is similar to the one made in Section~\ref{sec:D1} for the $D=1$ MPS, where the accuracy of the ground-state at fixed bond dimension decreases as the system size increases (for small $L$). For the 2D $J_1$-$J_2$ model, the choice of parameters $J_2/J_1=0.5$ coincides with a potential gapless ground-state~\cite{PhysRevB.88.060402,PhysRevB.98.241109}, which violates the area law, and thus correctly representing the ground state (to a given fixed energy accuracy per site) is expected to require a PEPS whose bond dimension increases with lattice size. 

Fig.~\ref{fig:D4} plots the SR and RGN optimization trajectories for the $8\times8$, $8\times10$ and $10\times10$ lattices with PEPS of $D=4$. These results further confirm our previous observations: to obtain significant RGN acceleration, (i) the $D=4$ PEPS requires significantly larger sample size than the $D=8$ PEPS for the same lattice size, and (ii) the sample size requirement also grows sharply with system size. Finally, comparing the relative convergence rates of SR and RGN for the $D=4$ and $D=8$ cases, we see that the improvement of RGN over SR for a fixed sample size is larger for the more expressive $D=8$ case. 

\section{Summary}

\begin{figure}[htb]
    \centering
    \begin{subfigure}{0.49\linewidth}
        \centering
        \includegraphics[width=1\linewidth]{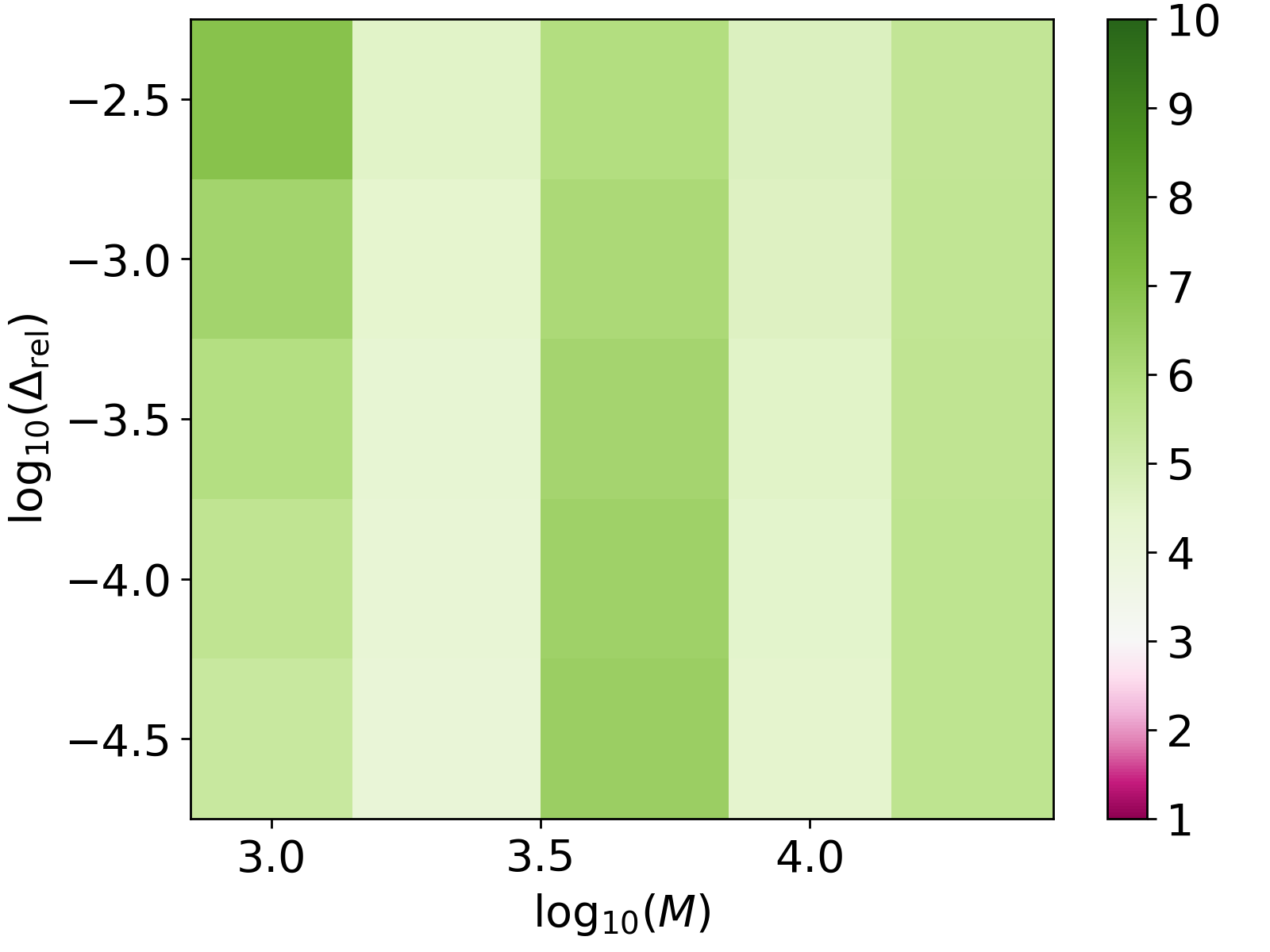}
        \caption{}
        \label{fig:nstep_L100_2}
    \end{subfigure}
    \centering
    \begin{subfigure}{0.49\linewidth}
        \centering
        \includegraphics[width=1\linewidth]{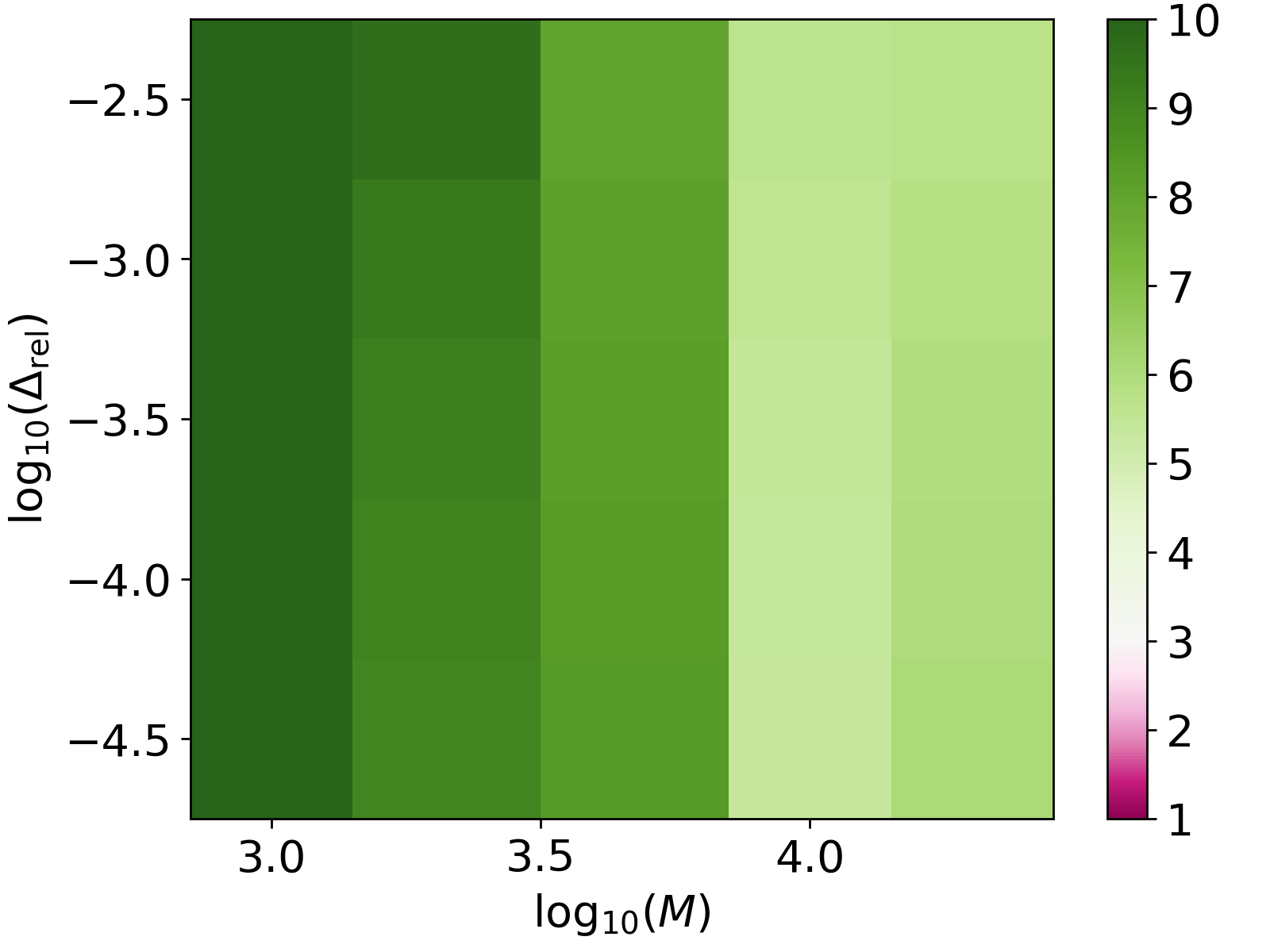}
        \caption{}
        \label{fig:nstep_L200_2}
    \end{subfigure}  
    \centering
    \begin{subfigure}{0.49\linewidth}
        \centering
        \includegraphics[width=1\linewidth]{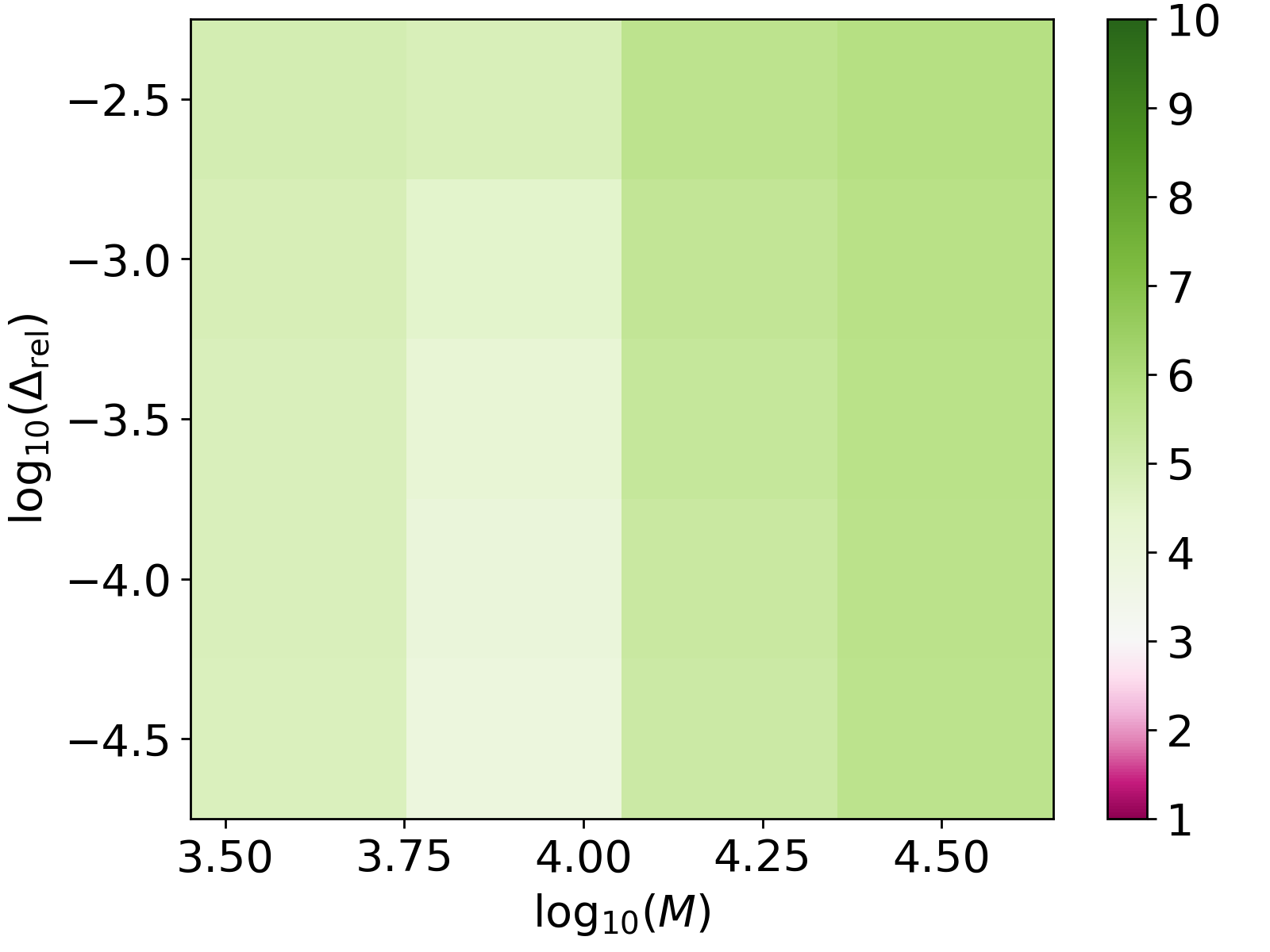}
        \caption{}
        \label{fig:nstep_L100_3}
        \centering
    \end{subfigure}  
        \begin{subfigure}{0.49\linewidth}
        \centering
        \includegraphics[width=1\linewidth]{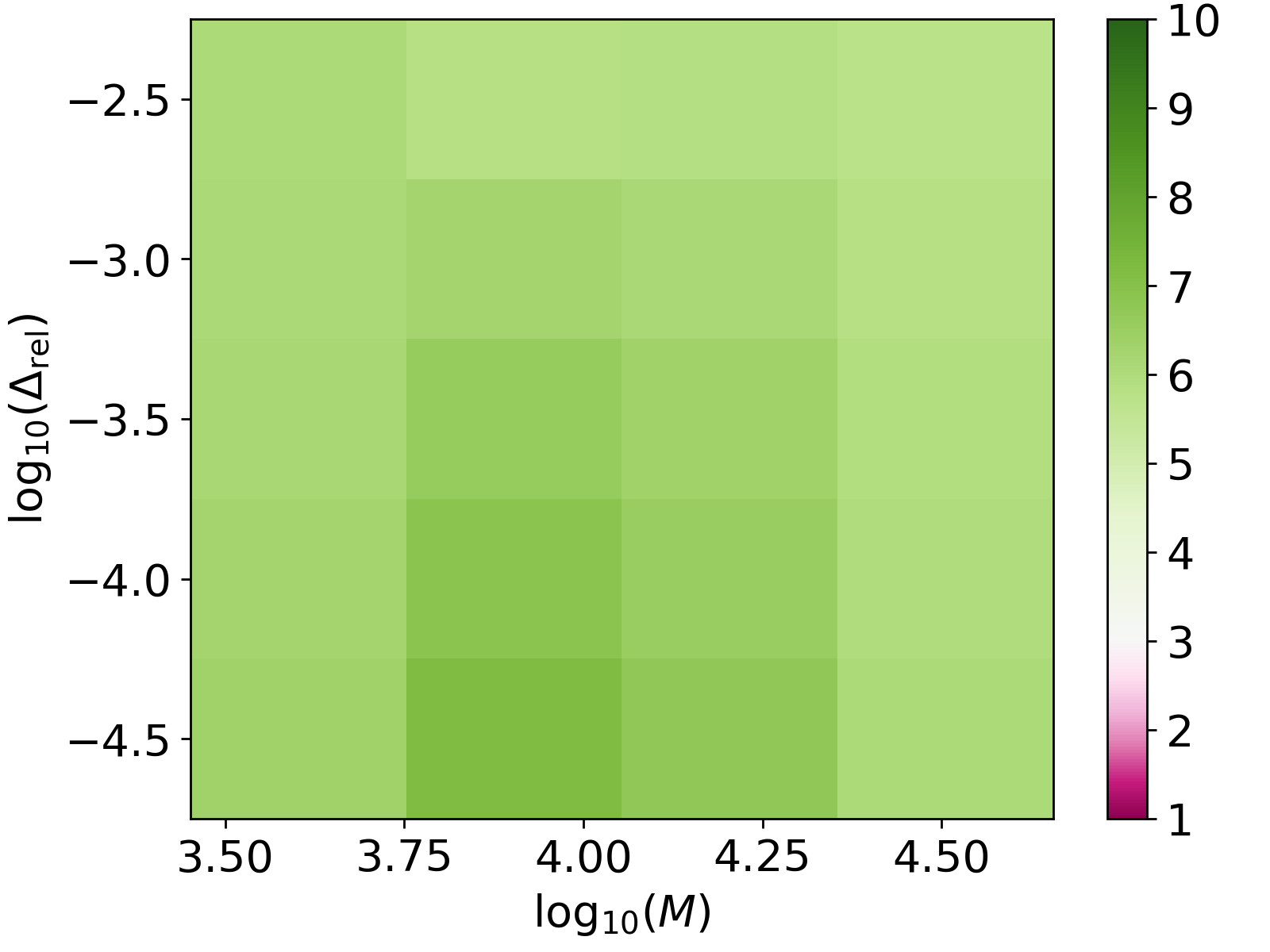}
        \caption{}
        \label{fig:nstep_L200_3}
        \end{subfigure}
    \caption{Ratio of steps $n_{\rm{SR}}/n_{\rm{RGN}}$ as a function of $(\Delta_{\rm{rel}},M)$ for the 1D $J_1$-$J_2$ model with $D=5$ MPS. (a) $L=100$, $\delta=0.01$. (b) $L=200$, $\delta=0.01$. (c) $L=100$, $\delta=0.001$. (d) $L=200$, $\delta=0.001$. }
    \label{fig:nstep_1D}
\end{figure}

\begin{figure}[htb]
    \centering
    \begin{subfigure}{0.49\linewidth}
        \centering
        \includegraphics[width=1\linewidth]{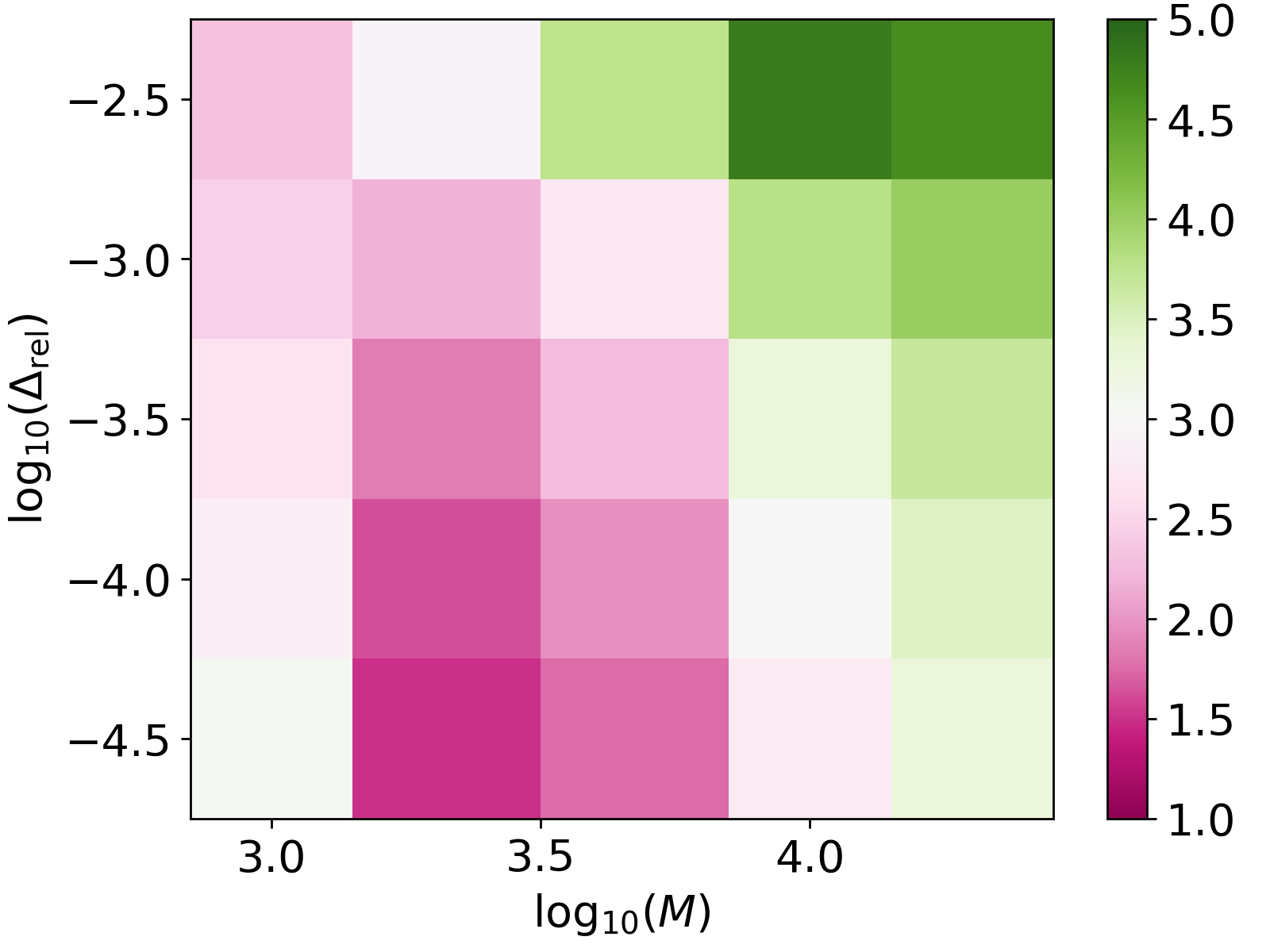}
        \caption{}
        \label{fig:nstep_6x6_D4}
    \end{subfigure}
    \centering
    \begin{subfigure}{0.49\linewidth}
        \centering
        \includegraphics[width=1\linewidth]{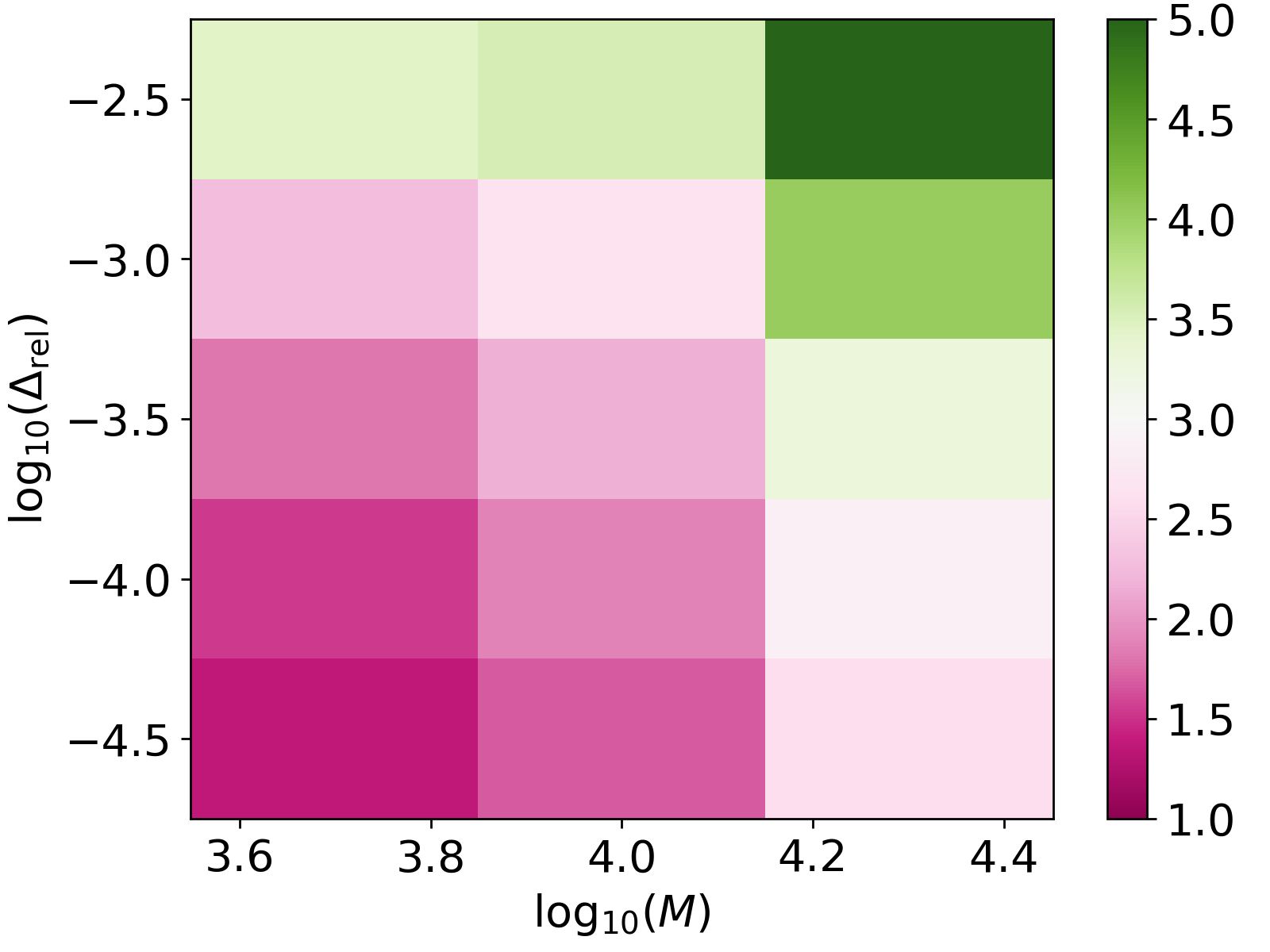}
        \caption{}
        \label{fig:nstep_8x8_D4}
    \end{subfigure}  
    \centering
    \begin{subfigure}{0.49\linewidth}
        \centering
        \includegraphics[width=1\linewidth]{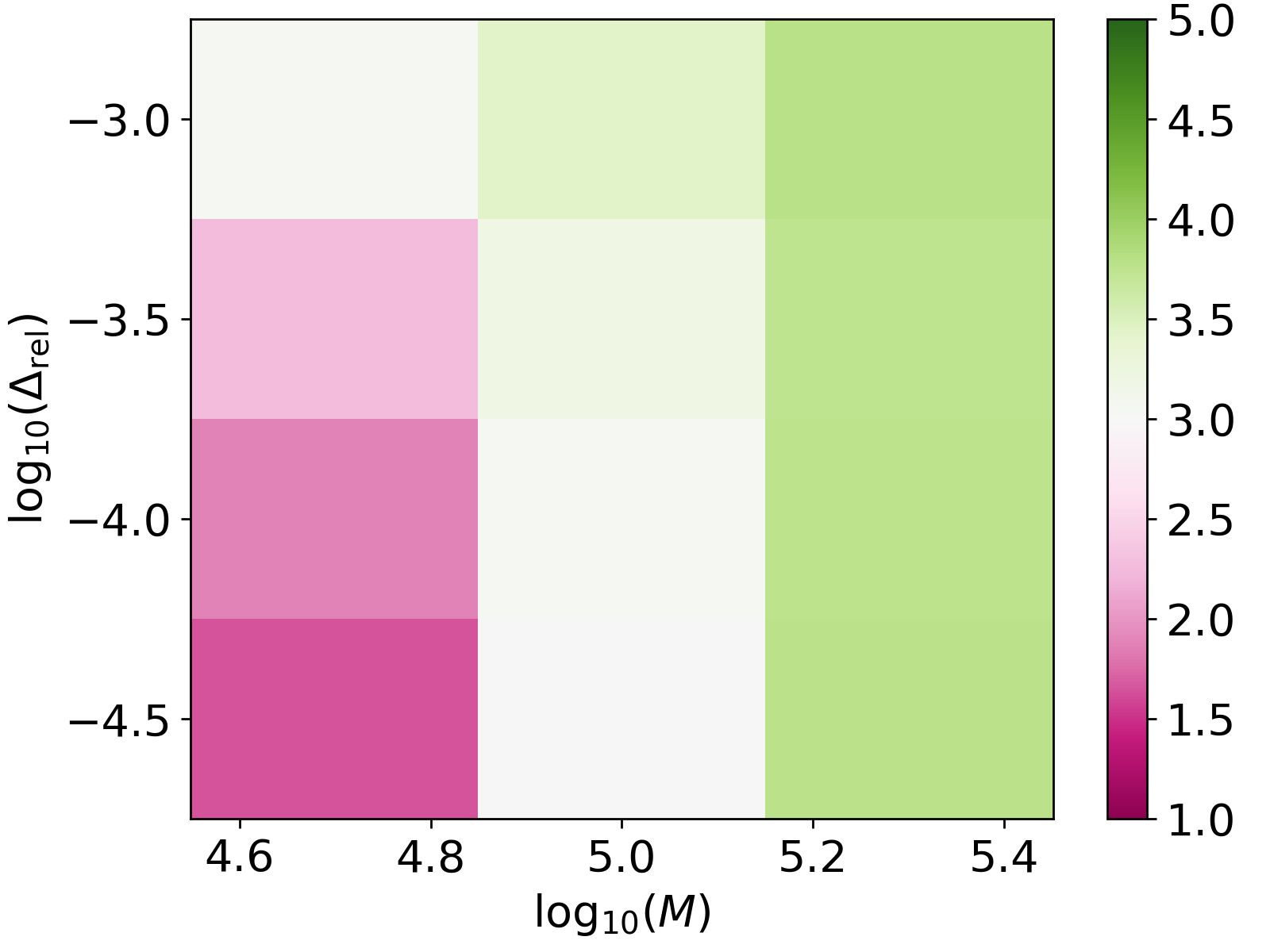}
        \caption{}
        \label{fig:nstep_8x10_D4}
        \centering
    \end{subfigure}  
        \begin{subfigure}{0.49\linewidth}
        \centering
        \includegraphics[width=1\linewidth]{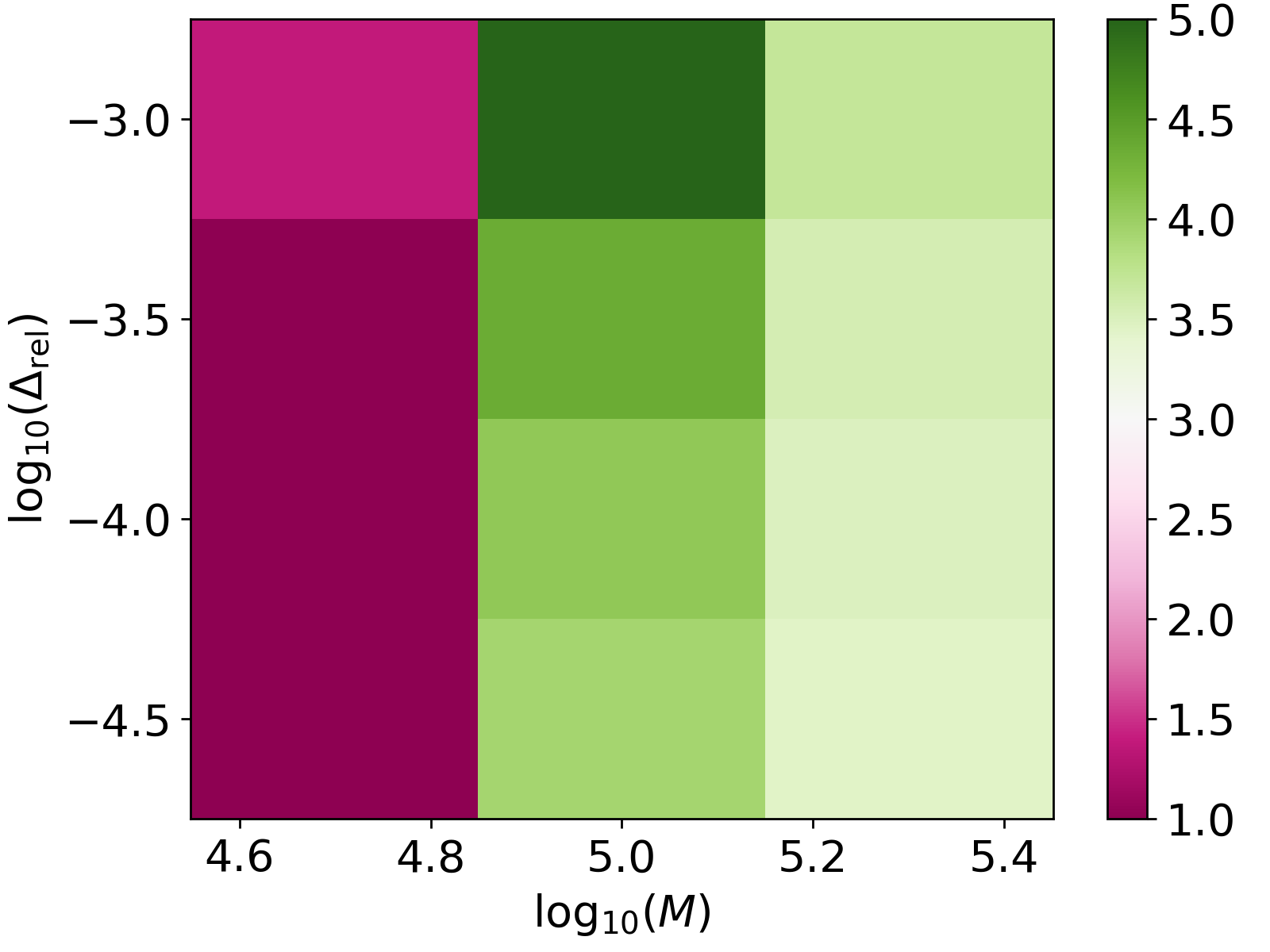}
        \caption{}
        \label{fig:nstep_10x10_D4}
        \end{subfigure}
    \caption{Ratio of steps $n_{\rm{SR}}/n_{\rm{RGN}}$ as a function of $(\Delta_{\rm{rel}},M)$ for the 2D $J_1$-$J_2$ model with $D=4$ PEPS. (a) $6\times6$. (b) $8\times8$. (c) $8\times10$. (d) $10\times10$. }
    \label{fig:nstep_D4}
\end{figure}

\begin{figure}[htb]
    \centering
    \begin{subfigure}{0.49\linewidth}
        \centering
        \includegraphics[width=1\linewidth]{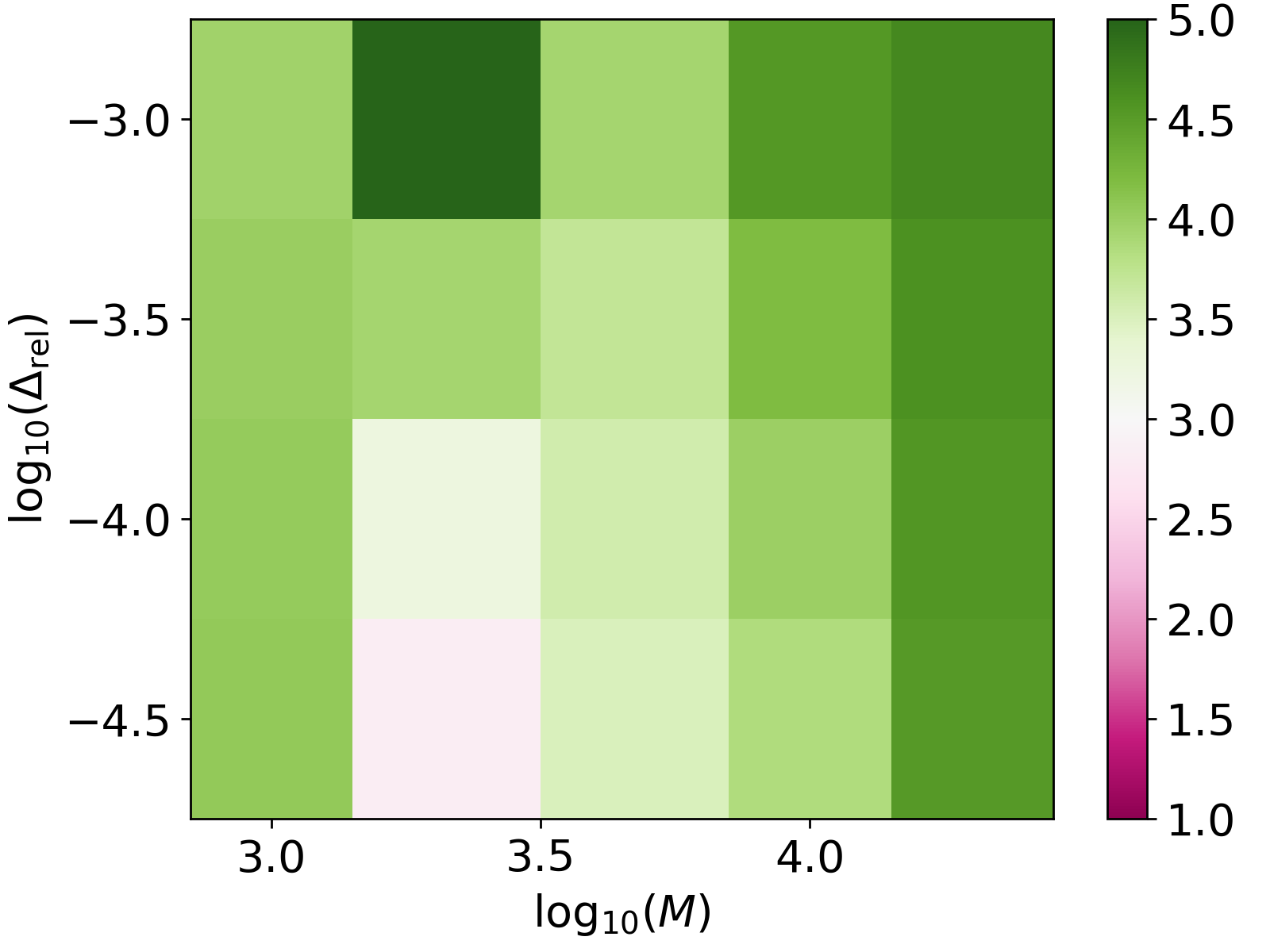}
        \caption{}
        \label{fig:nstep_6x6_D8}
    \end{subfigure}
    \centering
    \begin{subfigure}{0.49\linewidth}
        \centering
        \includegraphics[width=1\linewidth]{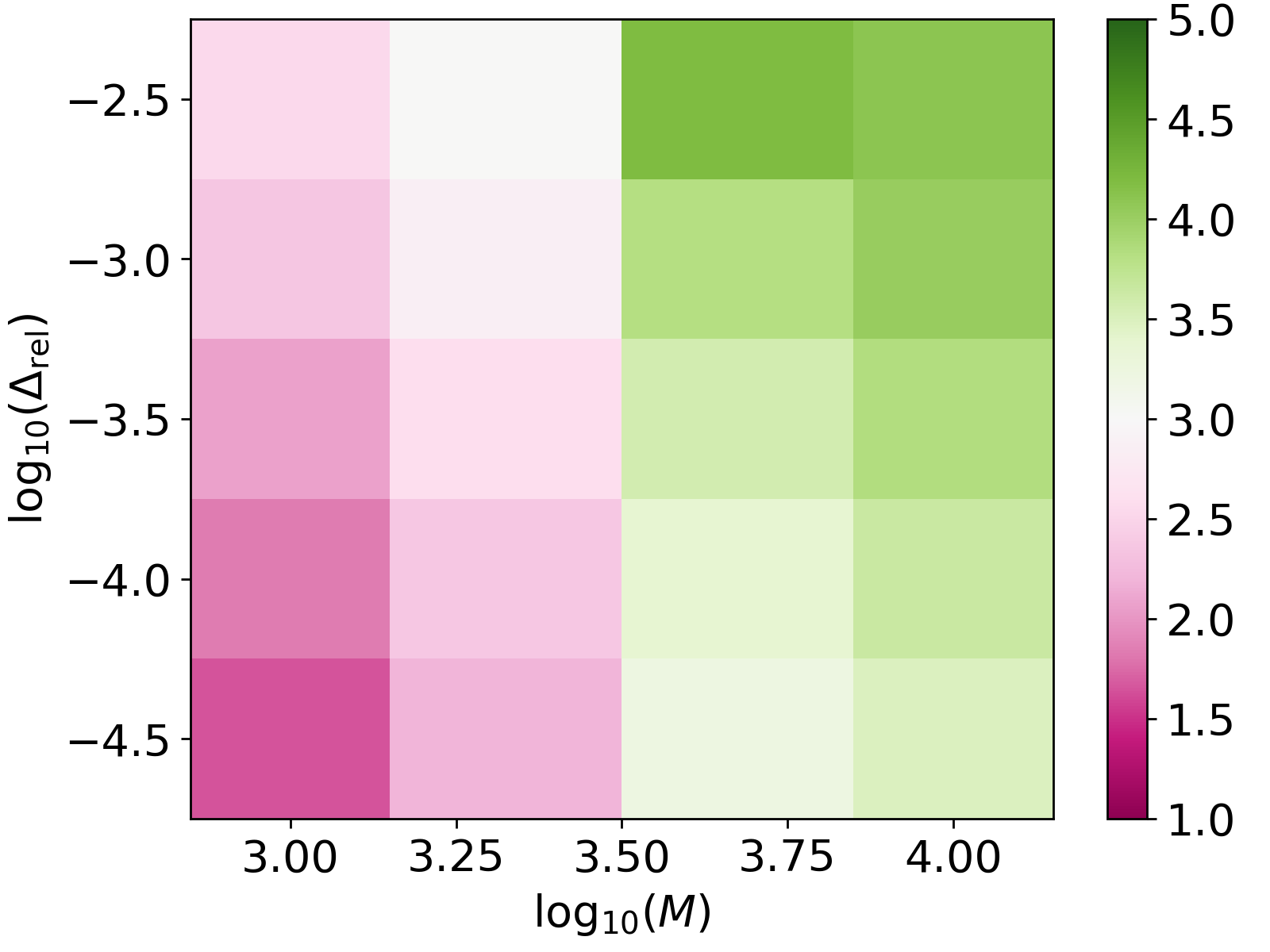}
        \caption{}
        \label{fig:nstep_8x8_D8}
    \end{subfigure}  
    \centering
    \begin{subfigure}{0.49\linewidth}
        \centering
        \includegraphics[width=1\linewidth]{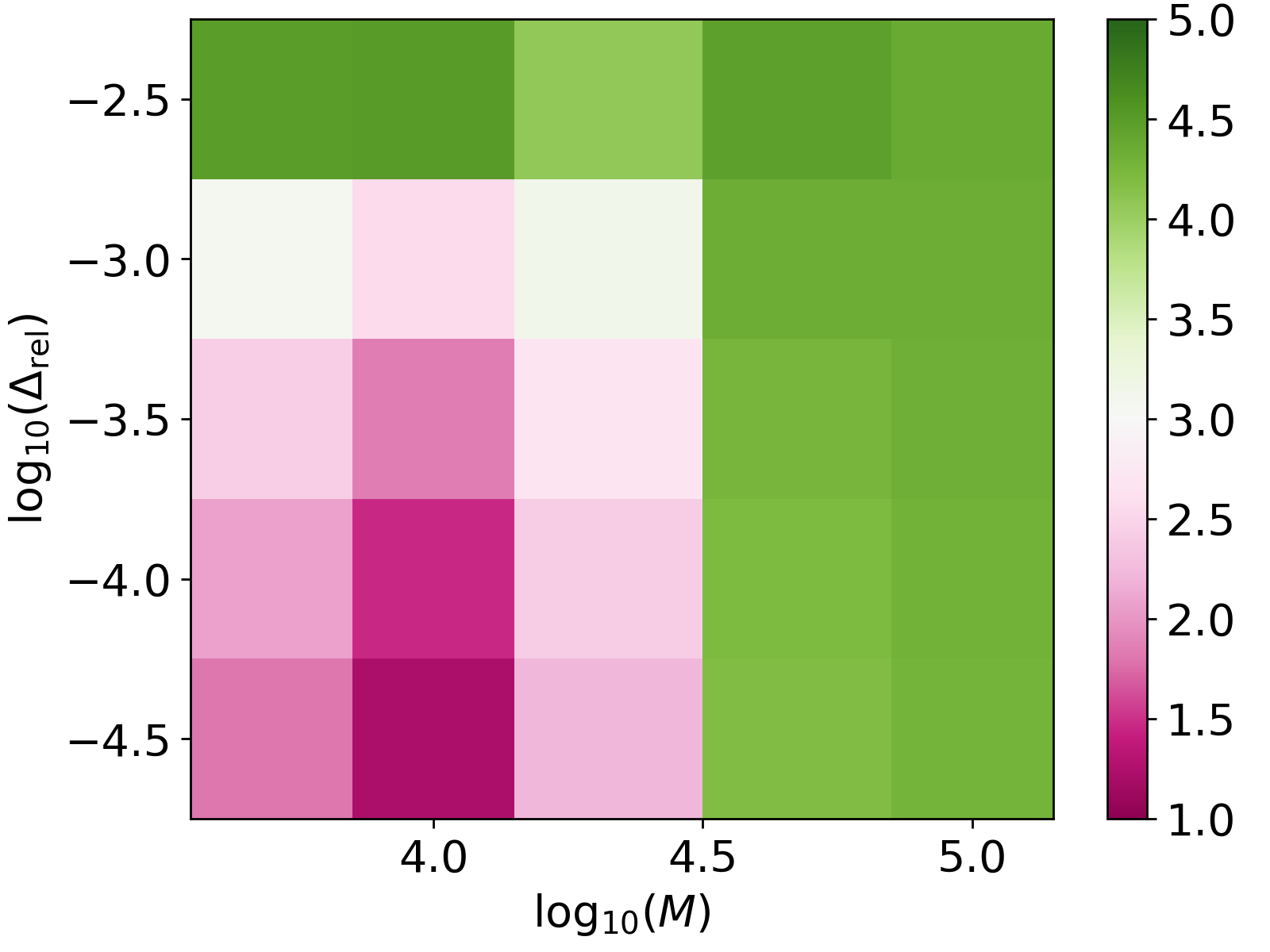}
        \caption{}
        \label{fig:nstep_8x10_D8}
        \centering
    \end{subfigure}  
        \begin{subfigure}{0.49\linewidth}
        \centering
        \includegraphics[width=1\linewidth]{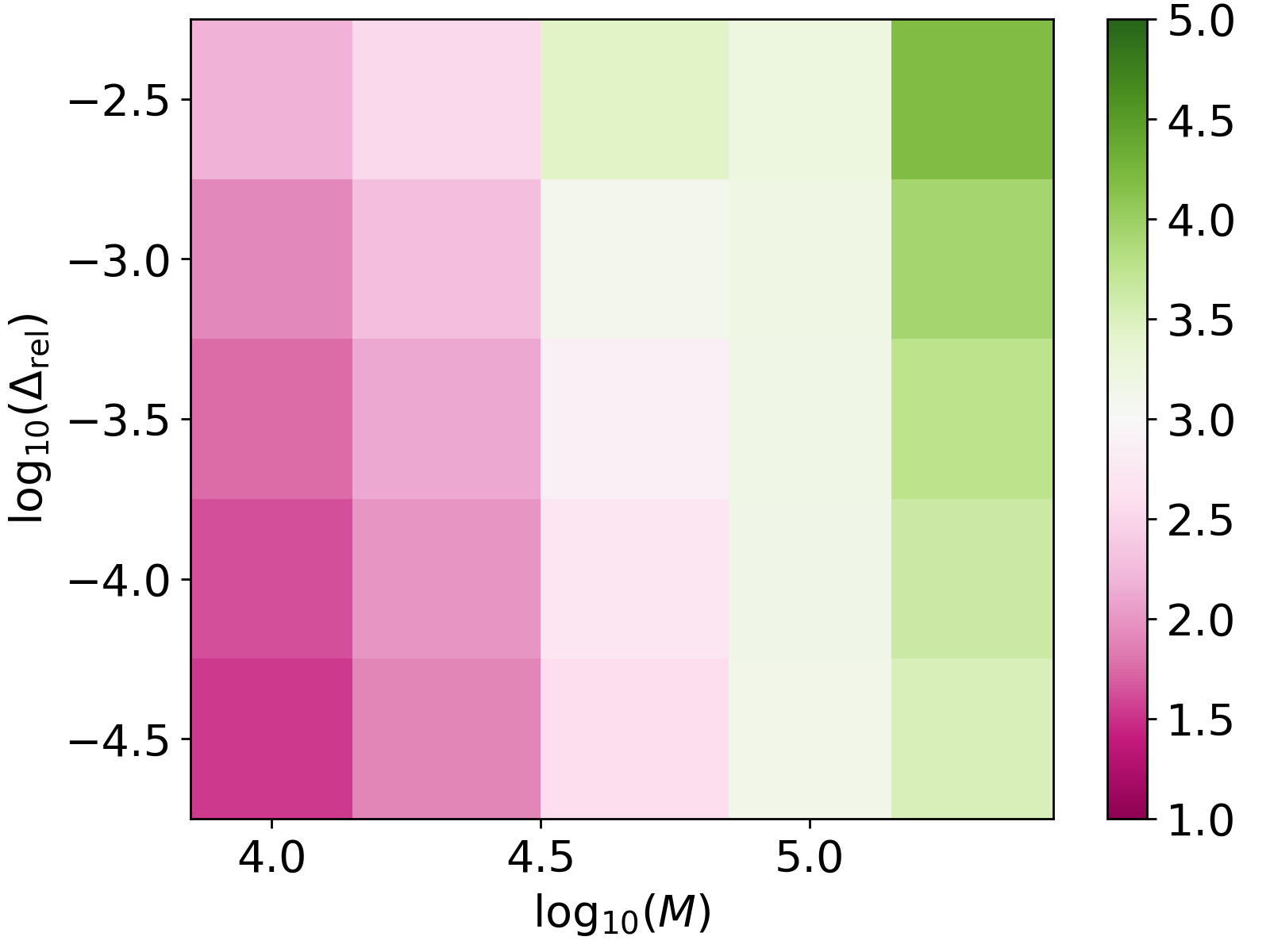}
        \caption{}
        \label{fig:nstep_10x10_D8}
        \end{subfigure}
    \caption{Ratio of steps $n_{\rm{SR}}/n_{\rm{RGN}}$ as a function of $(\Delta_{\rm{rel}},M)$ for the 2D $J_1$-$J_2$ model with $D=8$ PEPS. (a) $6\times6$. (b) $8\times8$. (c) $8\times10$. (d) $10\times10$. }
    \label{fig:nstep_D8}
\end{figure}

We now summarize our observations in relation to the questions raised in Section~\ref{sec:vmc} regarding the effect of wavefunction parametrization, wavefunction quality, and system size, on the sample size requirements and performance of RGN and SR updates. 

\begin{enumerate}
    \item Due to the intricate dependence of the energy landscape on the variational parameters, as well as the missing contribution from the wavefunction second derivatives, RGN usually cannot be expected to achieve superlinear convergence in energy even without sampling error, as can be seen from, e.g.,  Fig.~\ref{fig:1d_opt_exact} for the 1D $J_1$-$J_2$ model. Therefore, assuming that both SR and RGN have sufficient samples, although RGN will usually converge more quickly, the number of steps required by SR is expected to only be a constant factor times that of RGN, even as we increase the convergence threshold. 
    
    \item In the case of the sampling requirements, we first consider the case when the wavefunction ansatz is sufficiently expressive to contain the true ground state. Then, for an efficient RGN update, the sample size requirement is significantly affected by the closeness of the current wavefunction to the ground state (as measured by the relative energy error $\Delta_{\rm{rel}}=|(E-E_{\rm{g.s.}})/E_{\rm{g.s.}}|$). When $\Delta_{\rm{rel}}$ is large, the stochastic error in the sampled approximate Hessian determines the RGN performance. Assuming a variance in the eigenvalues $\lambda_i(H)$ of $O(L^2)$ (as we explicitly computed in the noninteracting model, and as seems reasonable in the other lattice models), an efficient RGN update then requires a sample size of $M\sim L^2$, as seen in Fig.~\ref{fig:1step_model} for the product model and Fig.~\ref{fig:1step_J1J2} for the 1D $J_1$-$J_2$ model. As $\Delta_{\rm{rel}}$ decreases, the sample size requirement for an efficient RGN update also significantly decreases, as seen in the major portion of the optimization trajectories for all 3 models studied (Fig.~\ref{fig:model300_opt}, Fig.~\ref{fig:J1J2_opt_3}, and Fig.~\ref{fig:6x6_D8}), until the RGN update becomes limited by the stochastic error in energy.
    
    On the other hand, {again }assuming the wavefunction is sufficiently expressive to contain the true ground state, the sample size requirement for an efficient SR update does not appear to depend on the closeness of the current wavefunction to the ground state (as measured by $\Delta_{\rm{rel}}$). 
    
    \item The sample size requirements for both SR and RGN behave quite differently, however, if the wavefunction ansatz is not sufficiently expressive to contain  the exact ground-state. This can be understood in terms of the zero-variance principle, which no longer applies, as well as from the scaling of the variance of eigenvalues $\lambda_i(S)$ and $\lambda_i(H)$ which may contain additional system size dependence (as explicitly worked out in the noninteracting model). This is numerically demonstrated in Fig.~\ref{fig:D1} (1D $J_1$-$J_2$ model with $D=1$ MPS), where the variational minimum of the ansatz is very far from the ground state. In this case, both SR and RGN require sharply increasing sample size with system size for stable updates. Fig.~\ref{fig:6x6}, Fig.~\ref{fig:D8}, and Fig.~\ref{fig:D4} (2D $J_1$-$J_2$ model with PEPS) consider the case when the variational minimum of the ansatz is closer to, but still insufficient to contain the ground state. In this case, the RGN performance strongly degrades if the sample size does not satisfy the required sharp system size scaling, whereas the SR performance has a milder dependence on sample size. 
   
\end{enumerate}

In a practical setting, the most pressing question to answer would be when to choose a first-order or a quasi-second-order method. One way to summarize the information we have gathered is shown in Fig.~\ref{fig:nstep_1D}, Fig.~\ref{fig:nstep_D4}, and Fig.~\ref{fig:nstep_D8}, which use the optimization trajectories computed above to estimate the ratio of the steps $n_\mathrm{SR}/n_{\rm{RGN}}$ needed to reach a given energy accuracy $\Delta_{\rm{rel}}=|(E-E_{\rm{min}})/E_{\rm{min}}|$ (measured relative to the variational minimum for the given parametrization), for different numbers of samples $M$. If we take the cost per RGN step to be roughly 3 times that of an SR step, this ratio corresponds to white in the chosen color scheme. Red then indicates that, for a given $M$, the total cost of the number of SR steps is less than that of the RGN steps, and green, vice versa. Naturally, we want to use as few samples as possible. Therefore, if the left side of the plot (corresponding to the smallest number of samples) is green, then RGN gives a speedup over SR, otherwise SR should be used. 

We see across the cases shown that if the wavefunction is sufficiently expressive, it is always cheaper (in our examples) to use RGN. This is the case for the 1D $J_1$-$J_2$ model (Fig.~\ref{fig:nstep_1D}) where the variational wavefunction includes the exact ground-state, and also in the 2D $J_1$-$J_2$ model for $D=8$ for the smaller lattice sizes (Fig.~\ref{fig:nstep_6x6_D8}). Once the variational wavefunction becomes less expressive, the leftmost part of the plots is always red, indicating that it is always more cost effective to use SR. Thus, expressivity may be taken as the determining factor for when to use quasi-second-order versus first-order methods. Note that requiring the wavefunction to be expressive is not the same as requiring that we start very close to the ground-state; as we have observed, when the wavefunction is expressive, even when starting far from the minimum, subsequent optimization steps quickly take one into a regime where the RGN update outperforms the SR update. The importance of the expressivity in distinguishing the methods is reminiscent of the strong zero-variance principle of the Linear Method described earlier~\cite{toulouse2007,PhysRevLett.98.110201,nightingale2001,toulouse2008,neuscamman2016,becca_sorella_2017}, which achieves an exact step so long as the current wavefunction and space of derivatives contains the exact ground-state. 

\section{Conclusions}\label{sec:conclusion}

In this work, we investigated the performance of first and quasi-second-order update methods in VMC, represented by the stochastic reconfiguration (SR) and Rayleigh-Gauss-Newton (RGN) methods respectively, across a variety of systems, and with respect to multiple factors, including the sample size requirements of the different algorithms and the dependence on wavefunction quality and wavefunction expressivity. We note that the absence of the expensive wavefunction second derivative, together with the iterative formulation used in this work, mean that the cost of a RGN step is only a small constant factor (e.g. 3 times) larger than that of a SR step with the same sample size. Although the presence of multiple factors led to a somewhat complex analysis (e.g. is it better to perform a noisy RGN step versus a well converged, with respect to sample size, SR step?), the ultimate picture that emerges is simple: (1) quasi-second-order methods achieve faster convergence than first-order methods, but do not achieve second-order convergence in practical settings, but at best a constant factor reduction in the number of steps. 
(2) Computing a faithful quasi-second-order step carries higher sampling requirements, due to the variance properties of the Hessian. However, quasi-second-order methods benefit greatly from an analog of the zero-variance principle for the Hessian; as shown explicitly in an analytical model, the Hessian variance becomes independent of system size for the exact ground-state. This property of the Hessian manifests rapidly even when starting far away from the ground-state, so long as the wavefunction is expressive enough to contain the ground-state, as the optimization steps rapidly decrease the variance. 
(3) Because the cost of a well sampled quasi-second-order step is then only a constant factor larger than that of a well sampled first-order step, the (constant) factor faster convergence of the quasi-second-order method wins out. Thus, the primary criterion for when to use a quasi-second-order method is not closeness to the variational minimum (as is the case for deterministic methods), but the overall expressivity of the wavefunction. 
(4) For sufficiently expressive wavefunctions (i.e. those that include the ground-state in the manifold), quasi-second-order methods thus lead to an overall reduction of cost over first-order methods. 

The analysis and conclusions in this work were drawn from specific lattice models and with specific classes of variational states. However, we believe these conclusions extend more broadly, because they reflect the scaling behaviour of SR and RGN with system size, and the Hamiltonians (sums of local terms) and wavefunctions (with product structure) reflect the generic structure of Hamiltonians and wavefunctions of large physical systems. Nonetheless, it is an important future direction to empirically verify these findings in other quantum many-body problems.

More broadly, for wave functions with more than a few thousand variational parameters, genuine second-order methods have seen only limited usage in VMC calculations so far due to the perception of high cost per sample. In this respect, the development of quasi-second-order methods, which have a similar cost per sample to first-order methods, presents a favorable alternative. However, their practical utility has remained constrained by the substantial sampling requirement. The findings of this work suggest that the high sample requirement of quasi-second-order methods is only an issue if the wavefunction is itself not sufficiently expressive. Given the increasing interest in types of wavefunction ansatz, such as tensor networks and neural network states, that are arbitrarily improvable (and therefore arbitrarily expressive) we therefore see an increasing role for second-order optimization methods. It is also worth noting that our implementation has favoured simplicity and other algorithmic enhancements, for example, related to stepsize control, should be revisited in the large number of parameters setting. Finally, we note that variational quantum algorithms are closely related to VMC, thus the findings of this work are relevant to them also.

\section{Acknowledgments}
This work was primarily supported by the US National Science Foundation via Award No. CHE-2102505. RP received an Eddleman Graduate Fellowship. GKC is a Simons Investigator in Physics and a member of the Institute for Quantum Information and Matter. We thank Cyrus Umrigar for valuable discussions.

\section{Appendix}

\subsection{Iterative solution of linear equation for SR and RGN}\label{sec:iterative_linea_eqn}

{Here we briefly recapitulate the iterative solution of the linear equations (Eq.~\ref{eq:invert_sr} and Eq.~\ref{eq:invert_rgn}) for SR and RGN in the stochastic setting with a large number of parameters, following Ref.~\onlinecite{neuscamman2012}. For SR, we solve Eq.~\ref{eq:invert_sr} by the MINRES~\cite{doi:10.1137/0712047,doi:10.1137/100787921} algorithm, where each iteration requires one matrix-vector multiplication between the $S+\delta I$ matrix and update vector $p$. From Eq.~\ref{eq:ovlp_sampled}, this can be written as
\begin{align}\label{eq:sparse_Sp}
\sum_{j=1}^{n_v}S_{ij}p_j&=\sum_{j=1}^{n_v}\left(\frac{1}{n_s}\sum_{\vec{x}}v_i(\vec{x})v_j(\vec{x})-v_iv_j\right)p_j\notag\\
&=\frac{1}{n_s}\sum_{\vec{x}}v_i(\vec{x})\left(\sum_{j=1}^{n_v}v_j(\vec{x})p_j\right)-v_i\sum_{j=1}^{n_v}v_jp_j
\end{align}
where $n_v$ is the number of variational parameters, $n_s$ the number of samples (which can be much smaller than $n_v$), and $v_i=\langle v_i(\vec{x})\rangle_{|\psi^2|}$. The leading computational cost of the matrix-vector multiplication is $O(n_sn_v)$, and the overall scaling of the MINRES algorithm is thus $O(nn_sn_v)$, where $n$ is the number of iterations. 

Similarly for RGN, we solve Eq.~\ref{eq:invert_rgn} by the LGMRES~\cite{doi:10.1137/0907058,doi:10.1137/S0895479803422014} algorithm, where each iteration requires one matrix-vector multiplication between the RGN matrix $H+(S+\delta I)/\epsilon$ and $p$. From Eq.~\ref{eq:hess_sampled}, the matrix-vector multiplication $Hp$ can be written as 
\begin{align}
\sum_{j=1}^{n_v}H_{ij}p_j&=\sum_{j=1}^{n_v}\left(\frac{1}{n_s}\sum_{\vec{x}}v_i(\vec{x})h_j(\vec{x})\right)p_j\notag\\
&-\sum_{j=1}^{n_v}\left(v_ih_j+g_iv_j+ES_{ij}\right)p_j\notag\\
&=\frac{1}{n_s}\sum_{\vec{x}}v_i(\vec{x})\left(\sum_{j=1}^{n_v}h_j(\vec{x})p_j\right)\notag\\
&-v_i\sum_{j=1}^{n_v}h_jp_j-g_i\sum_{j=1}^{n_v}v_jp_j-E\sum_{j=1}^{n_v}S_{ij}p_j
\end{align}
where $h_j=\langle h_j(\vec{x})\rangle_{|\psi^2|}$ and matrix-vector multiplication $Sp$ can be computed as in Eq.~\ref{eq:sparse_Sp}. The dominant scaling of the LGMRES algorithm is thus $O(kmn_sn_v)$, where $k$ is the number of restarts, and $m$ the size of the Krylov subspace, assuming $m\ll n_s$. 
}

\subsection{Model exact variances}\label{sec:model_exact_variance}

Here we provide analytical expressions for the variance $\sigma^2[E]$, $\sigma^2[g_i]$, $\sigma^2[S_{ij}]$ and $\sigma^2[H_{ij}]$ of the model problem in Sec.~\ref{sec:model}. For the energy
\begin{align}
\sigma^2[E]&=\langle E_L(x)^2\rangle-E^2=\sum_i\left(1-e_i^2\right).
\end{align}
To compute analytical variances for $g_i$, $S_{ij}$ and $H_{ij}$, we use error propagation as introduced in Ref.~\onlinecite{becca_sorella_2017}. Let $f(\{\bar{y}_i\})$ be a function of sampled expectation values $\{\bar{y}_i\}$, and we expand around the exact expectation values $\{\mu_i\}$
\begin{align}
f(\{\bar{y}_i\})&=f(\{\mu_i\})+\sum_i\partial_if\Delta_i\notag\\&+\frac{1}{2}\sum_{ij}\partial_{ij}f\Delta_i\Delta_j+O(\Delta^3),
\end{align}
where $\Delta_i=\bar{y}_i-\mu_i\sim O(1/\sqrt{M})$ is the estimated statistical error in the sampled expectation value $\bar{y}_i$ with $M$ samples. The sample bias in $f$
\begin{align}
\langle f(\{\bar{y}_i\})\rangle-f(\{\mu_i\})=\frac{1}{2}\sum_{ij}\partial_{ij}f\langle\Delta_i\Delta_j\rangle+O(\Delta^3)
\end{align}
is of order $O(1/M)$ hence can be ignored. The sample variance in $f$
\begin{align}
\sigma_M^2[f]
&=\langle f(\{\bar{y}_i\})^2\rangle-\langle f(\bar{y}_i)\rangle^2\notag\\
&=\sum_{ij}\partial_if\partial_jf\langle\Delta_i\Delta_j\rangle+O(\Delta^3).
\end{align}
In the following, we define $\sigma^2[f]=M\sigma^2_M[f]$ and ignore the $O(\Delta^3)$ contribution. Therefore, rewriting Eq.~\ref{eq:grad_sampled} as 
\begin{align}
g_i=f(\bar{y}_1,\bar{y}_2,\bar{y}_3)&=\bar{y}_1-\bar{y}_2\bar{y}_3
\end{align}
where $\bar{y}_1=\langle E_L(x)\nu_i(x)\rangle$, $\bar{y}_2=E$ and $\bar{y}_3=\langle \nu_i(x)\rangle$ gives 
\begin{align}
\sigma^2[g_i]=\sum_k\left(1-e_k^2\right)+O(1).
\end{align}
Similarly, from Eq.~\ref{eq:ovlp_sampled} and Eq.~\ref{eq:hess_sampled}, one gets
\begin{align}
\sigma^2[S_{ii}]&=\frac{4}{e_i^2}-4\\
\sigma^2[S_{ij}]&=1
\end{align}
\begin{align}
\sigma^2[H_{ii}]&=\left(\frac{4}{e_i^2}-4\right)\sum_{k}\left(1-e_k^2\right)+O(1)\\
\sigma^2[H_{ij}]&=\sum_{k}\left(1-e_k^2\right)+O(1)
\end{align}

\subsection{1D $J_1$-$J_2$ model}

\subsubsection{Computational procedure for 1-step $\Delta_ME/\Delta E$}\label{sec:1step_J1J2}

\begin{figure*}[htb]
    \centering
    \begin{subfigure}{0.43\linewidth}
        \centering
        \includegraphics[width=1\linewidth]{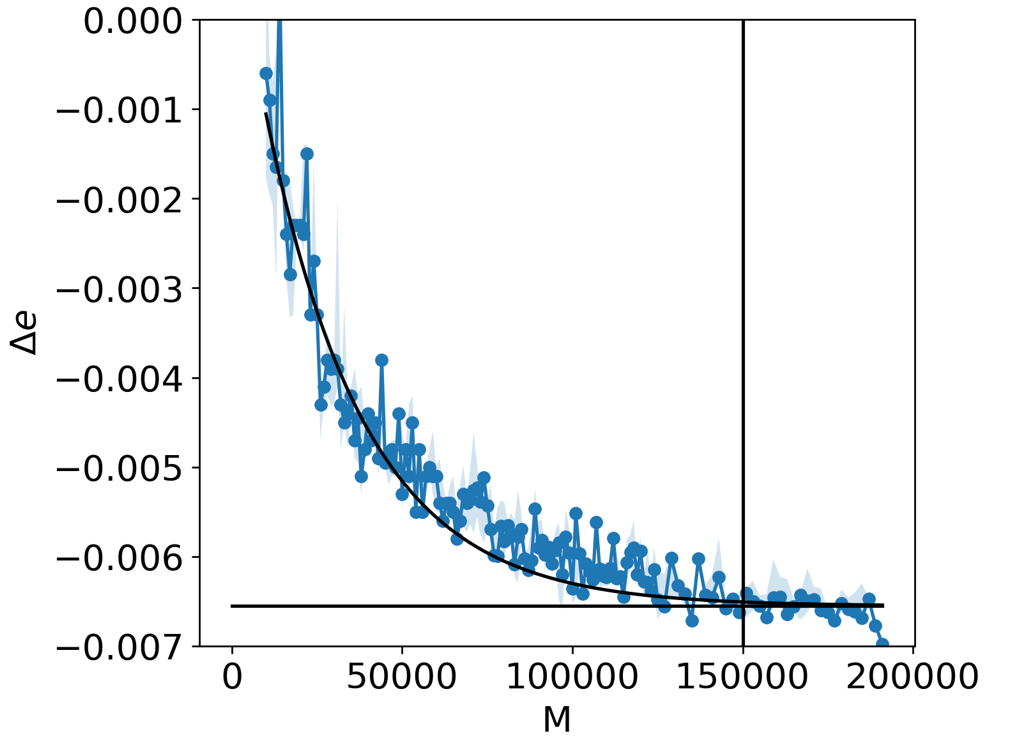}
        \caption{}
        \label{fig:p5_c3_rgn_L100_1}
    \end{subfigure}
    \centering
    \begin{subfigure}{0.43\linewidth}
        \centering
        \includegraphics[width=1\linewidth]{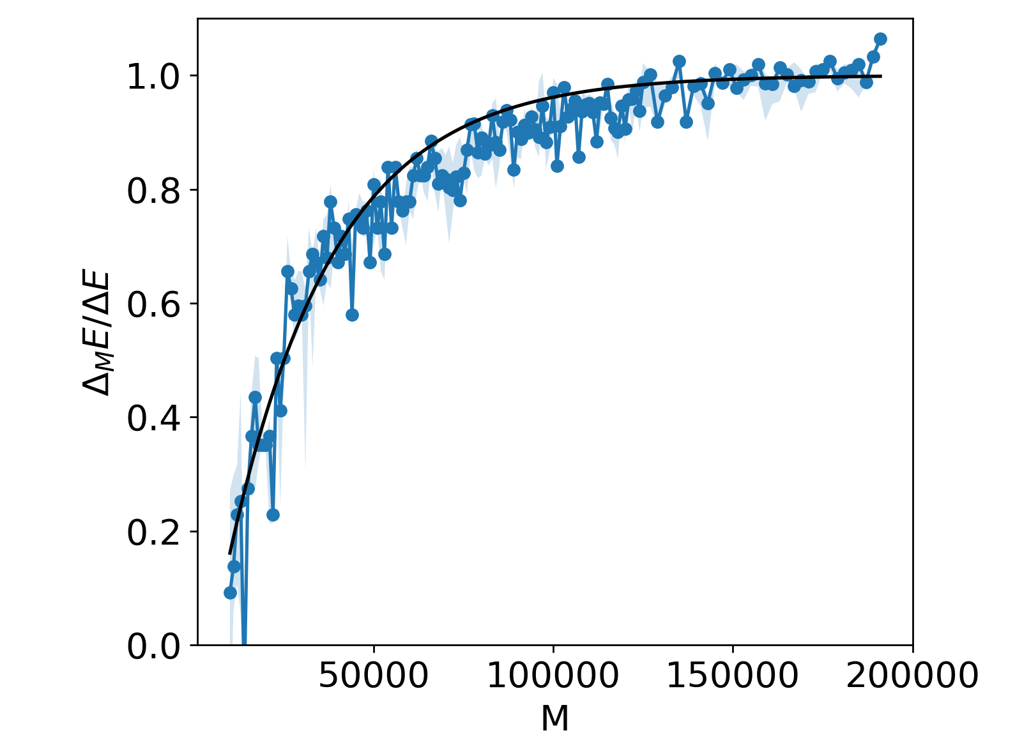}
        \caption{}
        \label{fig:p5_c3_rgn_L100_2}
    \end{subfigure}
    \caption{1D $J_1$-$J_2$ model 1-step energy difference at $\Delta_{\rm{rel}}=5.15\pm0.05\%$ for $L=100$ and $\delta=0.001$: (a) $\Delta_M e$ as a function of sample size $M$. The horizontal black line represents converged $\Delta_Me$ used to approximate the exact $\Delta e$. (b) $\Delta_ME/\Delta E$ as a function of $M$. The black curve represents the fitted $y=1-b\exp{(-aM)}$.}
    \label{fig:J1J2_c3_rgn_L100}
\end{figure*}

Here we provide the procedure to obtain Fig.~\ref{fig:1step_J1J2}. To obtain a wavefunction of a given per site energy, we randomly initialize a MPS of bond dimension $D=5$. We run imaginary time evolution on the MPS in the form of the simple update method \cite{jiang2008} with time step 0.1 for 20 steps, to obtain a state with roughly $\Delta_{\rm{rel}}=6\%$. We then use VMC with the SR update and adjust $\epsilon$ to obtain the MPS with the desired $\Delta_{\rm{rel}}=5.15\pm0.05\%$. We further note that we choose to only sample the $\sum_iS^z_i=0$ sector of the wavefunction in the VMC calculations. 

Given an MPS of length $L$ and desired quality, we compute the RGN $\Delta_Me=\Delta_ME/L$ for increasing $M$ with $\delta=0.01$ and $\delta=0.001$. Fig.~\ref{fig:p5_c3_rgn_L100_1} shows an example of $L=100$ and $\delta=0.001$. The data is very noisy due to the singularity in the $S$ and $H$ matrices as a result of the redundant degrees of freedom in the MPS. Furthermore, one cannot analytically compute the exact 1-step $\Delta e$. Therefore, we use large $M$ to converge the 1-step $\Delta_Me$, which we use to approximate the exact 1-step $\Delta e$. For instance, in Fig.~\ref{fig:p5_c3_rgn_L100_1}, we take $\Delta e$ as the median of $\Delta_Me$ for all $M\geq150000$, as labeled by the horizontal black line. The data points $(M,\Delta_ME/\Delta E)$ are then fitted to the functional form $y=1-b\exp{(-aM)}$ as shown in Fig.~\ref{fig:p5_c3_rgn_L100_2}. The fitted functions $y=1-b_L\exp{(-a_LM)}$ for each $L$ are used to generate Fig.~\ref{fig:1step_J1J2}. 

\subsubsection{Optimization trajectory with $\delta=0.01$}\label{sec:delta001_J1J2}

\begin{figure*}[htb]
    \centering
    \begin{subfigure}{0.43\linewidth}
        \centering
        \includegraphics[width=1\linewidth]{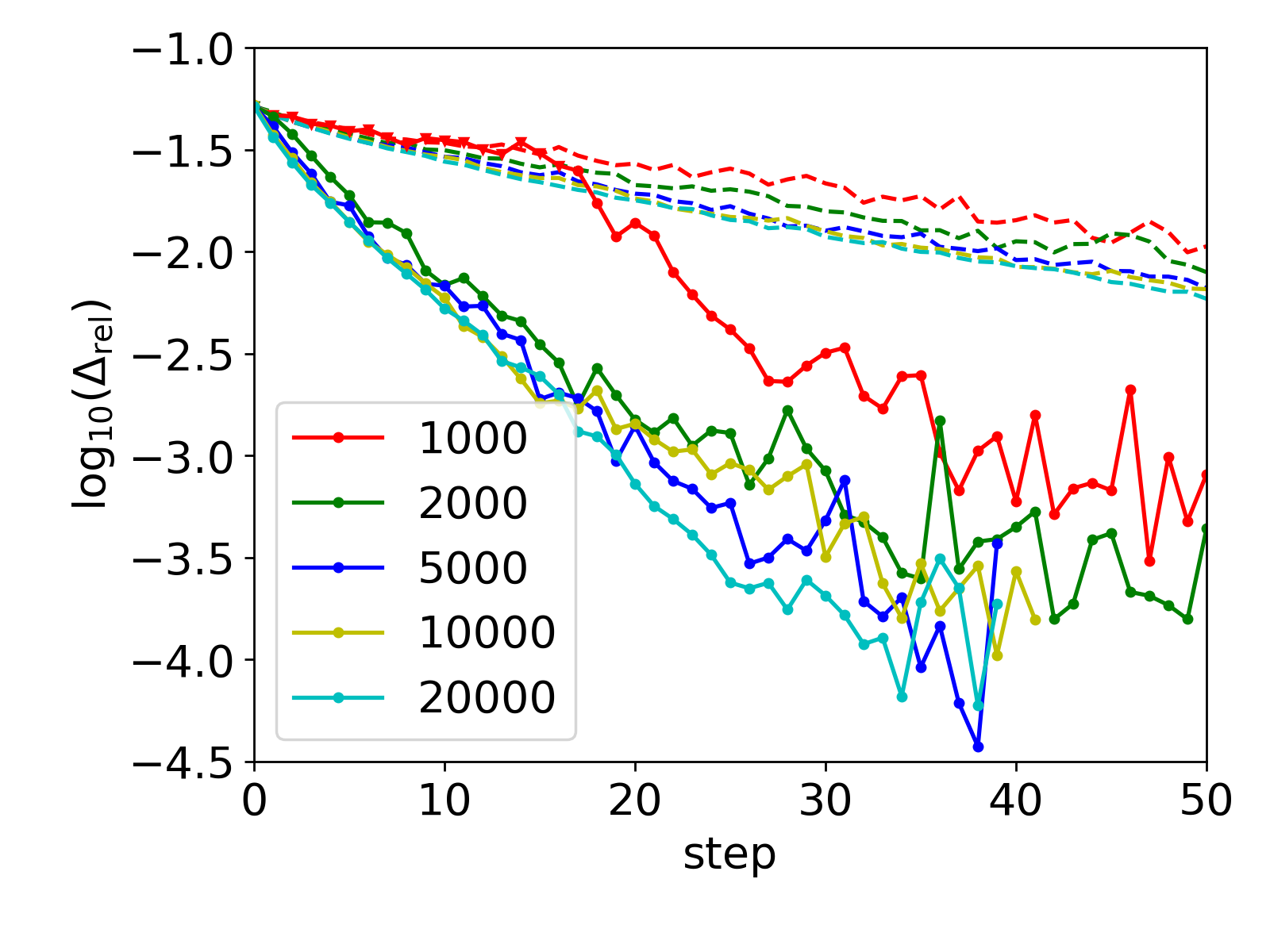}
        \caption{}
        \label{fig:J1J2_100_2}
    \end{subfigure}
    \centering
    \begin{subfigure}{0.43\linewidth}
        \centering
        \includegraphics[width=1\linewidth]{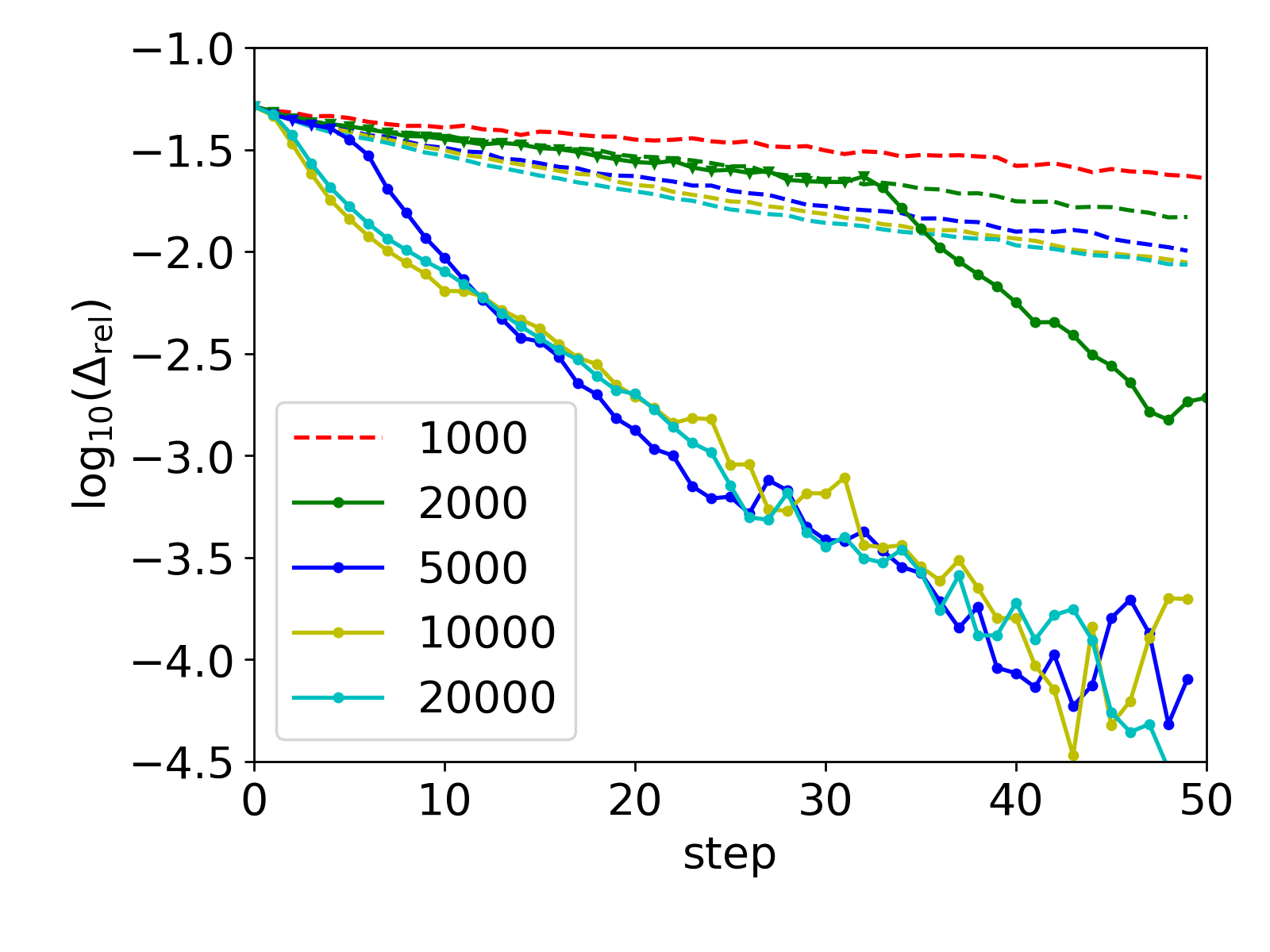}
        \caption{}
        \label{fig:J1J2_200_2}
    \end{subfigure}
        \centering
    \begin{subfigure}{0.43\linewidth}
        \centering
        \includegraphics[width=1\linewidth]{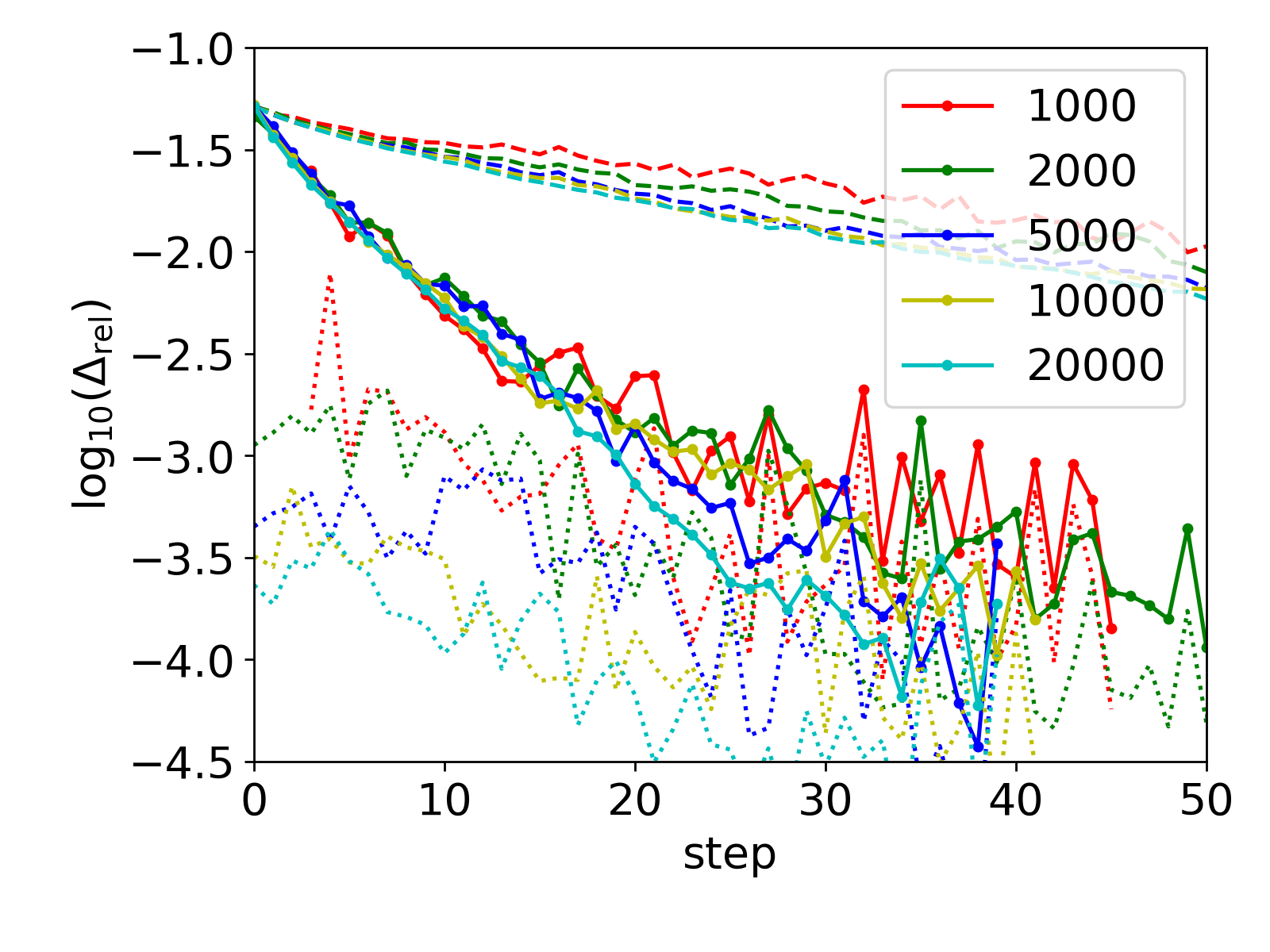}
        \caption{}
        \label{fig:J1J2_100_2_shift}
    \end{subfigure}
    \centering
    \begin{subfigure}{0.43\linewidth}
        \centering
        \includegraphics[width=1\linewidth]{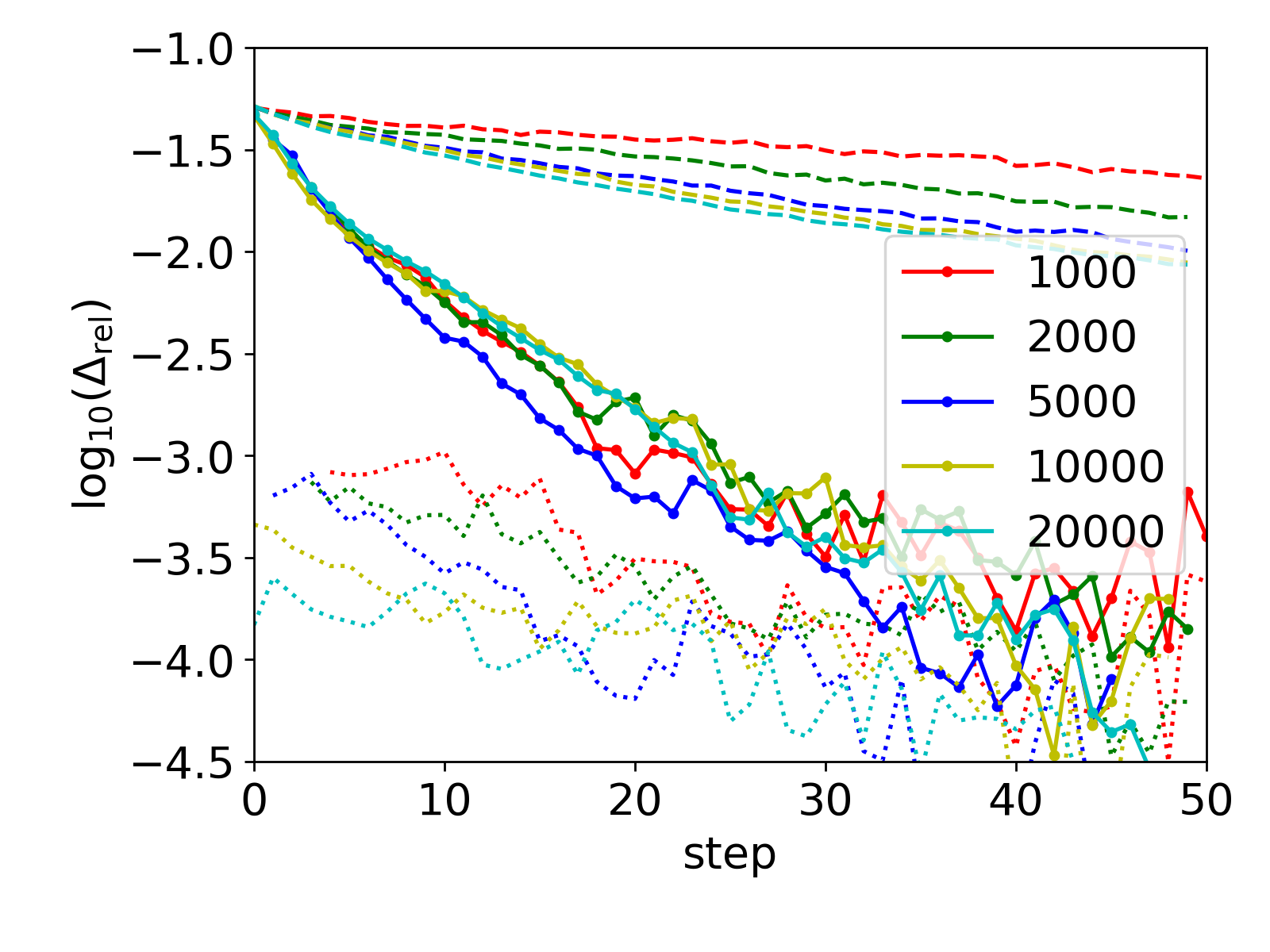}
        \caption{}
        \label{fig:J1J2_200_2_shift}
    \end{subfigure}
    \caption{Optimization trajectories for the 1D $J_1$-$J_2$ model at (a) $L=100$ and (b) $L=200$. In all plots, we use $\delta=0.01$, and each color corresponds to a sample size $M$. SR (RGN) results are plotted in dashed curves (solid curves with markers). (c) and (d) compares the slopes of the trajectories in (a) and (b) respectively, by plotting the same data as (a) and (b), but with the RGN trajectories shifted so that the first successful RGN step of each trajectory is aligned. The relative energy stochastic error $\sigma_M/|E_{\rm{g.s.}}|$ for each RGN trajectory is also plotted in the corresponding color with dotted lines.}
    \label{fig:J1J2_opt_2}
\end{figure*}

Fig.~\ref{fig:J1J2_100_2} and Fig.~\ref{fig:J1J2_200_2} plot the optimization trajectories for SR (dashed) and RGN (solid, with markers) at $L=100$ and $L=200$, where different sample sizes $M$ are shown in different colors. In both cases, we use $\delta=0.01$, and $\epsilon=0.1$ ($\epsilon=0.5$) for SR (RGN). Fig.~\ref{fig:J1J2_100_2_shift} and Fig.~\ref{fig:J1J2_200_2_shift} plot the shifted optimization trajectories in Fig.~\ref{fig:J1J2_100_2} and Fig.~\ref{fig:J1J2_200_2} respectively. We reiterate the observations made in Section~\ref{sec:opt_J1J2}: (i) in the low accuracy regime, i.e. $\Delta_{\rm{rel}}>0.01$, and for fixed $L$, the required sample size increases sharply to obtain a successful RGN update at larger $\Delta_{\rm{rel}}$, as can be seen from each of Fig.~\ref{fig:J1J2_100_2} and Fig.~\ref{fig:J1J2_200_2}. (ii) For a relatively constant $\Delta_{\rm{rel}}$, still in the low accuracy regime, the sample size requirement for a successful RGN update increases with system size, from comparing Fig.~\ref{fig:J1J2_100_2} and Fig.~\ref{fig:J1J2_200_2}. (iii) In the intermediate accuracy regime, i.e. $-3.5<\log_{10}(\Delta_{\rm{rel}})<-2$, the slope of RGN trajectories shows little sample size dependence. Furthermore, the same sample size shows no significant performance degradation with increasing system size, as can be seen from Fig.~\ref{fig:J1J2_100_2_shift} and Fig.~\ref{fig:J1J2_200_2_shift}. (iv) In the high accuracy regime, i.e. $\log_{10}(\Delta_{\rm{rel}})<3.5$, the converged $\Delta_{\rm{rel}}$ of each sample size saturates the relative energy stochastic error $\sigma_M/|E_{\rm{g.s.}}|$. This suggest that sampling in the high accuracy regime is limited by the stochastic error in energy, instead of in the approximate Hessian, which in turn suggests that optimization using SR at high accuracy would show similar sample size scaling. 

\subsection{2D $J_1$-$J_2$ model}\label{sec:6x8}

\begin{figure*}[htb]
    \centering
    \begin{subfigure}{0.43\linewidth}
        \centering
        \includegraphics[width=1\linewidth]{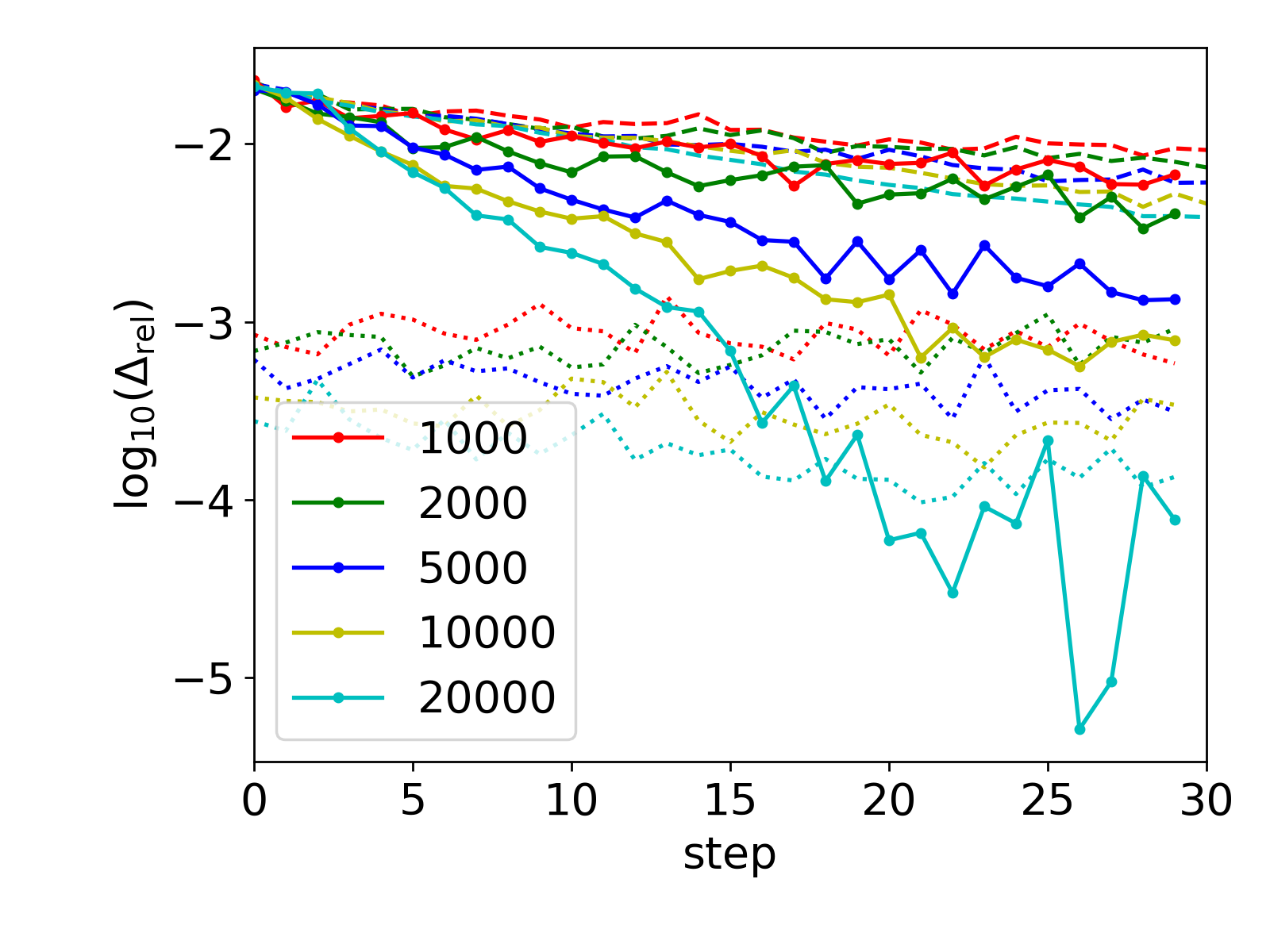}
        \caption{}
        \label{fig:8x6_D4}
    \end{subfigure}
    \centering
    \begin{subfigure}{0.43\linewidth}
        \centering
        \includegraphics[width=1\linewidth]{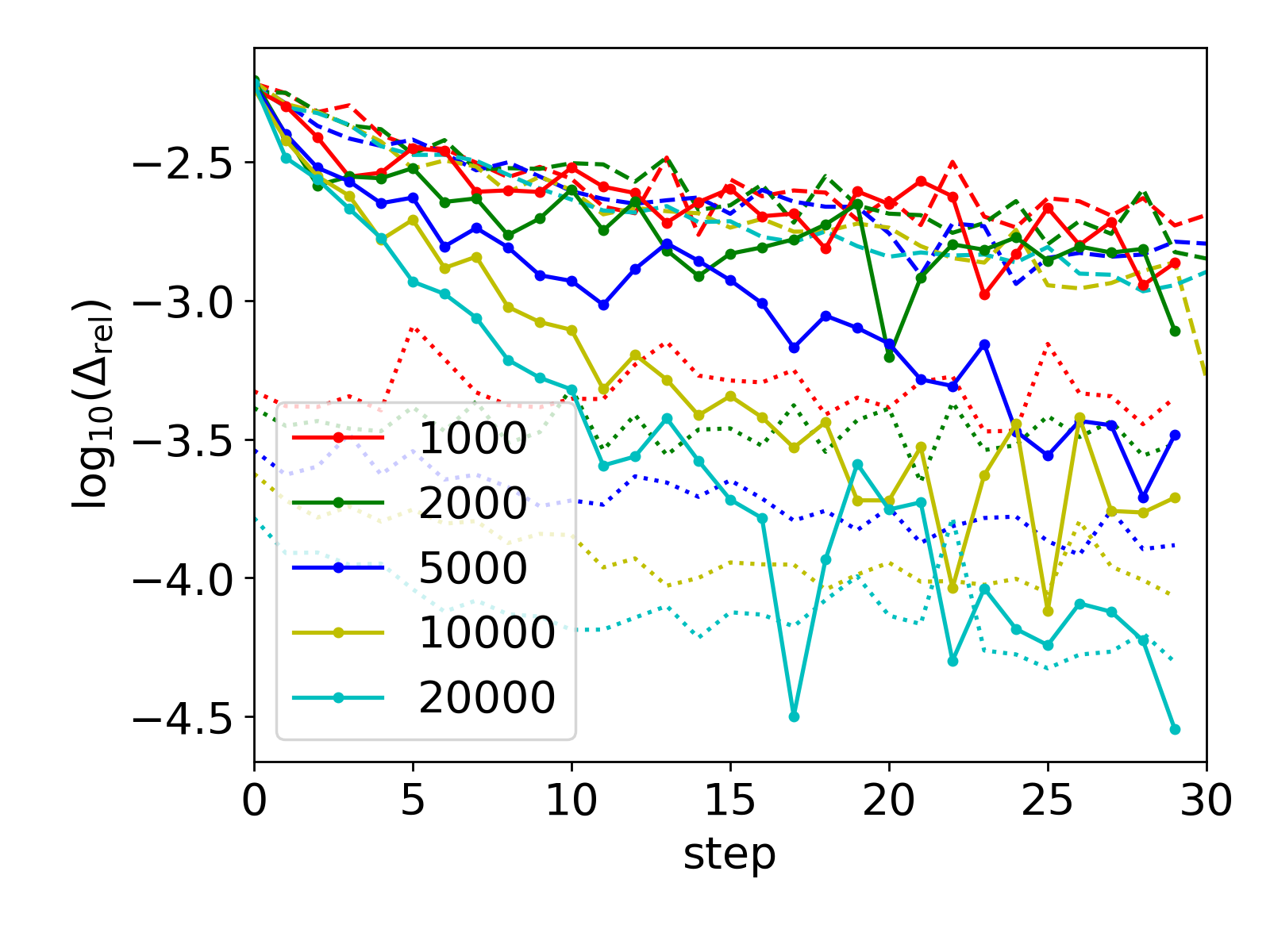}
        \caption{}
        \label{fig:8x6_D6}
    \end{subfigure}   
        \hfill
    \centering
    \begin{subfigure}{0.43\linewidth}
        \centering
        \includegraphics[width=1\linewidth]{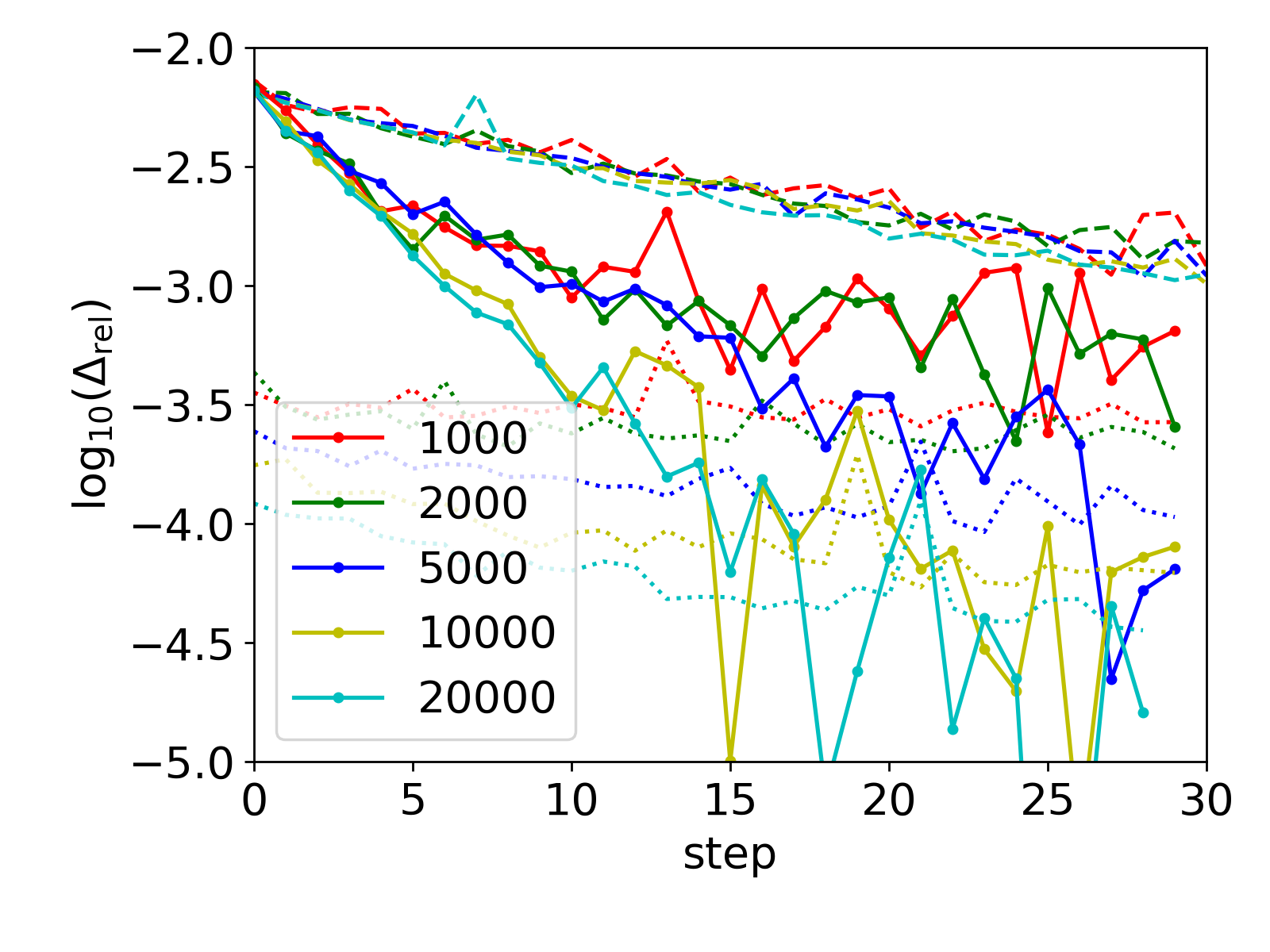}
        \caption{}
        \label{fig:8x6_D8}
    \end{subfigure}   
    \centering
    \begin{subfigure}{0.43\linewidth}
        \centering
        \includegraphics[width=1\linewidth]{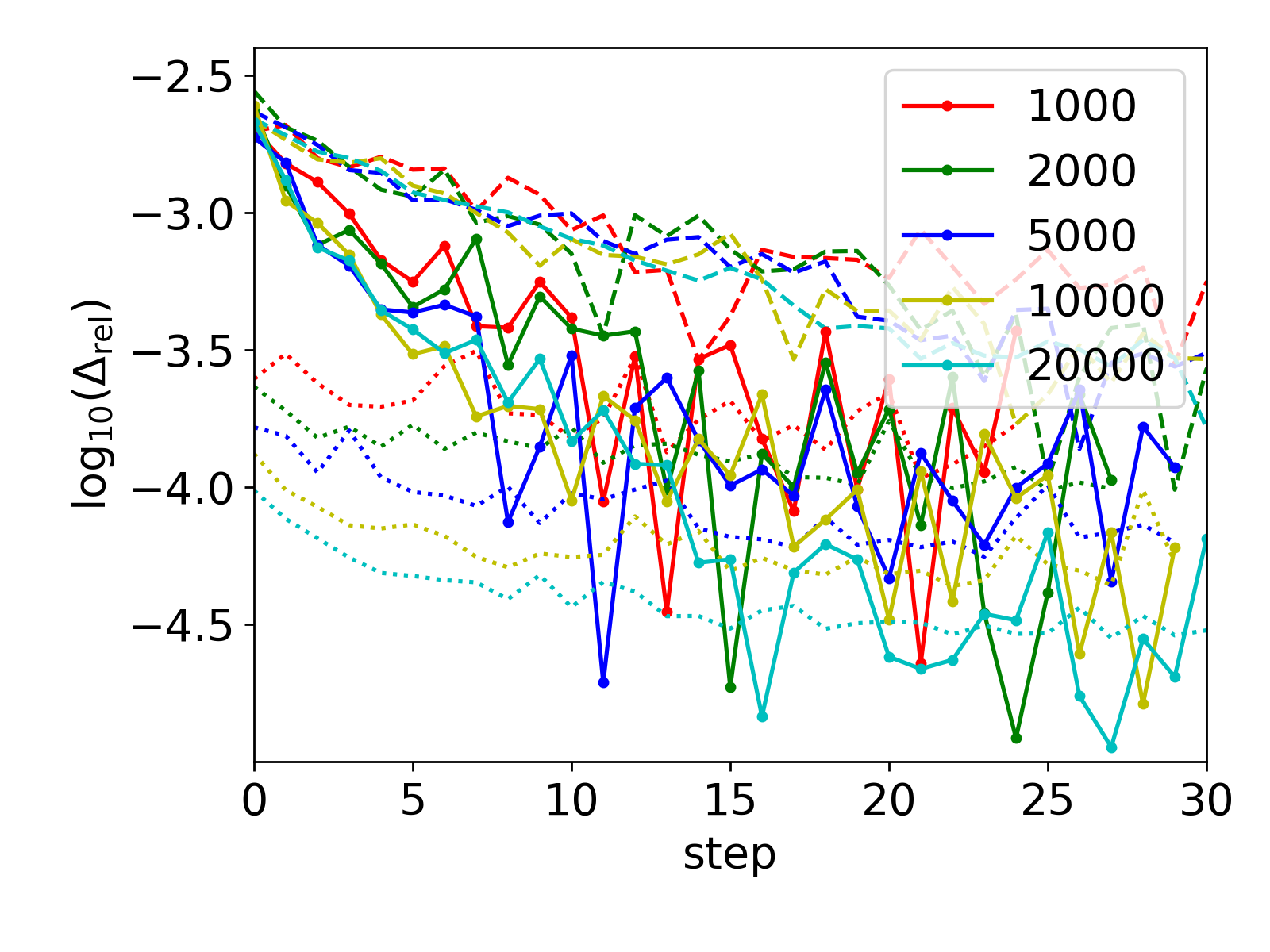}
        \caption{}
        \label{fig:8x6_D10}
    \end{subfigure}   
    \caption{$6\times8$ $J_1$-$J_2$ model optimization trajectories. with (a) $D=4$, $e_{\rm{min}}=-0.48095$. (b) $D=6$, $e_{\rm{min}}=-0.48144$. (c) $D=8$, $e_{\rm{min}}=-0.48147$. (d) $D=10$, $e_{\rm{min}}=-0.48153$. Different $M$ are labeled by their color. SR (RGN) results are shown in dashed curves (solid curves with markers). The dotted curves are the relative energy stochastic error $\sigma_M/|E_{\rm{min}}|$ of the RGN trajectories. }
    \label{fig:8x6}
\end{figure*}

Here we also supplement the observations in Section~\ref{sec:2D_D} with calculations on a $6\times8$ lattice using PEPS of bond dimension $D=4$, $D=6$, $D=8$ and $D=10$, as shown in Fig.~\ref{fig:8x6}, where the variational minimum in each case is reported in the caption and the ground state per site energy $e_{\rm{g.s.}}=-0.48156$ from DMRG~\cite{10.1063/5.0180424} calculation. We reiterate our observations in Section~\ref{sec:2D_D}: (i) when the variational minimum of the ansatz contains or is sufficiently close to the true ground state (Fig.~\ref{fig:8x6_D10} for $D=10$ PEPS), there is an intermediate accuracy region ($-3.5<\log_{10}(\Delta_{\rm{rel}})<-2.5$) along the RGN optimization trajectories where the different trajectories using different sample sizes $M$ have very similar slope. (ii) as the variational minimum of the wavefunction becomes further from the ground state (for PEPS of $D=8$, $D=6$ and $D=4$), the sample size has an increasingly important effect on the performance of both SR and RGN optimization. For RGN in particular, trajectories with different $M$ start to diverge at larger $\Delta_{\rm{rel}}$ ($\log_{10}(\Delta_{\rm{rel}})\approx-2.7$ for $D=8$, $\log_{10}(\Delta_{\rm{rel}})\approx-2.5$ for $D=6$, and the common region vanishes for $D=4$), so that a large sample size is needed for a larger portion of the optimization trajectory. 

\clearpage

\bibliography{export}

\end{document}